\documentclass[fleqn,usenatbib]{mnras}
\usepackage{newtxtext,newtxmath}
\usepackage[T1]{fontenc}
\DeclareRobustCommand{\VAN}[3]{#2}
\let\VANthebibliography\thebibliography
\def\thebibliography{\DeclareRobustCommand{\VAN}[3]{##3}\VANthebibliography}
\usepackage{graphicx}
\usepackage{amsmath}	
\usepackage{pdflscape}
\usepackage{hyphenat}

\title[Chemodynamical study of two CEMP-no stars \thanks]{Chemodynamical study of two CEMP-no stars from the Hamburg/ESO Survey \thanks{Based on data collected using Mercator/HERMES}}

\author[Shejeelammal et al.]{J. Shejeelammal$^{1,2}$, Aruna Goswami$^{1,3}$ \\
$^{1}$Indian Institute of Astrophysics, Koramangala, Bangalore 560034,
India;  shejeela@usp.br, aruna@iiap.res.in \\ 
$^{2}$Universidade de São Paulo, Instituto de Astronomia, Geofísica e Ciências Atmosféricas, IAG, Departamento de Astronomia, \\ 
Rua do Matão 1226, Cidade Universitária, 05508-900, SP, Brazil \\
$^{3}$ Institute of Frontier Science and Application, Bangalore, India\\
 }

\date{Accepted XXX. Received YYY; in original form ZZZ}

\pubyear{2023}

\begin{document}
\label{firstpage}
\label{lastpage}
\pagerange{\pageref{firstpage}--\pageref{lastpage}}
\maketitle

\begin{abstract}
The Carbon-Enhanced Metal-Poor (CEMP) stars with no enhancement of neutron-capture elements, 
the so-called CEMP-no stars are believed to be the direct descendants of first-generation stars 
and provide a unique opportunity to probe the early Galactic nucleosynthesis. We present 
a detailed chemical and kinematic analysis for two extremely metal-poor stars  HE1243$-$2408 and HE0038$-$0345  
using high-resolution (R${\sim}$86,000) HERMES spectra. For the object HE1243$-$2408, we could 
make a detailed comparison with the available literature values; however, only limited information 
is available for the other object HE0038$-$0345. Our estimated metallicity for these two objects 
are $-$3.05 and  $-$2.92  respectively. With estimated [C/Fe]  (1.03 and 1.05) and [Ba/Fe] 
($-$0.18 and $-$0.11) respectively, the objects are found to be bonafide CEMP-no stars.
From the observed abundances of C, Na, Mg, and Ba (i.e., A(C), A(Na), A(Mg), A(Ba)), 
the objects are found to belong to  Group II CEMP-no stars.
A detailed abundance profile analysis indicates that the objects 
are accreted from dSph satellite galaxies that support hierarchical Galaxy assembly. Further, 
our analysis shows that the progenitors of the stars are likely Pop II Core-Collapse Supernovae. 
The object HE0038$-$0345 is found to be a high-energy, prograde, outer-halo object, 
and HE1243$-$2408 is found to be a high-energy, retrograde, inner-halo object. 
Our detailed chemodynamical analysis shows that HE1243$-$2408 is related to I'itoi structure, 
where as HE0038$-$0345 is likely related to Sgr or GSE events. 
The mass of the progenitor galaxies of the program stars inferred from their dynamics 
is at par with their likely origin in massive dSph galaxies.
\end{abstract}
 
 
\section{Introduction} \label{section_introduction}
The CEMP-no stars, a sub-class of Carbon-Enhanced Metal-Poor (CEMP) stars with enhanced carbon 
abundance and  no signatures of heavy neutron-capture elements, are the most  chemically 
primitive  objects presently known \citep{Noris_2019, Yoon_2020}. These stars occupy the lowest 
metallicity tail ([Fe/H]$<$$-$3) of the Metallicity Distribution Function (MDF) of the Galaxy \citep{Yong_2013, Bennassuti_2017, Yoon_2019, Noris_2019}, 
and their fraction increases with decreasing metallicity, constituting $\sim$20\% of very metal-poor 
stars ([Fe/H]$<$$-$2), $\sim$40\% of extremely metal-poor stars ([Fe/H]$<$$-$3), and $\sim$75\% of ultra metal-poor 
stars ([Fe/H]$<$$-$4) \citep{Frebel_2006, Carollo_2012, Lee_2013, Placco_2014, Banerjee_2018}. Among the fourteen known CEMP stars
with [Fe/H]$<$$-$4.5, twelve stars belong to the CEMP-no group, and they might have been formed a few million years after 
the Big Bang  \citep{Noris_2019}. 
They are formed from the ejecta of the first-generation, massive, short-lived 
Pop III stars \citep{Frebel_2008, Nomoto_2013, Spite_2013, Placco_2014, Choplin_2017, Ezzedine_2019}. 
The indirect studies of the formation and nucleosynthesis pathways of the first stars, 
and their contribution to the early chemical enrichment of the Universe, can be achieved 
through the analysis of chemical imprints of their direct descendants, the CEMP-no stars \citep{Hansen_2016a, Yoon_2019}.
Hence, the CEMP-no stars are important probes in studies of Galactic archaeology that aims to 
understand the nature of the first generation stars, the associated nucleosynthesis, the 
early star formation and galaxy formation, and  early Galactic chemical evolution etc. 
\citep{Frebel_2015, Bonifacio_2015, Hansen_2016a, Yoon_2020}. Further, these stars are 
expected to shed light on the physical processes  resulting in the transitions from 
massive Pop III stars to normal Pop II stars \citep{Salvadori_2007, Hartwig_2015, Bennassuti_2017}. 

Different scenarios have been suggested by various authors for the origin of CEMP-no stars
(\citealt{Hansen_2015}, \citealt{Yoon_2016} and references there in), 
however, the exact origin of these stars  still remains unknown. According to some authors, the 
CEMP-no stars are generally 
single stars, suggesting an intrinsic origin, ie, formed from a pre-enriched natal cloud 
of ISM \citep{Spite_2013, Starkenburg_2014, Bonifacio_2015, Hansen_2016b, Yoon_2016}. 
The suggested progenitors for the CEMP-no stars include: (i) faint supernovae that underwent mixing and fall back. 
Due to the low explosion energy, such SNe could not expel all the synthesized element. Only the light elements
such as C, N, O, Na, Mg, Al, and Si are released to the ISM \citep{Umeda_2003, Umeda_2005, Tominaga_2014},
(ii) extremely metal-poor fast rotating massive stars or spinstars that enriched the ISM with C, N, and O 
through strong stellar winds \citep{Meynet_2006, Frischknecht_2012, Maeder_2015b, Choplin_2017}, and 
(iii) i-process in massive Pop III stars \citep{Clarkson_2018}. In addition to these scenarios, \cite{Arentsen_2019b} suggested that
the binary mass-transfer from extremely metal-poor AGB companion stars in the past might explain at least a few of the CEMP-no stars among the 
confirmed binary CEMP-no objects. If the binary-mass-transfer scenario holds true for 
the origin of CEMP-no stars, this will help to understand extremely metal-poor AGB stars and provide robust observational 
constraints to the AGB nucleosynthesis and mass-transfer models for metallicity [Fe/H]$<$$-$2. 

The differences in the morphology of CEMP-no stars on the Yoon - Beers diagram (A(C) - [Fe/H] diagram) 
leading to  two distinct groups, Group II and Group III CEMP-no stars, suggest
multiple formation 
pathways for them \citep{Yoon_2016}. In addition to these two groups of CEMP-no stars, a few of the 
CEMP-no stars ($\sim$ 15\%) are found to lie in the high-carbon region that is typically occupied 
by CEMP-s and CEMP-r/s stars (Group I), known as Group I CEMP-no stars. However, these CEMP-no stars 
show larger [Sr/Ba] values than those in the 
Group I CEMP-s and CEMP-r/s stars \citep{Noris_2019}. Eight objects among these Group I CEMP-no stars 
are confirmed binaries (\citealt{Yoon_2016}, \citealt{Arentsen_2019b} and references therein). 
Similarity in the morphology of the Yoon - Beers diagram for the CEMP-no stars in the 
Galactic halo and dwarf satellite galaxies suggested that the halo CEMP-no stars 
are accreted from the mini halos \citep{Bonifacio_2015, Yoon_2018, Yoon_2019, Yoon_2020}. 
Additional spectroscopic studies of Group I, Group II, and Group III CEMP-no stars in both the Galaxy and its satellite galaxies
are required to better constrain their origin, and hence the nature of first stars. 
In addition to this, such studies will help us to improve our understanding of Galactic assembly and Galactic halo formation. 
Further study is required to investigate  what fraction of these anomalous Group I CEMP-no stars are binaries,
and also to explain the origin of binary CEMP-no stars with low carbon abundance. 

Several cosmological simulations are in accordance with the hierarchical Galactic halo assembly 
through numerous complex mergers (eg., \citealt{Deason_2013, Phillepich_2014, Amorisco_2017}). 
The highly structured nature of the Galactic halo points at an accretion origin. 
This has simulated several high- and medium-resolution spectroscopic studies 
of different samples \citep{Myeong_2019, Naidu_2020, Woody_2021, Aguado_2021, Malhan_2022, Zepeda_2023}, 
and large-scale surveys (eg., GALAH; \citealt{Buder_2018}). Based on high-precision data from the Gaia Satellite,
these studies identified various structures arising from the clustering of satellite galaxies, stellar streams, and globular clusters
in the dynamic (Energy - Action) space, and attempted to uncover the Milky Way mergers. 
These structures in various chemical and dynamical planes, along with the observations of r-process stars in the halo, 
revealed the signatures of remnants of dwarf satellite galaxies accreted in the past by the Galaxy, 
and unambiguously confirmed the importance of accretion for its 
formation \citep{Myeong_2018, Roderer_2018, Dietz_2020, Helmi_2020, Woody_2021, Malhan_2022, Naidu_2022, Zepeda_2023, Frebel_2023}.  
It has been shown that the massive progenitor dwarf galaxies (M$_{\star}$ $\geq$ 10$^{8}$ - 10$^{9}$ M$_{\odot}$) contribute primarily 
to the accreted halo, with sub-dominant contributions from the low-mass satellite galaxies \citep{Dietz_2020, Naidu_2020}.  
Major massive dwarf satellite galaxy accretions that built the Galactic halo are Gaia-Sausage-Enceladus (GSE; \citealt{Helmi_2018, Belokurov_2018}), 
Kraken/Koala/Heracles \citep{Kruijssen_2019, Forbes_2020, Pfeffer_2021, Horta_2021}, and Sequoia \citep{Myeong_2019, Matsuno_2019}. 
These events happened $\sim$10 Gyrs ago at red-shifts z$\simeq$2, with more than 50\% of the halo 
stars in the solar neighbourhood coming from the GSE \citep{Helmi_2018, Kruijssen_2019, Frebel_2023}. 
Besides these major mergers, several other numerous minor accretions (both massive and low-mass) like Aleph, 
Arjuna, I'itoi, Wukong/LMS-1 \citep{Naidu_2020}, Thamnos \citep{Koppelman_2019} etc. have contributed to the Galactic halo.

Although it is known that numerous dwarf satellite galaxy accretions led to the formation of the present-day Milky Way halo, 
our understanding of this process is still lacking, particularly in terms of the total number of mergers, their
dynamical characteristics, and their contributions to various stellar populations of the halo \citep{Malhan_2022, Frebel_2023}. 
Since many metal-poor stars in the halo trace their origin to the dwarf satellite galaxies \citep{Frebel_2015, Frebel_2023}, 
they are ideal tools to study the Galactic halo assembly. 
Based on the kinematic, chemical, and dynamical properties of a sample of 644 CEMP stars, \cite{Zepeda_2023} have shown 
that the CEMP-no stars are not in-situ to the Galaxy, but were originally born in satellite galaxies that were 
accreted and disrupted by the Milky Way. This is also on par with the findings of \cite{Yoon_2019}. 
Several authors have discussed the relevance and scope of enhancing medium- and high-resolution 
spectroscopic studies of different sub-groups of CEMP-no stars to further our understanding of 
early Galactic chemical evolution and Galactic halo formation (eg. \citealt{Hansen_2015, Hansen_2016c, Noris_2019, Yoon_2019, Yoon_2020, Zepeda_2023}). 
The stars that were accreted very early into the Galaxy still retain their original chemical composition and a 
few kinematic properties such as energy, action, and angular momentum \citep{Frebel_2023}. 
\cite{Zepeda_2023} have discussed the possibility of recovering the Galactic accretion events 
through a clustering approach, and matching the chemical, orbital, and kinematic properties of CEMP stars with each other. 
Along this line, we have conducted a high-resolution spectroscopic and chemodynamical analysis of two CEMP-no stars, 
HE~0038$-$0345 and HE~1243$-$2408, with the aim of deriving clues to their origin.

This paper is arranged as follows. A brief discussion on previous studies of the stellar sample is given 
section \ref{section_previous_study}. Section \ref{section_sample} describes the method of sample selection, 
data acquisition and data reduction. Estimation of radial velocity is given in section \ref{section_RV}, 
followed by the description of stellar atmospheric parameter estimation and mass determination in section \ref{section_atmospheric_parameter}. 
Section \ref{section_abundance_determination} provides procedures adopted for abundance determination. 
Abundance uncertainty is given in section \ref{section_uncertainty}. A detailed discussion on the abundance pattern and the 
progenitors of the program stars is presented in section \ref{section_discussion}. 
Kinematic analysis of the sample is given in section \ref{section_kinematic_analysis}. 
A detailed discussion on the orbital and various dynamical properties of the program stars in the Galaxy is also 
provided in this section. In section \ref{section_structure_association}, we have discussed our chemodynamical analysis 
to assign the stars to known accretion structures of Milky Way. 
Finally, the conclusions are summarized  in  section \ref{section_conclusion}.

\section{Previous studies on the program stars: a brief summary} \label{section_previous_study}
 Based on an analysis of medium-resolution spectra ($\sim$ 2 \AA\ ) obtained with 2 - 4 m class telescopes, 
\cite{Frebel_2006} derived  metallicity 
of the program stars as $-$2.86 and $-$2.88 for HE~0038$-$0345 and HE~1243$-$2408 respectively, 
and carbon abundance with respect to iron [C/Fe], as  0.3  and 0.13 respectively. Their classification as 
metal-poor stars were determined by applying a selection procedure taking account the strength of 
the Ca II K line as a function of the metallicity and the effective  temperature. The strength of 
Ca II K line is measured following the procedure in \cite{Beers_1999}  by means of the line index KP, 
a pseudo-equivalent width measurement in {\rm \AA}. Ca II K line is known 
 as the strongest metal absorption line in the optical wavelength range. The optical photometry by \cite{Beers_2007}
 gave B$-$V = 0.767 with E(B-V) = + 0.043 for  the object  HE~0038$-$0345. \cite{Beers_2017} also derived 
 the metallicity [Fe/H] and carbonicity [C/Fe] of these objects as ($-$2.51, 0.76) and 
 ($-$2.65, 0.75) for HE~0038$-$0345 and HE~1243$-$2408 respectively, based on medium-resolution 
 spectral analysis, and classified them as CEMP-no stars. The wavelength regions  used as inputs 
 to the n-SSPP (non-SEGUE Stellar Parameter Pipeline) for determination of the parameters  span a range 
 of 3600 - 4400 {\rm \AA}, 3600 - 4800 {\rm \AA}, or 3600 - 5250 {\rm \AA}. 
 
Using spectral synthesis calculations, \cite{Holmbeck_2020} derived the metallicity [Fe/H] ($-$ 2.53) 
and abundances of  C, Sr, Ba, and Eu for the object HE~ 0038$-$0345, based on a snap-shoot high-resolution 
spectrum of HE 0038-0345 at a resolution$\sim$30,000 with a SNR$\sim$30. While carbon is found 
to be mildly 
enhanced with [C/Fe] = 0.46, Sr and Ba are found  below solar with [Sr/Fe] = $-$0.16 and 
[Ba/Fe] = $-$ 0.49. The abundance of Eu is however found as near-solar with [Eu/Fe] = 0.12.
The object is classified  as a non-RPE (no r-process enhancement) star \citep{Holmbeck_2020}. 
Our estimates of effective temperature and surface gravity match closely with those of 
\cite{Holmbeck_2020} within the error limits. However, we got lower values for the 
micro-turbulent velocity ($\Delta$$\sim$1.43 km s$^{-1}$) and metallicity ($\Delta$$\sim$0.39 dex) 
compared to them. 

Among the Fe I and II lines used for stellar atmospheric parameter determination, four Fe I lines 4889.74, 
5339.93, 5393.17, 5586.76 {\rm \AA}, and one Fe II line 4520.224 {\rm \AA} are common with those 
used by \cite{Holmbeck_2020}. In \cite{Holmbeck_2020}, the measured equivalent widths of these Fe I 
lines are almost double, and that of the Fe II line is almost 7 times as large as that measured by us. 
Besides, among the 84 Fe I lines used by them, 26 lines are with excitation potentials, E$_{low}$ $<$ 1.2 eV.  
In EMP stars, the Fe I lines with E$_{low}$ $<$ 1.2 eV are affected by 3D hydrodynamic effects and 
show significant deviation from 1D \citep{Tafelmeyer_2010, Hayek_2011, Dobrovolskas_2013, Jablonka_2015}. 
The abundances estimated from these lines are shown to be over-estimated \citep{Collet_2007, Dobrovolskas_2013, Mashonkina_2017}. 
For instance, in the analysis of \cite{Collet_2007}, in the T$\rm_{eff}$/log \,g/[Fe/H] $\sim$ 4858/2.2/$-$3 model, 
the Fe I line with E$_{low}$ = 0 eV gave an abundance $\sim$0.8 dex higher compared to the Fe I line with E$_{low}$ = 4 eV, 
whereas \cite{Mashonkina_2017} noted a difference of $\sim$0.36 dex, \cite{Dobrovolskas_2013} noted $\sim$0.5 dex. 
Since 30\% of the Fe I lines used by \cite{Holmbeck_2020}  are low E$_{low}$ lines, the Fe abundance could likely 
be over-estimated. 

Based on a VLT/UVES spectrum at resolution 40,000 and SNR$\sim$130, \cite{Lombardo_2022}
have performed an analysis of the object HE 1243-2408, and presented the stellar atmospheric parameters and 
elemental abundances of 15 elements. Our estimates of stellar atmospheric parameters are lower compared to 
\cite{Lombardo_2022}; $\Delta$(T$\rm_{eff}$, log \,g, $\zeta$, [Fe/H]) $\sim$ (223 K, 0.95, 0.74 km s$^{-1}$, 0.2). 
Though the analysis in both cases is based on 1D LTE, the adopted procedures and methodologies are different. 
The differences in the method of estimation of stellar atmospheric parameters and elemental abundances, 
differences in the choices of atomic data, solar abundance values, Fe partition functions etc. could affect
the derived values \citep{Roderer_2018}. While we have derived the stellar atmospheric parameters 
spectroscopically using the traditional excitation, ionization balances of Fe I – Fe II lines, 
\cite{Lombardo_2022} used the Gaia EDR3 photometry and parallax values, following the procedure in \cite{Koch-Hansen_2021}. 
The micro-turbulent velocity is estimated using a calibration equation from \cite{Mashonkina_2017}. 
The micro-turbulent velocity estimated this way is shown to exhibit a discrepancy of $\pm$0.5 km s$^{-1}$ 
from the spectroscopic value, reaching up-to a value of $\sim$0.7 km s$^{-1}$ \citep{Mashonkina_2017}. 
The analysis of \cite{Mucciarelli_2020} has shown that the photometric estimates of atmospheric parameters are 
higher than the spectroscopic ones, and the deviation increases with decreasing metallicity. 
At [Fe/H]$\sim$$-$2.5, the photometric temperature and surface gravity are higher than the spectroscopic values by 
$\sim$350 K and $\sim$1.0 dex, respectively \citep{Mucciarelli_2020}. This is in agreement with the 
discrepancy observed in the estimates of \cite{Lombardo_2022} and other studies in the literature. 
This could be likely due to the 3D and/or NLTE effects in the lower metallicity, which leads to 
the non-zero slope in the excitation potential and observed Fe abundances when using the photmetric 
temperature \citep{Mucciarelli_2020, Koch-Hansen_2021, Lombardo_2022}. 
However, current modelling of the giant stars incorporating these effects could not remove this 
slope completely \citep{Mucciarelli_2020}. 
Finally, the discrepancy in the estimates of stellar atmospheric parameters between our study 
 and that of \cite{Lombardo_2022} is reflected in the estimates of elemental abundances as well.

\section{STELLAR Spectra: Object SELECTION, Observation AND DATA REDUCTION} \label{section_sample}
\begin{figure}
\centering
\includegraphics[width=\columnwidth]{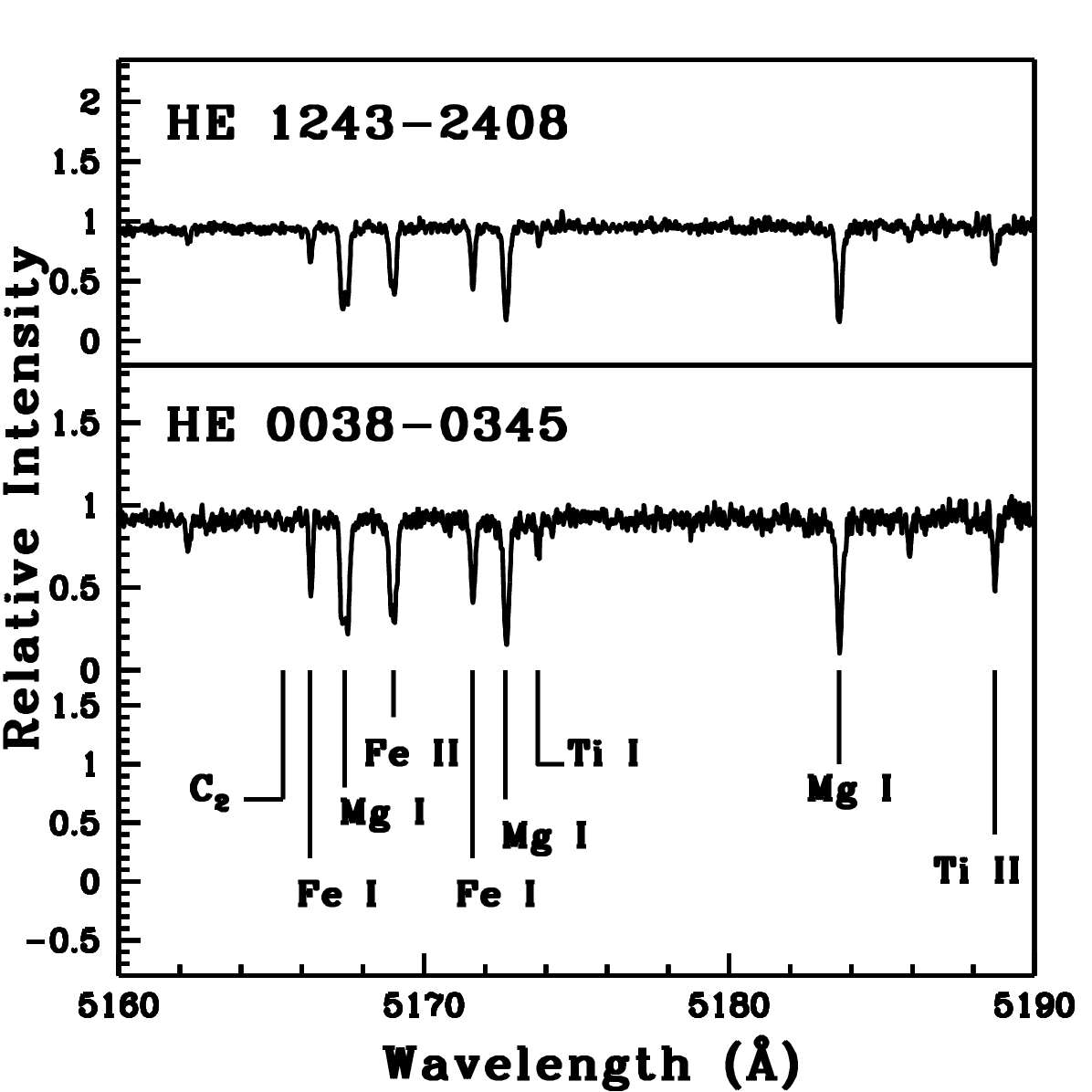}
\caption{Sample spectra of the program stars in the wavelength region 
5160 to 5190  {\rm \AA}.}\label{sample spectra}
\end{figure}

The program stars, HE~0038$-$0345 and HE~1243$-$2408, are selected from the the catalog of bright (13 $<$ B $<$ 14.5)
metal-poor candidates \citep{Frebel_2006} from the Hamburg/ESO Survey (HES). The high-resolution spectra
of the objects were acquired using the High-Efficiency and high-Resolution Mercator Echelle Spectrograph (HERMES)
attached to the 1.2m Mercator telescope at the Roque de los Muchachos Observatory in La Palma, 
Canary Islands, Spain \citep{Raskin_2011}. The spectra cover the wavelength range 
 3770 - 9000 {\rm \AA}, and have resolution R$\sim$86,000. 
For the object HE~0038$-$0345, three frames with exposures 2000, 700, and 1600 sec were taken on 
10/12/2018, 29/01/2019, and 31/10/2019, respectively. In the case of HE~1243$-$2408, two frames of 1400 sec 
exposures were taken on 02/02/2019. The spectra were reduced using the HERMES pipeline. 
Multiple frames for each object are co-added after the Doppler correction, to enhance the S/N ratio. 
The  co-added spectra are continuum fitted using 
\texttt{IRAF} (Image Reduction and Analysis Facility) software. 
The basic data of the program stars are given in Table \ref{basic data of program stars}.
A small portion of the continuum-fitted, radial-velocity-corrected spectra of the stars are shown in Figure \ref{sample spectra} as an example.

 {\footnotesize
\begin{table*}
\caption{\textbf{Basic information of the program stars.}\label{basic data of program stars}}
\resizebox{\textwidth}{!}{
\begin{tabular}{lcccccccccr}
\hline
Star      &RA$(2000)$       &Dec.$(2000)$    &B       &V       &J        &H        &K     &          & S/N       &         \\
          &                 &                &        &        &         &         &      & 4200 \AA &  5500 \AA & 7700 \AA \\
\hline
HE~0038$-$0345  & 00 41 09.29     & $-$03 28 59.51   & 11.99   & 11.30    & 9.886    & 9.419    & 9.293 & 15.08 & 58.67 & 74.86  \\ 
HE~1243$-$2408  & 12 45 53.85     & $-$24 25 02.42   & 11.66   & 10.97    & 9.075    & 8.551    & 8.434 & 13.55 & 56.41 & 65.01 \\ 
\hline

\end{tabular}}
\end{table*}
}

\section{Radial velocity} \label{section_RV}
The radial velocities of the program stars are estimated from the measured shift 
in wavelengths of several clean and unblended spectral lines, using the standard procedure as
discussed in \cite{Shejeelammal_2022}. The average of the radial velocities for each object 
corrected for heliocentric motion is taken as the radial velocity of the object. 
The detailed radial velocity data of the objects  will be published in a summary paper 
(Jorissen et al., in preparation) on the  orbits of CEMP stars. 
Our radial velocity estimates, along with those 
from Gaia DR3 \citep{Gaia_2022K} are presented in Table \ref{atmospheric parameters}. 
Our radial velocity values  for these two objects match quite closely with  Gaia values.

\section{STELLAR ATMOSPHERIC PARAMETERS} \label{section_atmospheric_parameter}
\begin{figure}
\centering
\includegraphics[width=\columnwidth]{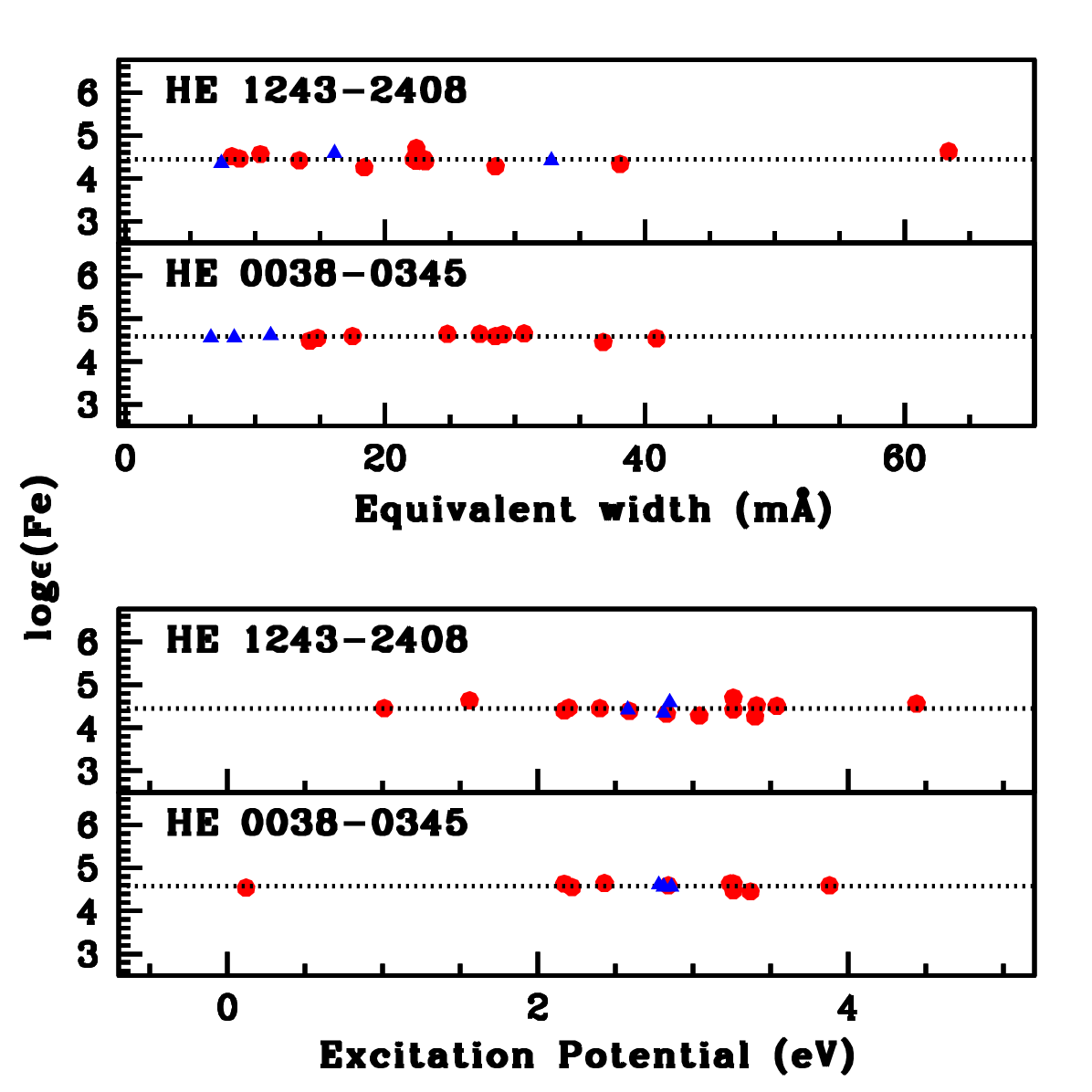}
\caption{The estimated Fe I (solid circles) and Fe II (solid triangles) abundances of the program stars 
with respect to (i) equivalent width (upper panel), and (ii) excitation potential (lower panel). 
The dotted lines in each panel is the adopted value of Fe abundance for the program stars. } \label{ep_ew}
\end{figure}

The stellar atmospheric parameters are derived using the standard excitation balance and 
ionization balance methods from a set of clean lines of Fe I and Fe II (Table \ref{linelist1}). 
We performed the analysis 
under the assumption of local thermodynamic equilibrium (LTE), using the code 
\texttt{MOOG} (version 2013, \citealt{Sneden_1973}). The selected spectral lines have excitation potentials in the 
range 0 - 6 eV and equivalent widths in the range 5 - 180 m{\rm \AA}. The model atmospheres are 
selected from the Kurucz grid of model atmospheres (\citealt{Castelli_2003}, \url{http://kurucz.harvard.edu/grids.html}). 
We have followed the 
detailed procedure for stellar atmospheric parameter estimation as discussed 
in our earlier papers \cite{Shejeelammal_2020, Shejeelammal_2021a, Shejeelammal_2021b}. 
By setting the slope of the estimated Fe I abundances with excitation potentials to zero, 
the effective surface temperature is obtained. The micro-turbulent velocity is then derived by establishing 
a zero trend between the abundances of Fe I lines and their reduced equivalent widths. 
Lastly, by adjusting its value until the abundances of the Fe I and Fe II lines coincide, the surface gravity is determined.
To derive the final stellar atmospheric parameters, an iterative process is 
performed until these three conditions are simultaneously accomplished.
The derived stellar atmospheric parameters of the stars, along with the literature values are 
given in Table \ref{atmospheric parameters}. The iron abundances derived from different 
Fe I and Fe II lines are shown in Figure \ref{ep_ew}. 

{\footnotesize
\begin{table*}
\caption{Derived atmospheric parameters and radial velocity of the program stars.} \label{atmospheric parameters}
\begin{tabular}{lcccccccc}
\hline
Star                &T$\rm_{eff}$  & log g      &$\zeta$         & [Fe I/H]          &[Fe II/H]          & V$_{r}$$^{\dagger}$             &  V$_{r}$   & Reference \\
                    &    (K)       & cgs        &(km s$^{-1}$)   &                   &                   & (km s$^{-1}$)       & 
                    (km s$^{-1}$)  &   \\
                    & $\pm$100     & $\pm$0.2   & $\pm$0.2       &                   &                   & (This work)         & (Gaia DR3)      &    \\  
\hline
HE~0038$-$0345      & 4830      & 1.75  & 0.78  & $-$2.92$\pm$0.07  & $-$2.92$\pm$0.03  & $-$43.66$\pm$0.05               & $-$44.42$\pm$0.84    &  1  \\
                    & --        & --    & --    & $-$2.86           & --                & --                              & --                   & 2  \\
                    & 5010      & 1.82  & --    & $-$2.51           & --                & --                              & --                   & 3 \\
                    & 4898      & 1.78  & 2.21  & $-$2.53           & --                & --                              & --                   & 4  \\
HE~1243$-$2408      & 4800      & 1.40  & 0.98  & $-$3.05$\pm$0.12  & $-$3.05$\pm$0.12  & +213.61$\pm$0.03                & +213.22$\pm$0.21     & 1  \\
                    & --        & --    & --    & $-$2.88           & --                & --                              & --                   & 2 \\
                    & 5293      & 2.13  & --    & $-$2.65           & --                & --                              & --                   & 3 \\
                    & 5023      & 2.35  & --    & $-$2.85           & --                & --                              & --                   & 5 \\
\hline
\end{tabular}
 
$^{\dagger}$ The numbers with the radial velocity values are standard deviations. \\
References: 1. Our work, 2. \cite{Frebel_2006}, 3. \cite{Beers_2017}, 4. \cite{Holmbeck_2020}, 5. \cite{Lombardo_2022} \\
\end{table*}
}

\begin{figure}
\centering
\includegraphics[width=\columnwidth]{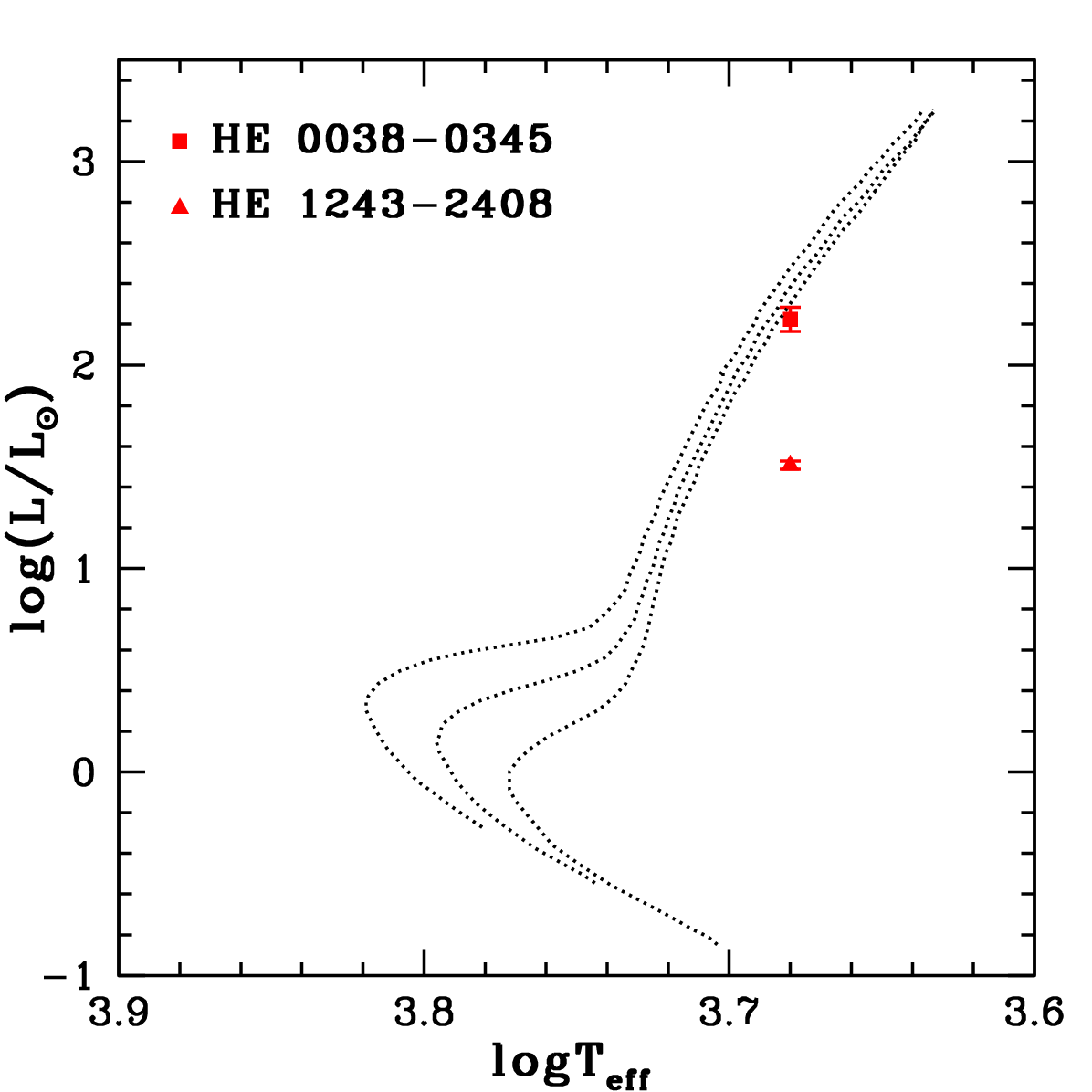}
\caption{The evolutionary tracks for 0.6, 0.7, and 0.8 M$_{\odot}$ for z = 0.0004 
are shown from bottom to top.} \label{tracks}
\end{figure}

The surface gravity of the program stars are estimated from the parallax method as well. 
The log\, g values are calculated using the equation 
log (g/g$_{\odot}$)= log (M/M$_{\odot}$) + 4 log (T$\rm_{eff}$/T$\rm_{eff\odot}$) - log (L/L$_{\odot}$), 
where T$\rm_{eff\odot}$ = 5770 K, log g$_{\odot}$ = 4.44, and M$\rm_{bol\odot}$ = 4.74 mag \citep{Bessel_1998}.
The stellar masses are determined from the position on the H-R diagram. 
We have adopted the stellar evolutionary tracks of \cite{Girardi_2000} to generate the H-R diagrams. 
The detailed procedure followed is discussed in \cite{Shejeelammal_2020}. 
The positions of the program stars on the H-R diagram are shown in Figure \ref{tracks}. 
We have used z = 0.0004 tracks for this H-R diagram. We could not estimate the mass of the star HE~1243$-$2408 
since the evolutionary tracks corresponding to its luminosity and temperature are not available. 
The estimated mass of the object HE~0038$-$0345 is 0.60$\pm$0.05 M$_{\odot}$ and the log\, g value is 
1.68. The spectroscopic surface gravity value of this star match closely with that 
derived from the parallax method within 0.07 dex. We have adopted the spectroscopic surface gravity 
value for the analysis. 

\section{Abundance determination} \label{section_abundance_determination}
Abundances of different elements are determined using the measured equivalent widths of
spectral lines as well as the spectrum synthesis method. For the elements with hyper-fine splitting (HFS) 
as well as for molecular bands, we have used spectrum synthesis calculation. 
Different line parameters are taken from \texttt{linemake} (\url{https://github.com/vmplacco/linemake}) 
atomic and molecular line database \citep{Placco_2021}. The abundances are estimated 
relative to solar values \citep{Asplund_2009}. The NLTE corrections are applied to 
the derived elemental abundances wherever applicable. The derived elemental abundances of the 
program stars are given in Table \ref{abundance_table1}. The lines used to derive the
abundances are given in Tables \ref{linelist1} - \ref{linelist2}.

{\footnotesize
\begin{table*}
\caption{Elemental abundances in HE~0038$-$0345 and HE~1243$-$2408} \label{abundance_table1}
\resizebox{\textwidth}{!}
{\begin{tabular}{lcccccccc}
\hline
       &    &                                &                    & HE~0038$-$0345 &        &                    & HE~1243$-$2408 &                    \\ 
\hline
       & Z  & solar log$\epsilon^{\ast}$   & log$\epsilon$      & [X/H]    & [X/Fe]     & log$\epsilon$      & [X/H]    & [X/Fe]    \\ 
\hline 
C (CH band 4315 {\rm \AA})$^{\dagger}$	&	6	&	8.43	&	6.43(syn)	&	$-$2.00	& 0.92 (1.05)	&	5.95(syn)	&	$-$2.48	&	0.57 (1.03)			\\
N ($^{12}$CN band 4215 {\rm \AA})	&	7	&	7.83	&	$<$7.83(syn)	&	$<$0.0	&	$<$2.92	&	$<$8.33(syn)	&	$<$0.5	&	$<$3.55			\\
Na I$_{NLTE}$	&	11	&	6.24	&	 4.17 (syn)(2)	&	 $-$2.07 	&	 0.85 	&	3.62$\pm$0.09(syn)(2)	&	$-$2.62	&	0.43			\\
Na I$_{LTE}$	&	11	&	6.24	&	 4.20 (syn)(2)	&	 $-$2.04 	&	 0.88 	&	--	&	--	&	--			\\
Mg I	&	12	&	7.6	&	5.35$\pm$0.13(2)	&	$-$2.25	&	0.67	&	 5.39$\pm$0.13(2) 	&	 $-$2.21 	&  0.84 		\\
Si I	&	14	&	7.51	&	 5.41(1) 	&	 $-$2.10 	&	 0.82 	&	--	&	--	&	--			\\
Ca I	&	20	&	6.34	&	 3.89$\pm$0.19(5) 	&	 $-$2.45 	&	 0.47 	&	3.60$\pm$0.09(5) & $-$2.74	& 0.31		\\
Sc II	&	21	&	3.15	&	--	&	--	&	--	&	 0.25(syn)(1) 	&	 $-$2.90  &	 0.15 			\\
Ti I	&	22	&	4.95	&	2.56$\pm$0.15(3)	&	$-$2.39	&	0.53	&	2.34$\pm$0.15(3) & $-$2.61	& 0.44	\\
Ti II	&	22	&	4.95	&	 2.58$\pm$0.12(4) 	&	 $-$2.37 	&	 0.55 	&	 2.30$\pm$0.16(3)  &  $-$2.65 	&  0.40 	\\
Cr I	&	24	&	5.64	&	 2.83$\pm$0.20(2) 	&	 $-$2.81 	&	 0.11 	&	 2.48$\pm$0.07(4)  &  $-$3.16 	&  $-$0.11     \\
Fe I	&	26	&	7.5	    &	4.58$\pm$0.07(10)	&	$-$2.92	&	-	&	4.45$\pm$0.12(14)	&	$-$3.05	&	-			\\
Fe II	&	26	&	7.5	    &	4.58$\pm$0.03(3)	&	$-$2.92	&	-	&	4.45$\pm$0.12(3)	&	$-$3.05	&	-			\\
Ni I	&	28	&	6.22	&	 3.42$\pm$0.06(2) 	&	 $-$2.80 	&	 0.12 	&  3.17$\pm$0.10(3) 	&  $-$3.05  &	 0.0 	\\
Zn I	&	30	&	4.56	&	 2.20(1) 	&	 $-$2.36 	&	 0.56 	&	 1.88(1) 	&	 $-$2.68 	&	 0.37 			\\
Sr II	&	38	&	2.87	& $-$0.87(syn)(1)	&	$-$3.74	&	$-$0.82	&	--	&	--	&	--	\\
Y II	&	39	&	2.21	&	$<$$-$1.31(syn)(1)	&	$<$$-$3.52	& $<$$-$0.60	&	$-$1.23(syn)(1) & $-$3.44	& $-$0.39			\\
Zr I	&	40	&	2.58	&	--	&	--	&	--	&	$<$$-$0.58(syn)(1)	& $<$$-$3.16	& $<$$-$0.11		\\
Ba II	&	56	&	2.18	&	$-$0.85(syn)(1)	&	$-$3.03	&	$-$0.11	&	$-$1.05(syn)(1)	& $-$3.23 &	$-$0.18	 	\\
Eu II	&	63	&	0.52	&	$<$$-$1.82(syn)(1)	&	$<$$-$2.34	&	$<$0.58	&	--	& --	& --		\\
\hline
\end{tabular}}

$\ast$  \cite{Asplund_2009}, The number of lines used to derive the abundance are given in parenthesis. \\
\textit{NLTE} refers to the abundance obtained after applying the NLTE correction. \\
$^{\dagger}$ The values in the parenthesis for [C/Fe] are obtained after applying corrections for evolutionary effect.
\end{table*}
}

\subsection{Light elements: C, N, Na, $\alpha$-, and Fe-peak elements}
\begin{figure}
\centering
\includegraphics[width=\columnwidth]{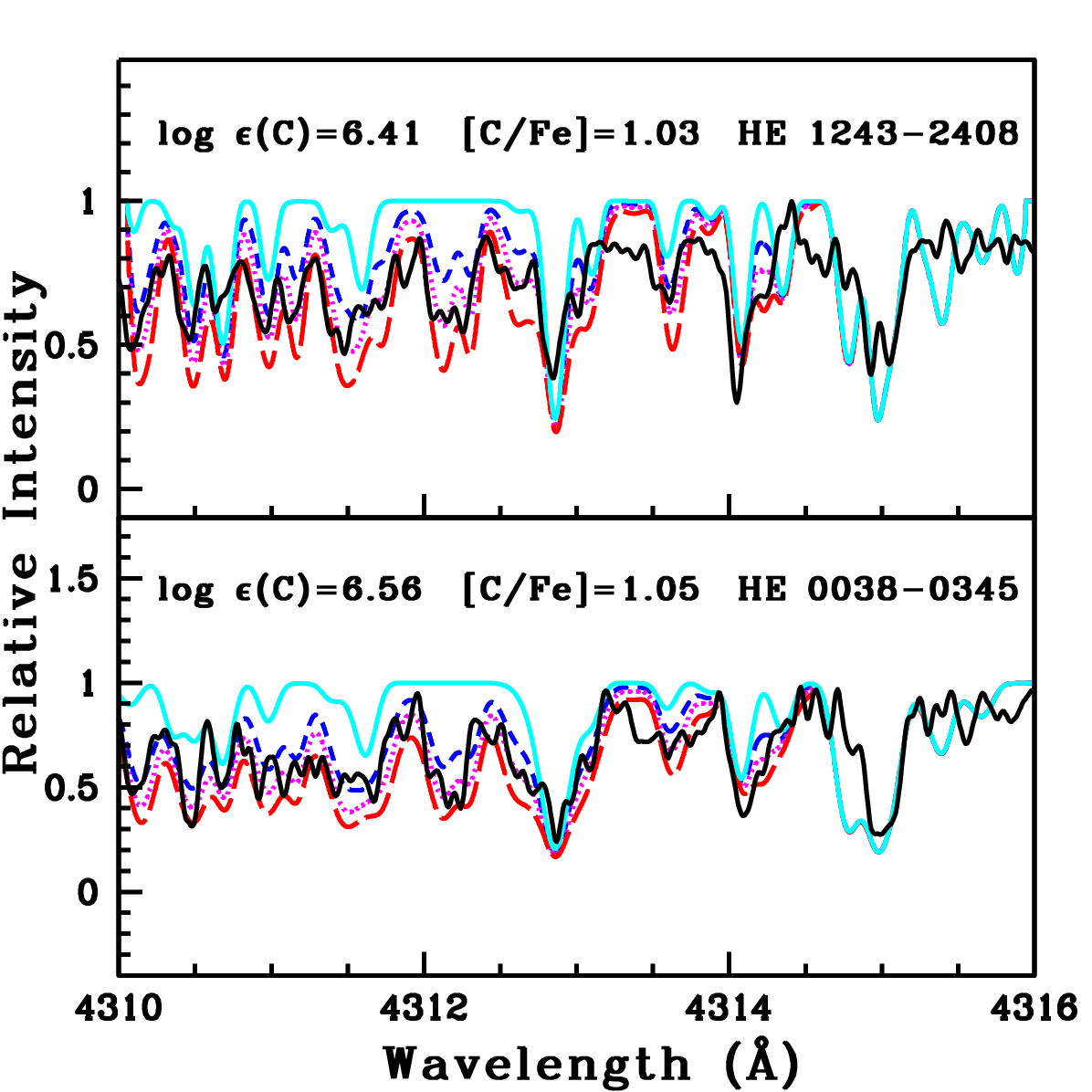}
\caption{Spectrum synthesis fits of CH G-band around 4315 {\rm \AA}. 
Observed and synthetic spectra are represented by solid and dot lines, respectively. 
The synthetic spectra for $\Delta$ [C/Fe] = $-$0.3 and +0.3, respectively, are 
represented by short- and long-dashed lines. The cyan solid lines are the synthesized spectra without carbon.} \label{carbon}
\end{figure}

The carbon abundances are derived from the spectral synthesis calculation of 
CH G-band at 4315 {\rm \AA}. The derived [C/Fe] values 
with evolutionary corrections are 1.05 and 1.03 for 
HE~0038$-$0345 and HE~1243$-$2408, respectively. The other C$_{2}$ molecular bands 
are very weak and marginally detectable in the program stars, and hence, could not be used for abundance analysis.
The spectral synthesis fits for the CH band at 4315 {\rm \AA} are shown in Figure \ref{carbon}.

The nitrogen abundance is derived from the spectrum synthesis calculation of $^{12}$CN band at 4215 {\rm \AA}. 
Since this region of the spectra is noisy, the derived abundances are taken to be upper limits.
Both the program stars show enhanced abundances of nitrogen with [N/Fe] values $<$2.92 in HE~0038$-$0345,  and $<$3.55 in HE~1243$-$2408. 

The sodium abundances are derived from the spectrum synthesis calculation of Na I 5889.951, 5895.924 {\rm \AA} lines.
The NLTE corrections to the abundances derived from these lines are adopted from \cite{Andrievsky_2007}, which are
$-$0.53 and $-$0.41  for HE~0038$-$0345 and HE~1243$-$2408, respectively. 
In HE~0038$-$0345, we could derive the abundance of sodium from Na I 5682.63 {\rm \AA} line as well. 
While the object HE~0038$-$0345 shows a large enhancement in Na with [Na/Fe]$\sim$0.85, 
the object HE~1243$-$2408 shows a  moderate enhancement with [Na/Fe]$\sim$0.43. 

The abundances of Mg, Si, Ca, Ti, Cr, Ni, and Zn are derived from the measured equivalent widths of different 
spectral lines listed in Table \ref{linelist2}. 
The abundance of magnesium is derived from the Mg I 4702.991, 5528.405 {\rm \AA} lines in both the
program stars. While Mg is  enhanced in HE~0038$-$0345 with [Mg/Fe]$\sim$0.67,  the object HE~1243$-$2408 
shows [Mg/Fe]$\sim$0.84. The Si I 5948.541 {\rm \AA} line is used to derive the silicon abundance in HE~0038$-$0345 and 
found to be enhanced with [Si/Fe]$\sim$0.82.  No good Si lines  usable for the abundance determination could 
be found in HE~1243$-$2408.  In both stars, five Ca I lines (Table \ref{linelist2}) are used 
to derive the calcium abundances. While the object HE~0038$-$0345 shows an abundance of Ca with [Ca/Fe]$\sim$0.47, 
the object HE~1243$-$2408 shows moderate enhancement with [Ca/Fe]$\sim$0.31. 
The abundance of scandium could only be estimated in the object HE~1243$-$2408, where we have used the spectral synthesis 
calculation of Sc II 5526.79 {\rm \AA} line. Scandium shows a mild  enhancement ([Sc/Fe]$\sim$0.15) in this object. 
We could not estimate the Sc abundance in the star HE~0038$-$0345 as there are no good lines. 
Both program stars show moderate enhancement of Ti  with values [Ti I/Fe]$\sim$0.53 (HE~0038$-$0345) and 
0.44 (HE~1243$-$2408). The abundance of Ti measured using Ti II lines also shows enhancement 
with [Ti II /Fe]$\geq$0.40. 

\begin{figure}
\centering
\includegraphics[width=\columnwidth]{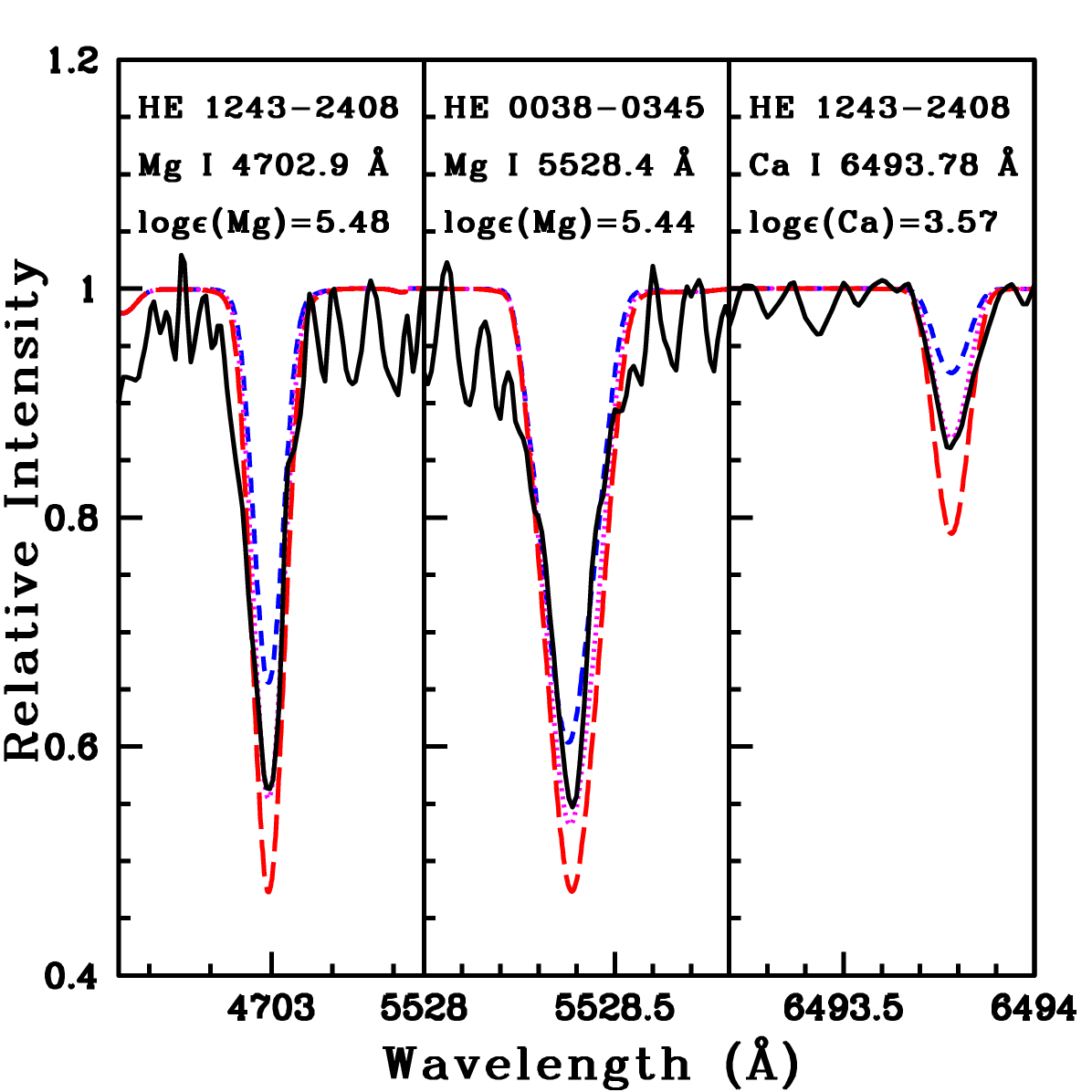}
\caption{Spectrum synthesis fits for Mg I 4702.9, 5528.4 {\rm \AA}  and Ca I 6493.78 {\rm \AA} lines. 
Observed and synthetic spectra are represented by solid and dot lines, respectively. 
The synthetic spectra for $\Delta$ [X/Fe] = $-$0.3 and +0.3, respectively, are 
represented by short- and long-dashed lines.} \label{alpha_elements}
\end{figure}

Chromium shows near-solar values in both the stars with [Cr/Fe]$\sim$0.11 and $-$0.11  for HE~0038$-$0345 and 
HE~1243$-$2408 respectively. Nickel is also found to show near-solar values with [Ni/Fe]$\sim$0.12 (HE~0038$-$0345) and [Ni/Fe]$\sim$0.00 (HE~1243$-$2408).
The Zn abundances are derived using the lines 
Zn I 4722.150 and 4810.530 {\rm \AA} for HE~0038$-$0345 and HE~1243$-$2408, respectively. Zinc is found to be moderately enhanced in 
both the objects with [Zn/Fe]$\sim$0.56 and 0.37 respectively for HE~0038$-$0345 and HE~1243$-$2408. 
The spectral synthesis fits for the alpha elements Mg and Ca are shown in Figure \ref{alpha_elements}.

\subsection{Heavy elements}

We have derived the abundances of heavy elements Sr, Y, Zr, Ba, and Eu in our program stars, whenever possible. 
The spectrum synthesis calculation is used for the abundance determination of all these elements. 
The Sr abundance could  be determined only for HE~0038$-$0345 as there are no good lines of Sr detected in the spectra of the other object. 
The abundance of Sr derived using the  Sr II 4215.519 {\rm \AA} line is found to be under-abundant in this object with  [Sr/Fe]$\sim$$-$0.82. 
The yttrium abundance is derived from the Y II 5087.420 {\rm \AA} line in HE~0038$-$0345, and
Y II 4883.684 {\rm \AA} line in HE~1243$-$2408. In the case of HE~0038$-$0345, we could only derive the 
upper limit to yttrium abundance. In both stars, yttrium is found to be under-abundant with values [Y/Fe]$\leq$$-$0.39. 
We could derive an upper limit to the Zr abundance in the object HE~1243$-$2408, where we have used Zr I 6134.999 {\rm \AA} line.
Zirconium is found to be under-abundant in this object with  [Zr/Fe]$<$$-$0.11.
The abundance of barium is derived in both the program stars. We have used Ba II 6141.7 {\rm \AA} line in HE~0038$-$0345, 
and Ba II 6496.9 {\rm \AA} line in HE~1243$-$2408. Ba is found to be under-abundant with values 
[Ba/Fe]$\sim$$-$0.11 (HE~0038$-$0345) and $-$0.18 (HE~1243$-$2408). 
We have derived an upper limit to the Eu abundance in 
HE~0038$-$0345. The europium abundance is derived from the Eu II 6645.07 {\rm \AA} line which returned a value [Eu/Fe]$<$0.58 for this object.

\section{ABUNDANCE UNCERTAINTIES} \label{section_uncertainty}
The uncertainty in the elemental abundances are calculated using the 
procedure discussed in \cite{Shejeelammal_2020}. 
The uncertainty in the elemental abundances as a result of the uncertainties in the 
stellar atmospheric parameters are calculated using, \\
\begin{center}
$\sigma_{log\epsilon}^{2}$ = $\sigma_{ran}^{2}$ + $(\frac{\partial log \epsilon}{\partial T})^{2}$ 
$\sigma_{T\rm_{eff}}^{2}$ + $(\frac{\partial log \epsilon}{\partial log g})^{2}$ $\sigma_{log g}^{2}$ + 
  $(\frac{\partial log \epsilon}{\partial \zeta})^{2}$ $\sigma_{\zeta}^{2}$ + 
  $(\frac{\partial log \epsilon}{\partial [Fe/H]})^{2}$ $\sigma_{[Fe/H]}^{2}$ \\    
\end{center} 

The uncertainties in the atmospheric parameters ($\sigma$'s) are 
$\Delta$(T$\rm_{eff}$, log \,g, $\zeta$, [Fe/H]) $\sim$ $\pm$ (100 K, 0.2, 0.2 km s$^{-1}$, 0.1). 
The partial derivatives ($\partial$) in the above equation are evaluated by changing the parameters 
by an amount equal to their respective uncertainties. 
The line-to-line scatter, standard deviation ($\sigma_{s}$), of the abundances of a 
specific element species determined from N number of lines gives rise to the random error 
in elemental abundance, which is calculated as $\sigma_{ran}$ = $\frac{\sigma_{s}}{\sqrt{N}}$. 
The cumulative uncertainty on the abundance of an element (say X) is computed from, \\
$\sigma_{[X/Fe]}^{2}$ = $\sigma_{X}^{2}$ + $\sigma_{Fe}^{2}$  \\

The uncertainty on [X/Fe] of 
each element is calculated in the specific case of HE~0038$-$0345. Since Sc II and Zr I 
could not be detected in this object, we used HE~1243$-$2408 to calculate the
uncertainty in the abundances of these two elements. Table \ref{uncertainty_abundance} 
gives the calculated uncertainty on [X/Fe], $\sigma_{[X/Fe]}$, of each element.

{\footnotesize
\begin{table*}
\caption{Variation in the abundances ($\Delta$log$\epsilon$) of various 
elements (of the star HE~0038$-$0345) with changes
in each stellar atmospheric parameters (columns 2 - 5). Total uncertainty 
on [X/Fe] of each element is given in the last column.}  
\label{uncertainty_abundance}
\begin{tabular}{lccccc}
\hline                       
Element & $\Delta$T$_{eff}$  & $\Delta$log g  & $\Delta$$\zeta$       & $\Delta$[Fe/H] &  $\sigma_{[X/Fe]}$  \\
        & ($\pm$100 K)       & ($\pm$0.2 dex) & ($\pm$0.2 km s$^{-1}$) & ($\pm$0.1 dex) &                \\
\hline
C	        & $\pm$0.20	       & $\mp$0.02	    & 0.00	                & 0.00	         	  & 0.23   \\
N	        & $\pm$0.30	       & $\mp$0.06	    & $\pm$0.02	            & 0.00	     	      & 0.32   \\
Na I	    & $\pm$0.05	       & $\mp$0.01	    & 0.00  	            & 0.00  	     	  & 0.12   \\
Mg I	    & $\pm$0.06	       & $\mp$0.02	    & $\mp$0.02	            & 0.00  	     	  & 0.15   \\
Si I	    & $\pm$0.04	       & 0.00	        & 0.00  	            & 0.00	     	      & 0.11  \\
Ca I	    & $\pm$0.07	       & $\mp$0.02	    & $\mp$0.03	            & $\mp$0.01  	   	  & 0.14   \\
Sc II	    & $\pm$0.05	       & $\pm$0.05	    & $\mp$0.03	            & 0.00  	     	  & 0.13   \\
Ti I	    & $\pm$0.13	       & $\mp$0.02	    & $\mp$0.02	            & 0.00  	     	  & 0.19   \\
Ti II	    & $\pm$0.04	       & $\pm$0.07	    & $\mp$0.02	            & $\pm$0.01	     	  & 0.14    \\
Cr I	    & $\pm$0.13	       & $\mp$0.03      & $\mp$0.08	            & $\mp$0.01	     	  & 0.21   \\
Fe I	    & $\pm$0.10	       & 0.00	        & $\mp$0.01	            & 0.00  	          & -- \\
Fe II	    & $\mp$0.03	       & $\pm$0.09	    & $\mp$0.01	            & $\pm$0.03 	      & -- \\
Ni I	    & $\pm$0.07	       & 0.00    	    & 0.00  	            & 0.00  	          & 0.13   \\
Zn I	    & $\pm$0.03	       & $\pm$0.03	    & $\mp$0.02	            & 0.00  	     	  & 0.11   \\
Sr II	    & $\pm$0.13	       & $\pm$0.05	    & $\mp$0.08	            & $\pm$0.02      	  & 0.19   \\
Y II	    & $\pm$0.05	       & $\pm$0.06	    & $\mp$0.01	            & $\pm$0.01	     	  & 0.13   \\
Zr I	    & $\pm$0.06	       & 0.00  	        & $\mp$0.05	            & $\mp$0.02   	      & 0.13   \\
Ba II	    & $\pm$0.07	       & $\pm$0.06	    & $\mp$0.04	            & $\pm$0.01	    	  & 0.14   \\
Eu II	    & $\pm$0.05	       & $\pm$0.07	    & $\mp$0.03	            & $\pm$0.01	      	  & 0.14   \\
\hline
\end{tabular}

\end{table*}
}

\section{abundance analysis and discussion} \label{section_discussion}
The two objects analyzed in this study, HE~0038$-$0345 and HE~1243$-$2408, are found to be extremely metal-poor stars with [Fe/H]$\sim$$-$3. 
The observed nitrogen abundances in them are [N/Fe] $<$ 2.92 (HE 0038$-$0345) and $<$ 3.55 (HE 1243$-$2408).  
The neutron-capture elements are found to be under abundant ($<$ $-$0.1) in both the objects. 
The stars on the Red-Giant Branch (RGB) undergo internal mixing and hence alters the 
surface CN abundances (eg. \citealt{Aoki_2007, Gratton_2000, Spite_2006, Placco_2014}). 
Based on the analysis of a sample of 505 metal-poor ([Fe/H] $\leq$ $-$2) stars with no enhancement 
of neutron-capture elements from the literature, \cite{Placco_2014} shown that the frequency of 
CEMP-no stars ([C/Fe] $\geq$ 0.7) in the Galaxy is increased by $\sim$ 11\% for [Fe/H] $\leq$ $-$3, 
when proper correction required for the carbon depletion is applied to the observed [C/Fe]. 

\begin{figure}
\centering
\includegraphics[width=\columnwidth]{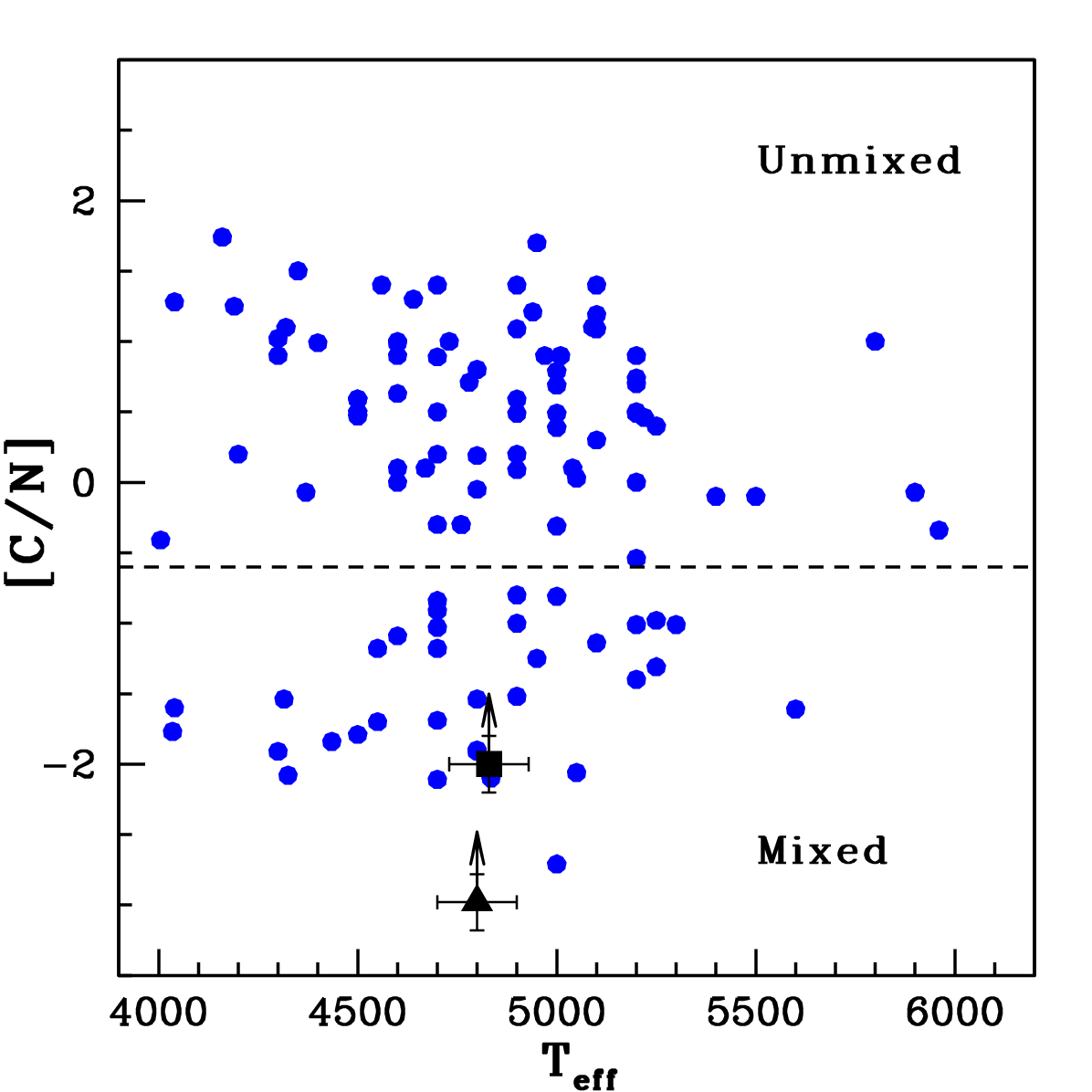}
\caption{Distribution of [C/N] ratio as a function of T$\rm_{eff}$ for CEMP stars from the 
literature (\citealt{Purandardas_2021a} and references therein, \citealt{Goswami_2021, Shejeelammal_2021a, Shejeelammal_2021b, Shejeelammal_2022}). 
Our program stars HE~0038$-$0345 (filled square) and HE~1243$-$2408 (filled triangle) are shown in black; 
these are the lower limits for the [C/N] ratio. } \label{CN_temp}
\end{figure}

\begin{figure}
\centering
\includegraphics[width=\columnwidth]{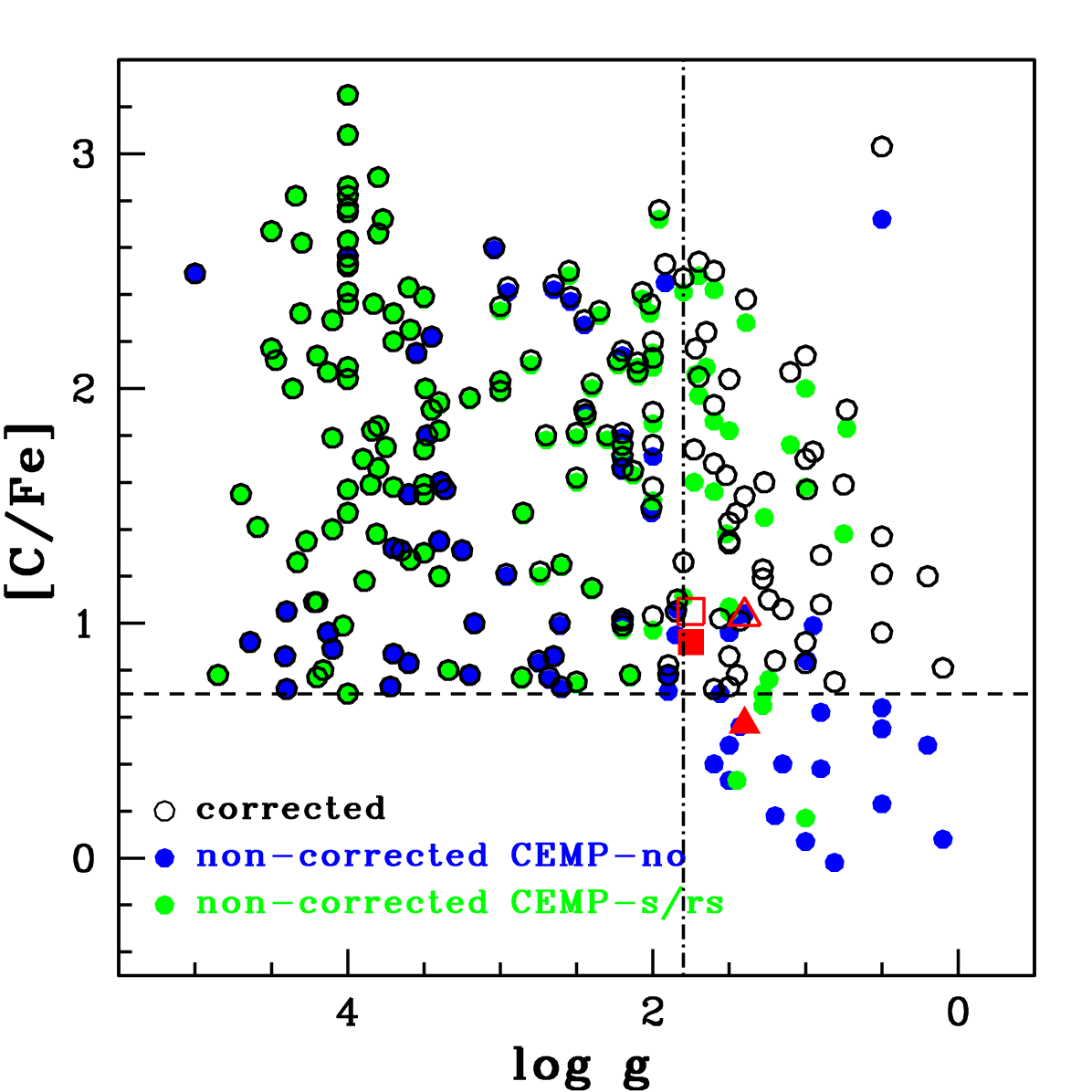}
\caption{Distribution of [C/Fe] ratio as a function of log \,g for CEMP stars from the 
literature (\citealt{Placco_2014} and references therein). The observed [C/Fe] of the CEMP-no stars are 
shown with blue filled circles, and CEMP-s and CEMP-r/s stars with green filled circles.  
The corrected [C/Fe] abundance for the evolutionary effect are shown with black open circles. 
The red symbols represent our program stars HE~0038$-$0345 (square), and HE~1243$-$2408 (triangle); 
where filled symbols are observed [C/Fe] and open symbols are corrected [C/Fe]. 
The dashed horizontal line corresponds to [C/Fe] = 0.7, and the vertical dot-dashed line
corresponds to log \,g = 1.8. } \label{CEMP_placco_correction}
\end{figure}

Since both our program stars are giants (Figure \ref{tracks} and Table \ref{atmospheric parameters}), 
it is likely that the stars have undergone internal mixing, altering the surface CN composition \citep{Spite_2005, Placco_2014}.
We have checked the [C/N] ratio, an indicator of mixing, of the program stars as discussed in \cite{Spite_2005}. 
The plot of [C/N] ratio with respect to effective temperature of the program stars is shown in Figure \ref{CN_temp}.
The [C/N] values of other known CEMP stars are also plotted in the same figure, for a comparison. 
From the figure, it is clear that both program stars  are mixed, with [C/N]$<$$-$0.6 \citep{Spite_2005}. 
Since CEMP-no stars are important objects for studying the first stellar populations and 
the Initial Mass Function (IMF), the identification of genuine CEMP-no stars is crucial. 
Hence, we have applied the necessary evolutionary correction to the observed carbon abundance values 
in order to take  account of any extra mixing on the giant branch that could have altered the carbon abundance. 
The corrections to the carbon abundances are calculated using the online tool publicly 
available at \url{http://vplacco.pythonanywhere.com/}, developed by \cite{Placco_2014}. 
The calculated correction factors are +0.13 and +0.46 respectively for HE~0038$-$0345 and HE~1243$-$2408, 
and the final corrected [C/Fe] values for them are 1.05 and 1.03. 
The observed and corrected [C/Fe] values of the program stars, along with other 
CEMP stars from \cite{Placco_2014} are plotted in Figure \ref{CEMP_placco_correction}. 
As we can see from this figure, for log \,g $<$ 1.8, compared to CEMP-s and CEMP-r/s stars,  
a larger number of CEMP-no stars lie below [C/Fe] $\sim$ 0.7, even closer to [C/Fe] $\sim$ 0. 
The amount of carbon depleted are noticeably larger for lower surface gravities ($<$ 1.8), 
and all the stars in this region are brought to [C/Fe] $>$ 0.7 when corrected for the mixing. 
With the corrected [C/Fe] values ($>$ 1) and [Ba/Fe] values ($<$ 0), our program stars are 
found to belong to CEMP-no classification \citep{Beers_2005}. 

Since the program stars are enhanced in nitrogen, we have examined if the stars are 
Nitrogen-Enhanced Metal-Poor (NEMP) star candidates. The NEMP stars are those with 
[N/Fe] $\geq$ 1 and [C/N] $\leq$ $-$0.5, according to \cite{Johnson_2007}. 
They generally show enhancement of s-process elements in addition to nitrogen and are mostly 
found at [Fe/H] $<$ $-$2.9. Binary mass transfer from intermediate-mass AGB stars 
with hot-bottom burning (HBB; M $\geq$ 5 M$_{\odot}$) are suggested to be the origin of these 
class of stars \citep{Johnson_2007, Pols_2012, Simpson_2019}. 
At lower metallicities ($\sim$ $-$2.3), the stars of lower masses ($\sim$ 3 M$_{\odot}$) 
could experience HBB, and hence the same mechanism that produce CEMP-s stars would produce 
NEMP stars as well \citep{Johnson_2007, Pols_2012}. 
However, they are rare and only $\sim$ 15 NEMP stars were identified 
till date \citep{Pols_2012, Simpson_2019}. Four out of these 15 objects 
show no evidence of s-element enhancement, similar to CEMP-no stars. 
According to \cite{Komiya_2007}, the binary mass transfer from low metallicity AGB 
stars ([Fe/H] $<$ $-$2.5) with masses in the range 3.5 - 6 M$_{\odot}$ could produce 
CEMP-no stars. In such a scenario, NEMP stars with no s-element enhancement are expected to be 
formed through the same process of mass-transfer that would produce CEMP-no stars \citep{Pols_2012}. 
We found that the program stars satisfy the conditions [N/Fe] $\geq$ 1 and [C/N] $\leq$ $-$0.5 to be NEMP stars.  
According to the definitions of CEMP and NEMP stars, they tend to overlap partially 
(some CEMP stars show [N/Fe] $\geq$ 1 and [C/N] $\leq$ $-$0.5), 
and the internal mixing on RGB could lead to the misinterpretation of CEMP stars as NEMP stars \citep{Pols_2012}.  
In the study of \cite{Pols_2012}, among the 24 NEMP candidates identified according to the 
above definition, ten objects are found to be internally mixed true CEMP stars on RGB. 
So, though the program stars follow the definition of NEMP stars, based on the discussion in the 
previous paragraph and from Figures \ref{CN_temp} and \ref{CEMP_placco_correction}, we conclude that our objects are  bonafide CEMP-no stars. 

Detailed abundance profile analysis of the program stars and the interpretation are presented in the
following sub-sections.

\subsection{Carbon abundances and location of the program stars on the Yoon - Beers A(C) - [Fe/H] diagram}
The identification and further studies on the existence of two carbon plateaus of CEMP stars 
in the distribution of absolute carbon abundance A(C) versus metallicity [Fe/H]
point at the likelihood of  different mechanisms for the production of carbon in CEMP stars \citep{Spite_2013, Hansen_2015, Bonifacio_2015, Yoon_2016}.
According to the extended literature compilation performed by \citealt{Yoon_2016},
the two plateaus are at carbon abundance values A(C)$\sim$7.96 and 6.28. 
Their analysis have further shown that different classes of CEMP stars exhibit different morphology on the
A(C) - [Fe/H] diagram, and based on the morphology the stars are classified into three groups. 
In this plot, the high-carbon band is occupied by the CEMP-s and CEMP-r/s stars, with a wide scatter in A(C) at any given [Fe/H]. 
The stars in this band is referred  as Group I stars. They show a very weak trend of A(C) with respect to [Fe/H].
The CEMP-no stars that occupy the low-carbon band exhibit two distinct behaviours.
One sub-set ($-$5$\leq$[Fe/H]$\leq$$-$2.5) shows a clear dependence 
of A(C) on [Fe/H], referred as Group II stars, and  another sub-set with no clear dependence of 
A(C) on [Fe/H], referred as 
Group III stars. Different morphology of the three groups of CEMP stars led to the 
interpretation that these stars have different origin. The radial velocity studies 
of different CEMP stars have shown that stars in the low-carbon band (CEMP-no)
are generally single stars and those in the high-carbon band (CEMP-s and CEMP-r/s) are generally binaries. 
Hence, binary mass-transfer from a companion AGB star is supported as the origin of CEMP-s and CEMP-r/s stars \citep{Starkenburg_2014, Bonifacio_2015}. 
The exact origin of CEMP-no stars has not been clearly understood yet, however, it is inferred that they 
are born from  
a natal cloud polluted by massive Pop III stars \citep{Cooke_2014, Frebel_2015, Norris_2013b}.

\begin{figure}
\centering
\includegraphics[width=\columnwidth]{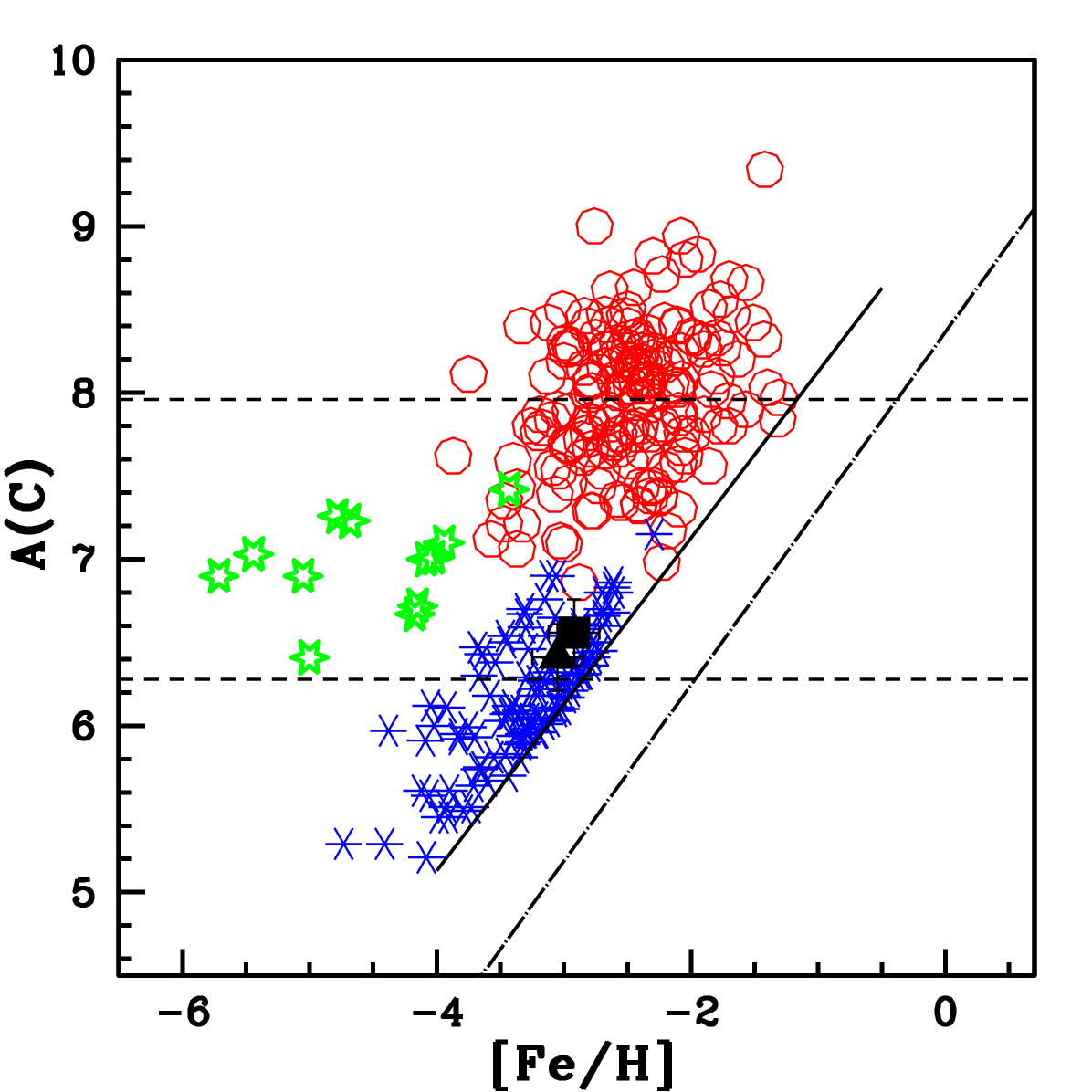}
\caption{Distribution of A(C) as a function of [Fe/H] for CEMP stars from the 
literature (\citealt{Yoon_2016, Purandardas_2019, Purandardas_2021a, Purandardas_2021b, Goswami_2021, Shejeelammal_2021a, Shejeelammal_2021b, Shejeelammal_2022}).
Red symbols are Group I CEMP stars.
Blue and green symbols represent Group II and Group III CEMP-no stars respectively. 
Our program stars HE~0038$-$0345 (filled square), and HE~1243$-$2408 (filled triangle) are shown in black.
The high- (A(C)$\sim$7.96) and low- (A(C)$\sim$6.28) 
carbon bands of CEMP stars are shown with dashed lines. The solid line is for [C/Fe] = 0.70 and 
long-dash dot line represent the solar carbon value, [C/Fe] = 0.} \label{carbon_cemp}
\end{figure}

The locations of our program stars on the A(C) - [Fe/H] diagram, along with other CEMP stars 
from the literature, are shown in the Figure \ref{carbon_cemp}. The carbon abundances corrected 
for the evolutionary effect are used in this plot. 
As it is clear from the Figure \ref{carbon_cemp}, the two program stars fall in the region of Group II CEMP-no stars. 
Radial velocity studies of the program stars are not available in the literature. However, our estimated 
radial velocities are similar to those listed in the Gaia. This may indicate that these stars are likely single, but not conclusively.

In the following sections we have discussed the  possible progenitors of the program stars based on a detailed analysis of their abundance patterns.

\subsection{Analysis of abundance patterns observed  in the program stars and comparison with the literature}
\begin{figure}
\centering
\includegraphics[width=\columnwidth]{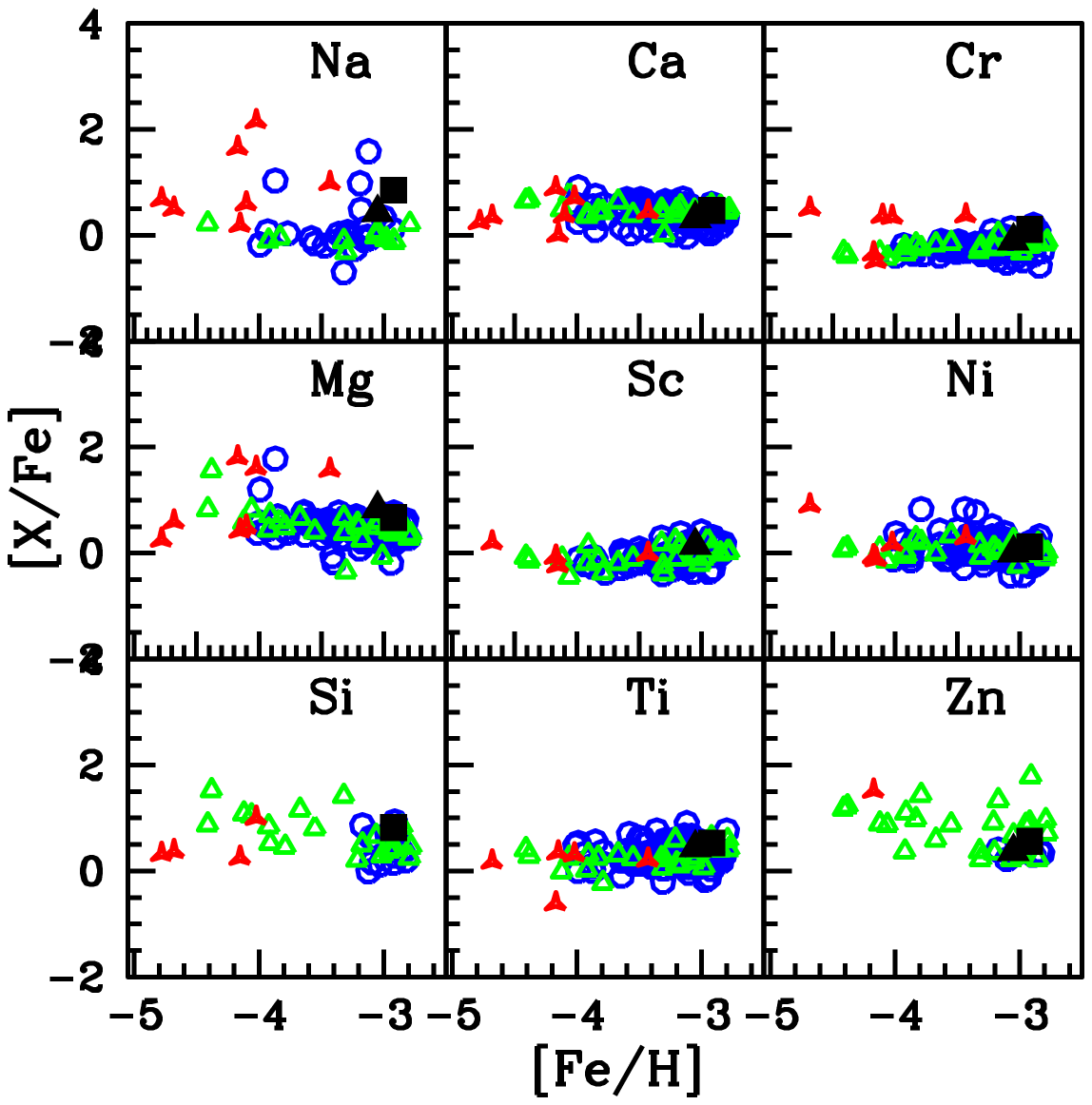}
\includegraphics[width=\columnwidth]{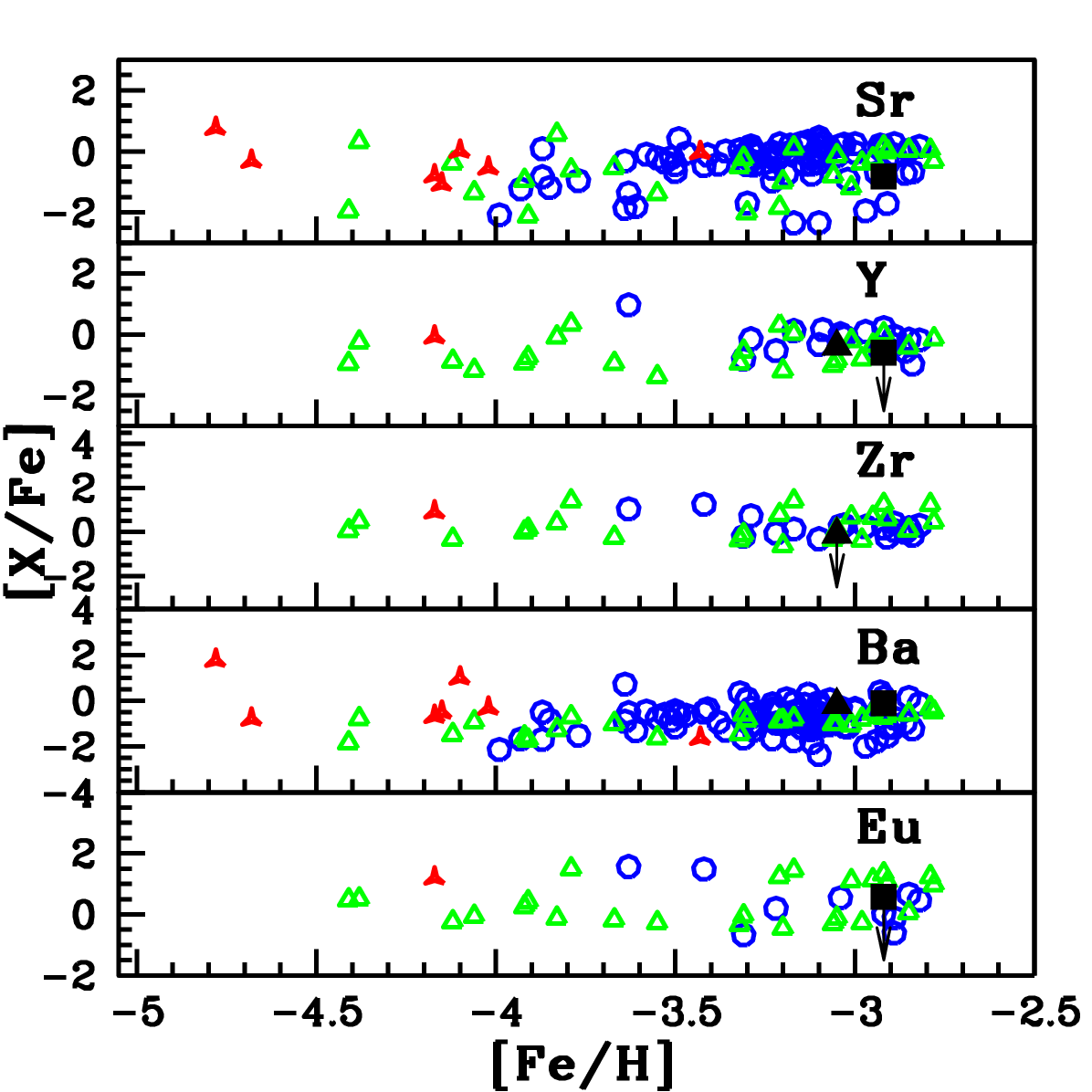}
\caption{Observed [X/Fe] ratios of the light elements (upper panel) and heavy elements (lower panel) 
in the program stars with respect to metallicity [Fe/H].   
Green triangles and red starred triangles correspond to Group II and III CEMP-no stars respectively from the literature 
(\citealt{Yoon_2016} and references therein, \citealt{Purandardas_2021a}). Blue open circles represent 
extremely metal-poor stars from the literature \citep{Venn_2004,Honda_2004, Aoki_2005, Aoki_2007, Roederer_2014}.
Our program stars HE~0038$-$0345 (filled square), and HE~1243$-$2408 (filled triangle) are shown in black. 
The NLTE Na abundances are plotted for the program stars.
The symbols with downward arrow indicate the derived upper limit for the abundances.  
} \label{elements_abundance_comparison}
\end{figure}

The observed abundances of different elements in the program stars are compared with those in other
CEMP-no stars and EMP stars from the literature. These comparisons are shown in Figure \ref{elements_abundance_comparison}. 
The estimated elemental abundances in the program stars match closely with the abundances of elements 
in other extremely metal-poor stars. Both program stars show enhancement of N, Na, and Mg. 
HE~0038$-$0345 show enhancement of Si as well, while Si abundance is not available for HE~1243$-$2408. 
These elements are found to be over abundant in majority of extremely metal-poor stars, indicating the 
signatures of first star nucleosynthesis \citep{Aoki_2002, Nomoto_2013, Roederer_2014, Aoki_2018}.

\begin{figure}
\centering
\includegraphics[width=0.8\columnwidth, height= 0.8\columnwidth]{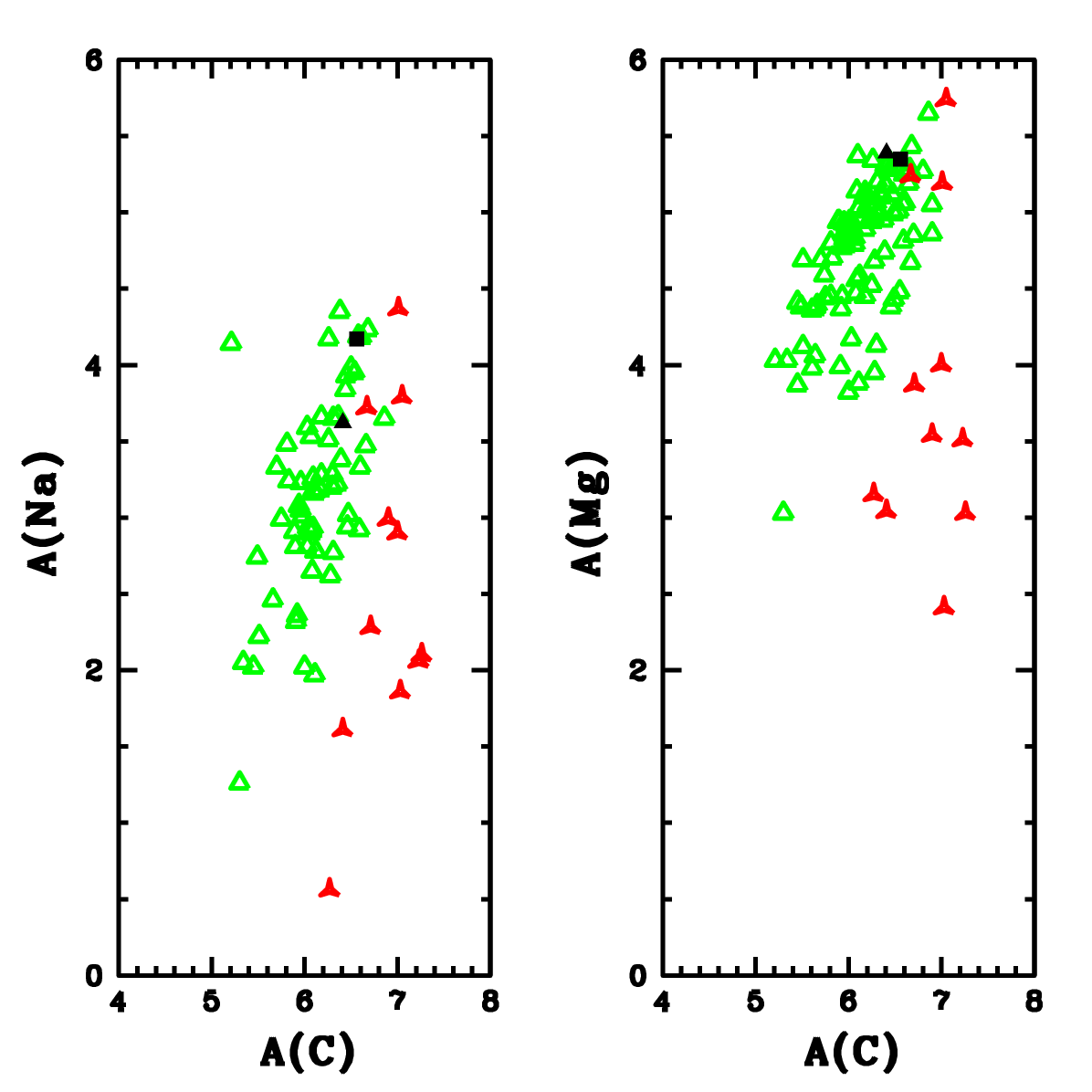}
\includegraphics[width=0.8\columnwidth, height= 0.8\columnwidth]{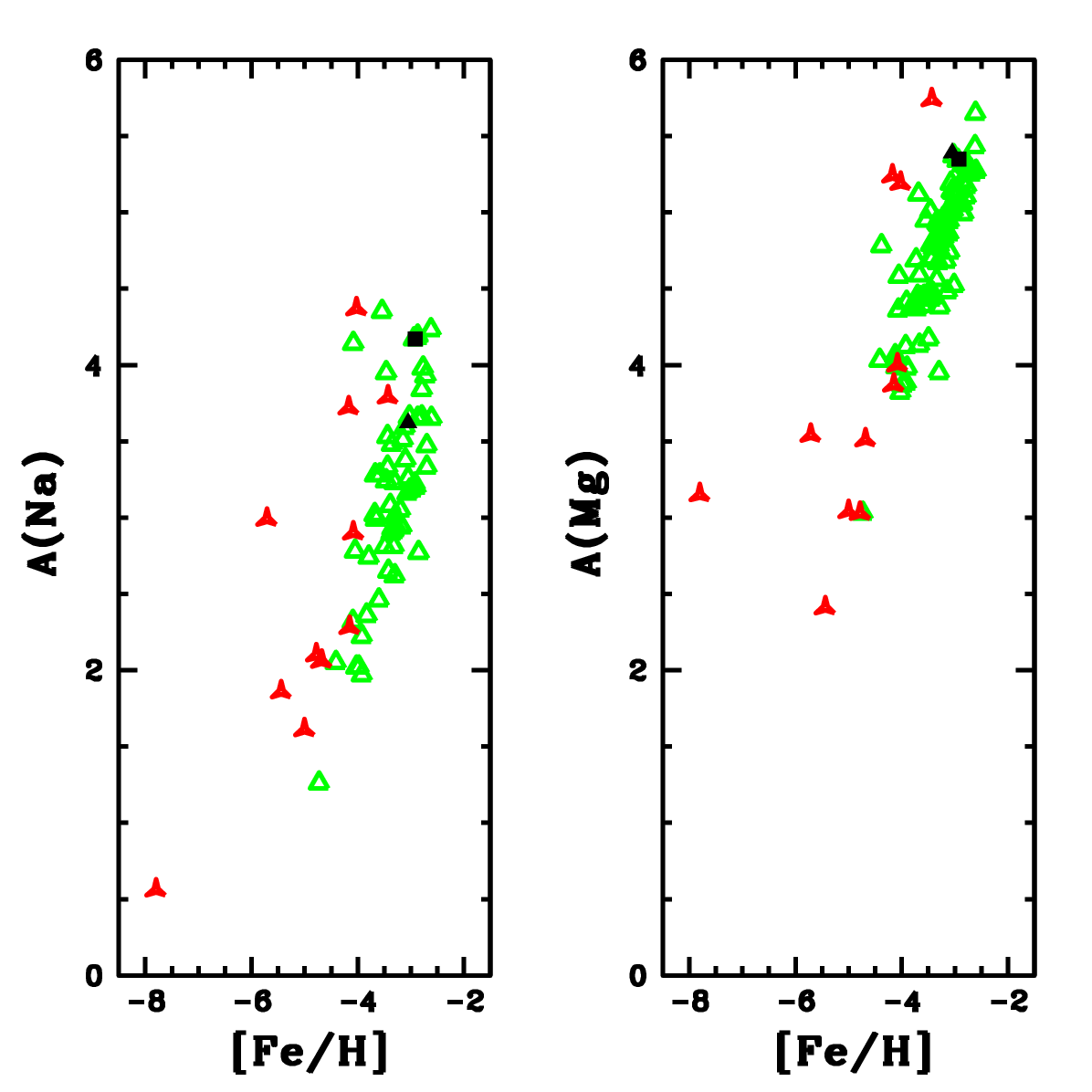}
\includegraphics[width=0.8\columnwidth, height= 0.8\columnwidth]{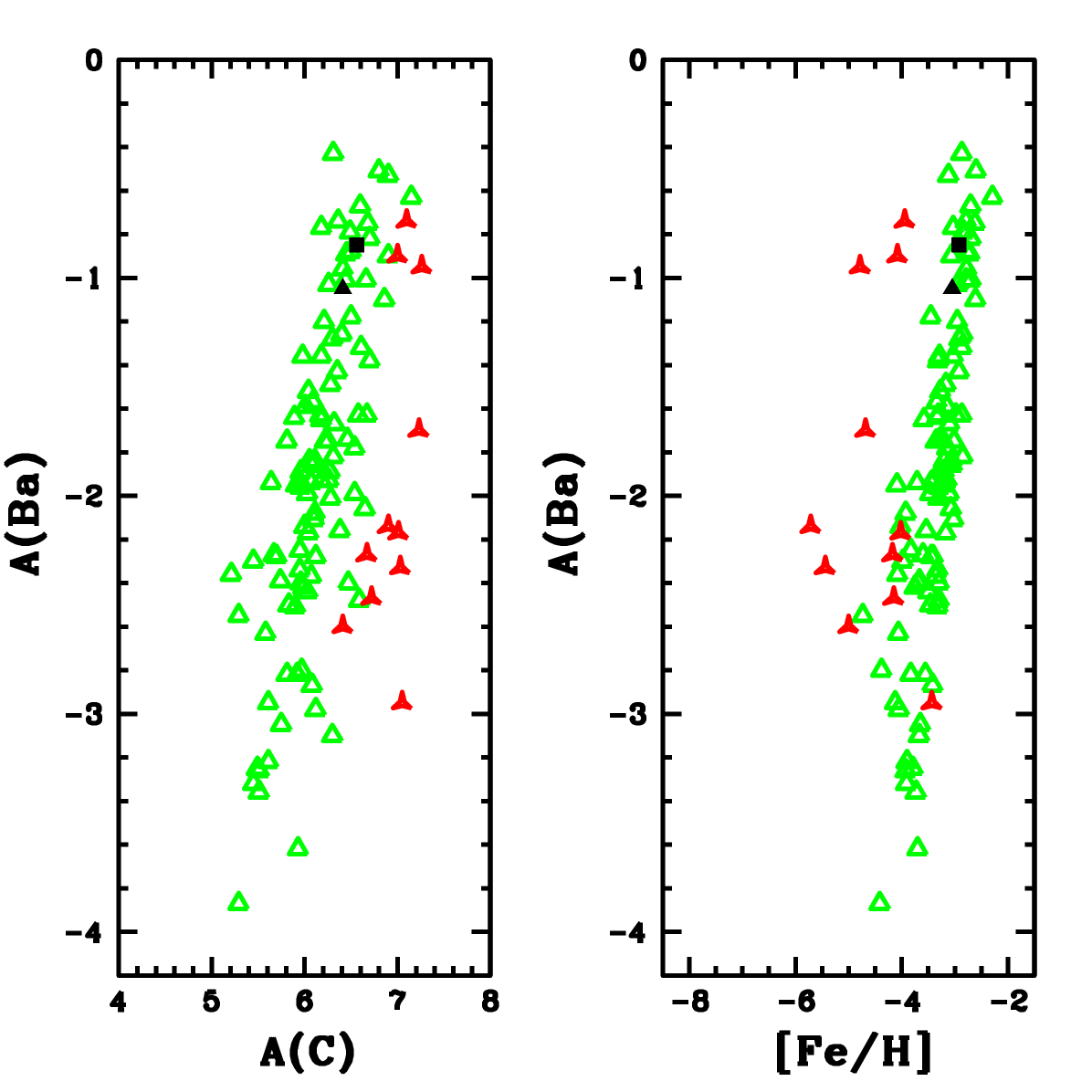}
\caption{Observed A(Na), A(Mg), and A(Ba) values with respect to A(C) and 
metallicity [Fe/H]. Symbols have same meaning as in Figure \ref{elements_abundance_comparison}} \label{Na_Mg_Ba_C_Fe}
\end{figure}

A close match of the abundance ratios of the program stars with their counterparts observed in 
the Group II CEMP-no stars (Figure \ref{elements_abundance_comparison}), indicates that the program stars 
likely belong to the same group. The two groups of CEMP-no stars show distinct behaviours in terms of the 
abundances of Na, Mg, and Ba, in addition to carbon explained in the previous section. There exist a strong 
correlation between A(C, Na, Mg, Ba) and [Fe/H] among Group II CEMP-no stars, where as there is no such 
dependence among the Group III CEMP-no stars \citep{Yoon_2016, Yoon_2019}. We have examined the abundances 
of Na, Mg, and Ba, with respect to A(C) and [Fe/H] and performed a comparison with Group II and Group III CEMP-no 
stars from the literature. These plots are shown in Figure \ref{Na_Mg_Ba_C_Fe}. 
The locations of the program stars in these plots suggest that they are Group II CEMP-no stars, similar to that 
inferred from Figures \ref{carbon_cemp} and \ref{elements_abundance_comparison}.

\subsection{Possible origin of the program stars} \label{Section_possible_origin}
Origin of CEMP-no stars is still an open question with several suggested scenarios
that include pre-enrichment of their natal cloud by 
faint SNe, spinstars, metal-free massive stars, or the contribution from all these 
sources (\citealt{Yoon_2016} and references therein, \citealt{Hansen_2016a}). However, the study of 
\cite{Arentsen_2019b} shown that the  binary fraction of CEMP-no stars is 
about 32\%  (11 out of 34 stars), 
where eight of these stars have high-carbon abundance (A(C) $>$ 6.6).
This led them to suggest that at least a few of these stars might have been polluted by
extremely metal-poor AGB stars that produced carbon but not a significant amount of s-process 
elements, like the abundances observed in these stars. 

From the complex morphology of A(C) - [Fe/H] space for CEMP-no stars and their distinct behaviour 
in terms of abundances of Na and Mg, \cite{Yoon_2016} presented the first evidence for the operation 
of at least two different classes of first-generation stellar progenitors for the CEMP-no stars. 
This hypothesis has been supported by \cite{Placco_2016} from the analysis 
of 12 Ultra Metal-Poor (UMP) CEMP-no stars. Likelihood of existence of multiple formation pathways for CEMP-no stars has
been confirmed consequently by several other authors \citep{Arentsen_2019b, Yoon_2019, Yoon_2020, Jeon_2021, Dietz_2021, Purandardas_2021a}. 
It may also be possible that, a single yet to be identified process, could account for the 
abundance pattern of all UMP stars \citep{Placco_2016}. On the other hand, more than one 
progenitors could also contribute to the observed abundance pattern of the same CEMP-no star \citep{Placco_2016, Arentsen_2019b, Yoon_2019}. 

From the abundance characteristics of the two groups of CEMP-no stars, and previous studies on 
BD+44 493 (Group II star)  and  HE~1327-2326 (Group III star), \cite{Yoon_2016} had suggested that 
the progenitors of Group II stars may be faint SNe and that of Group III stars may be spinstars. 
\cite{Placco_2016} performed an analysis to identify the progenitors of 12 UMP stars 
in their sample. They have considered the non-rotating, metal-free massive star models 
of \cite{Heger_2010} for a mass  range 10 - 100 M$_{\odot}$, and compared these yields with the 
light element abundance patterns of their sample. The abundance patterns of only five out of 
twelve stars could be reproduced with the SN models of \cite{Heger_2010}.
Among these five stars, three objects are Group II stars, and the rest two are Group III stars. 

In the context of a hierarchical Galactic assembly, \cite{Yoon_2019} performed a comparison of 
the morphology of Galactic halo CEMP-no stars on the A(C) - [Fe/H] diagram with those of CEMP-no stars
in the dwarf satellite galaxies - ultra-faint Dwarfs (UFDs) and dwarf spheroidal (dSph) galaxies, 
of the Milky Way (MW). A very similar behaviour of A(C) - [Fe/H] as that of halo stars, with two distinct groups, 
is found among the CEMP-no stars of the UFD and dSph galaxies. This similarity in the 
morphology among these stars is a strong evidence that CEMP-no halo stars are accreted from 
the mini-halos, supporting various studies of Galactic assembly and halo 
formation \citep{Frebel_2015, Starkenburg_2017, Amorisco_2017, Spite_2018}. 
The characteristics of each group of CEMP-no stars depend on the nature of the host mini-halos. 
While the UFD galaxies host both Group II and III stars, the dSph galaxies host only Group II stars, 
due to several distinct properties of these galaxies such as mass, star formation history (SFH), 
metal-enrichment/pollution pathways, cooling agents, etc (\citealt{Yoon_2018, Yoon_2019, Yoon_2020} and references therein). 
According to the study by \cite{Yoon_2019},  the majority of Group II CEMP-no stars in 
the Galactic halo are likely accreted from dwarf spheroidal-like systems, and  the Group III CEMP-no 
stars are likely accreted from  ultra-faint dwarf-like satellite galaxies. 

The dSphs are massive ($\geq$10$^{9}$ M$_{\odot}$) with more gas content than UFDs that result in 
extended star formation, leading to the formation of Pop II stars that explode as core collapse supernovae, 
in addition to the Pop III supernovae. This results in a more metal-rich environment. 
The extended star formation history (SFH)  leading  to the strong chemical evolution is reflected
as strong correlation among A(C), A(Na,Mg,Ba), and [Fe/H]. The next generation stars are formed from 
the well-mixed ISM enriched by multiple SNe, both from Pop II and III \citep{Salvadori_2015, Bennassuti_2017, Chiaki_2018}. 

In contrast, UFDs are less massive ($\leq$10$^{6}$ M$_{\odot}$) with dark-matter dominated halos.
These mini-halos likely undergo only one or a few generations of star formation due to low star 
formation efficiency  and supernovae feed back \citep{Salvadori_2009, Vincenzo_2014, Jeon_2017}. As 
a result, the stars in these systems are  primitive and likely mono-enriched, as evident from 
their low metallicity and [Mg/C] value ($<$$-$0.5) \citep{Hartwig_2018}. 
Only one Pop III progenitor - CCSNe, faint SNe, and/or spinstars, had enriched the natal clouds
of these Group II and Group III stars \citep{Salvadori_2015, Bennassuti_2017}. 

The distinct abundance patterns of Group II and Group III stars support these different formation 
environments.  The abundances of elements C, Na, Mg, and Fe suggest that Group II stars are likely 
formed from the gas cooled  by the silicate grains, whose key elements are Mg and Si that are main 
products of CCSNe. The Group III stars are formed  from the gas cooled by the carbon grains, whose 
key element is carbon, which is mainly produced by spinstars and/or faint SNe 
(\citealt{Chiaki_2017}, \citealt{Yoon_2019, Yoon_2020} and references therein). 
The simulations of \cite{Jeon_2021} have shown that
the Group II CEMP-no stars attribute their origin to normal SNe and Group III 
CEMP-no stars to faint SNe.

\begin{figure}
\centering
\includegraphics[width=\columnwidth]{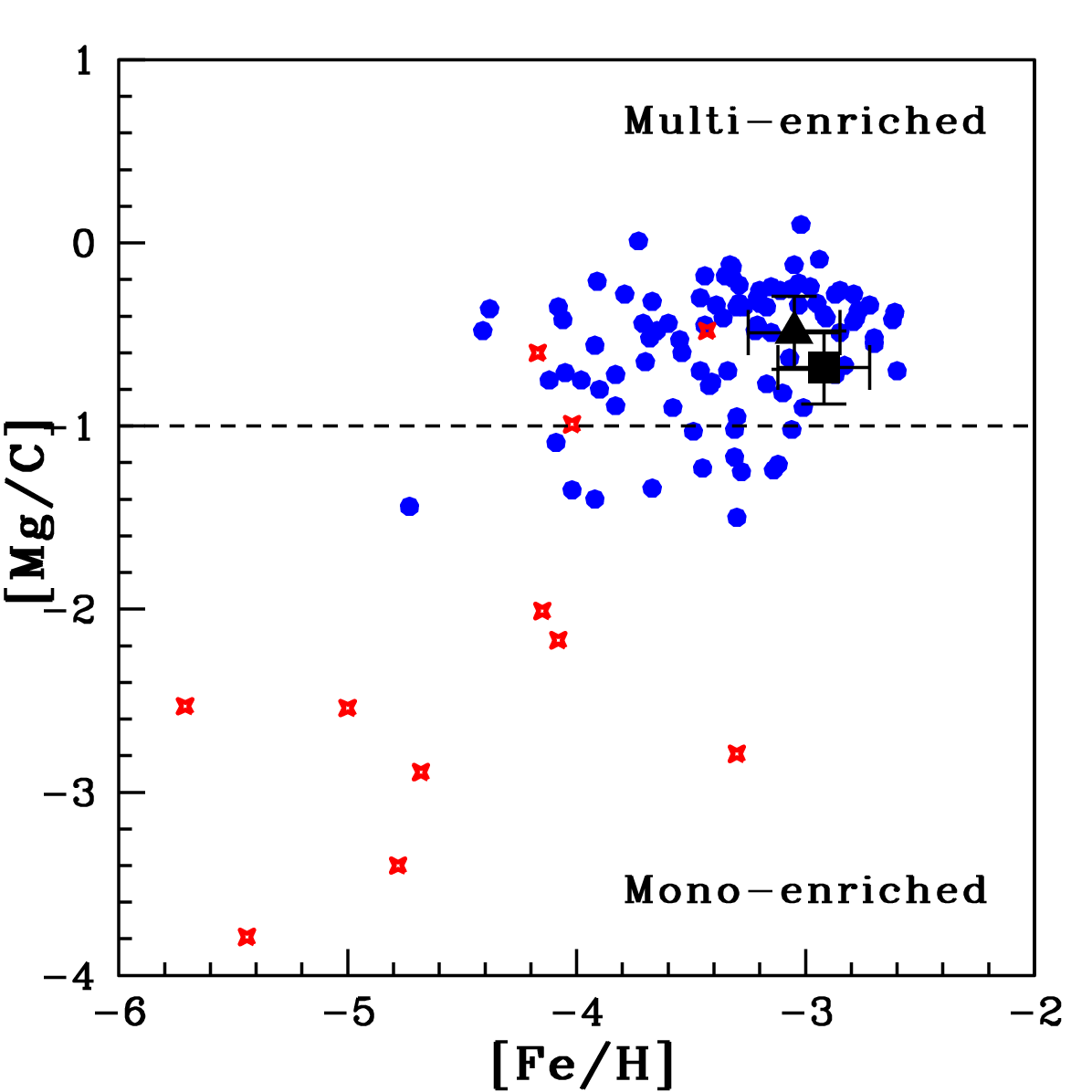}
\caption{Distribution of [Mg/C] as a function of [Fe/H] for CEMP-no stars. 
Blue circles are Group II CEMP-no stars and red stars are Group III CEMP-no stars  from the 
literature (\citealt{Purandardas_2021b} and references therein).  
Our program stars HE~0038$-$0345 (filled square), and HE~1243$-$2408 (filled triangle) 
are shown in black.} \label{Mg_C}
\end{figure}

We have found from the abundance analysis, as discussed in the previous sections, that our program stars 
belong to Group II CEMP-no stars. We have examined the [Mg/C] ratio of the program stars to see if 
they are mono-enriched or multi-enriched. 
According to \cite{Hartwig_2018}, the stars with [Mg/C]$<$$-$1 are mono-enriched. 
It should be noted that, like carbon, the abundances of the light elements
Li, Mg, and Fe are found to show significant changes in metal-poor stars as they evolve from the turn-off point to 
RGB, owing to atomic diffusion. In the case of Mg and Fe, the corrections can be as much as $-$0.3 dex; however, 
it has very little impact on [X/Fe] ratio of these elements \citep{Korn_2007}. We have applied the diffusion correction 
of $-$0.3 to the [Mg/H] values of the program stars before calculating [Mg/C] ratio.  
The calculated (diffusion corrected) values of this ratio in HE~0038$-$0345 and HE~1243$-$2408 respectively are $-$0.68 and 
$-$0.49, indicating multi-enrichment. The location of the program stars on the [Mg/C] - [Fe/H] plot is shown 
in Figure \ref{Mg_C}, along with other CEMP-no stars from the literature.   
The program stars are found to be 
multi-enriched based on  the [Ca/Fe] value ($<$2) as well. 
If the result of \cite{Yoon_2019} holds true, this multi-enrichment suggests that the program stars may 
owe their origin to the dSph satellite galaxies, as discussed above. Further, they might have formed 
from the gas cloud pre-enriched by CCSNe, from the ground of arguments discussed previously in this section. 
In the case of spinstar progenitor, the stars are expected to show [Sr/Ba]$>$0 \citep{Frischknecht_2012, Hansen_2019}. 
We could estimate this ratio in 
the object HE~0038$-$0345, it returned a value [Sr/Ba]$\sim$$-$0.71, discarding spinstar progenitor. 
In the case of faint SNe progenitor, the stars are expected to show [C/Fe]$\geq$2. The faint SNe are 
found to contribute in the metallicity range $-$5.5$\leq$[Fe/H]$\leq$$-$3.5. In the case of stars 
with [C/Fe]$<$1 and $-$4$\leq$[Fe/H]$\leq$$-$3, CCSNe are found to contribute to the abundance 
pattern \citep{Umeda_2003, Umeda_2005, Tominaga_2014, Mardini_2022}. 
Further, the CEMP-no stars with [C/Fe]$\geq$2 are originated from Pop III stars, 
and Pop II stars are responsible for the origin of CEMP stars with [C/Fe]$<$ 2 \citep{Jeon_2017, Jeon_2021}.
Hence, from the [C/Fe] values ($<$2) and [Fe/H] ($\sim$$-$3), the progenitors of our program stars 
are likely Pop II CCSNe.

\begin{figure}
\centering
\includegraphics[width=\columnwidth]{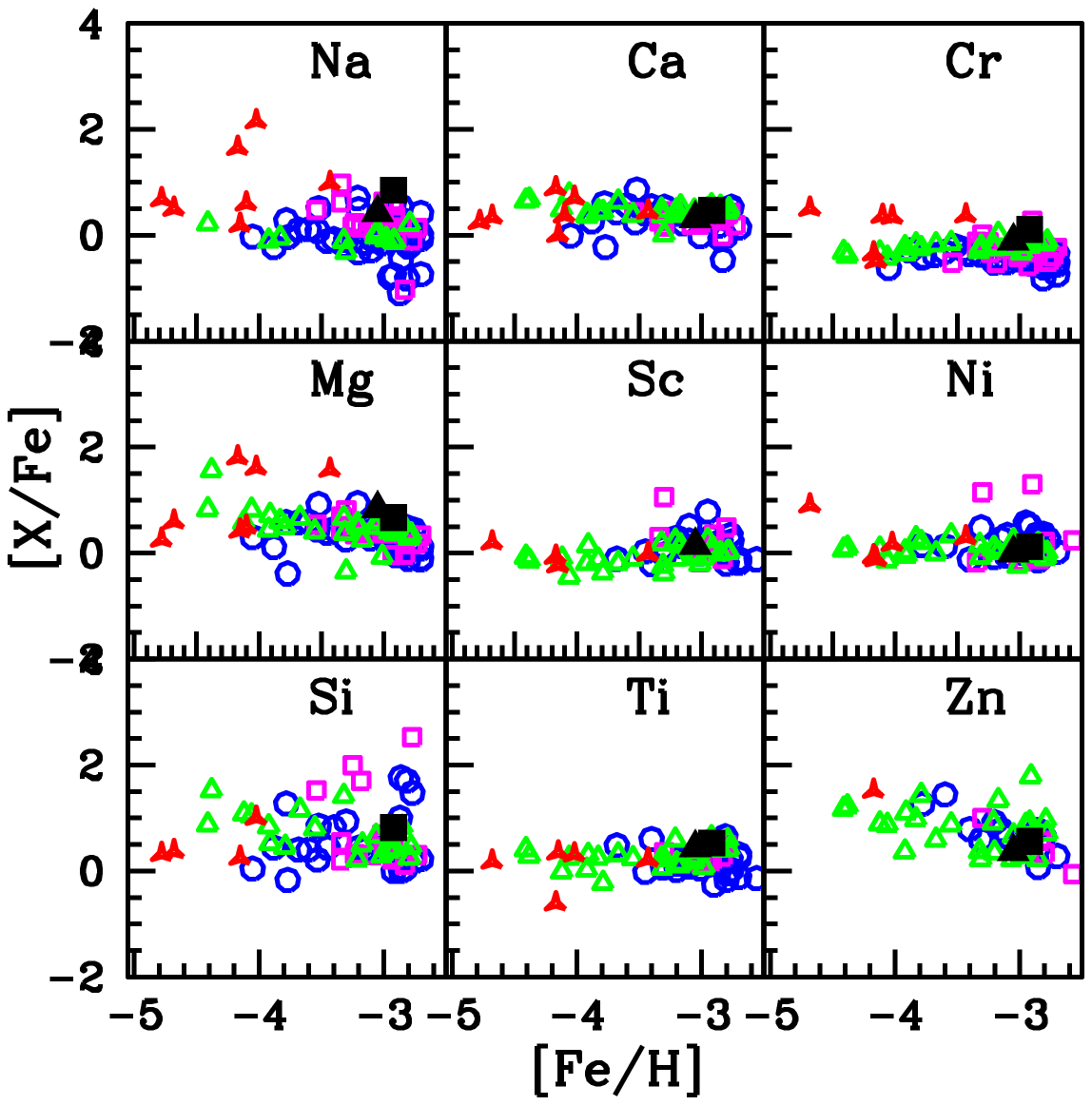}
\includegraphics[width=\columnwidth]{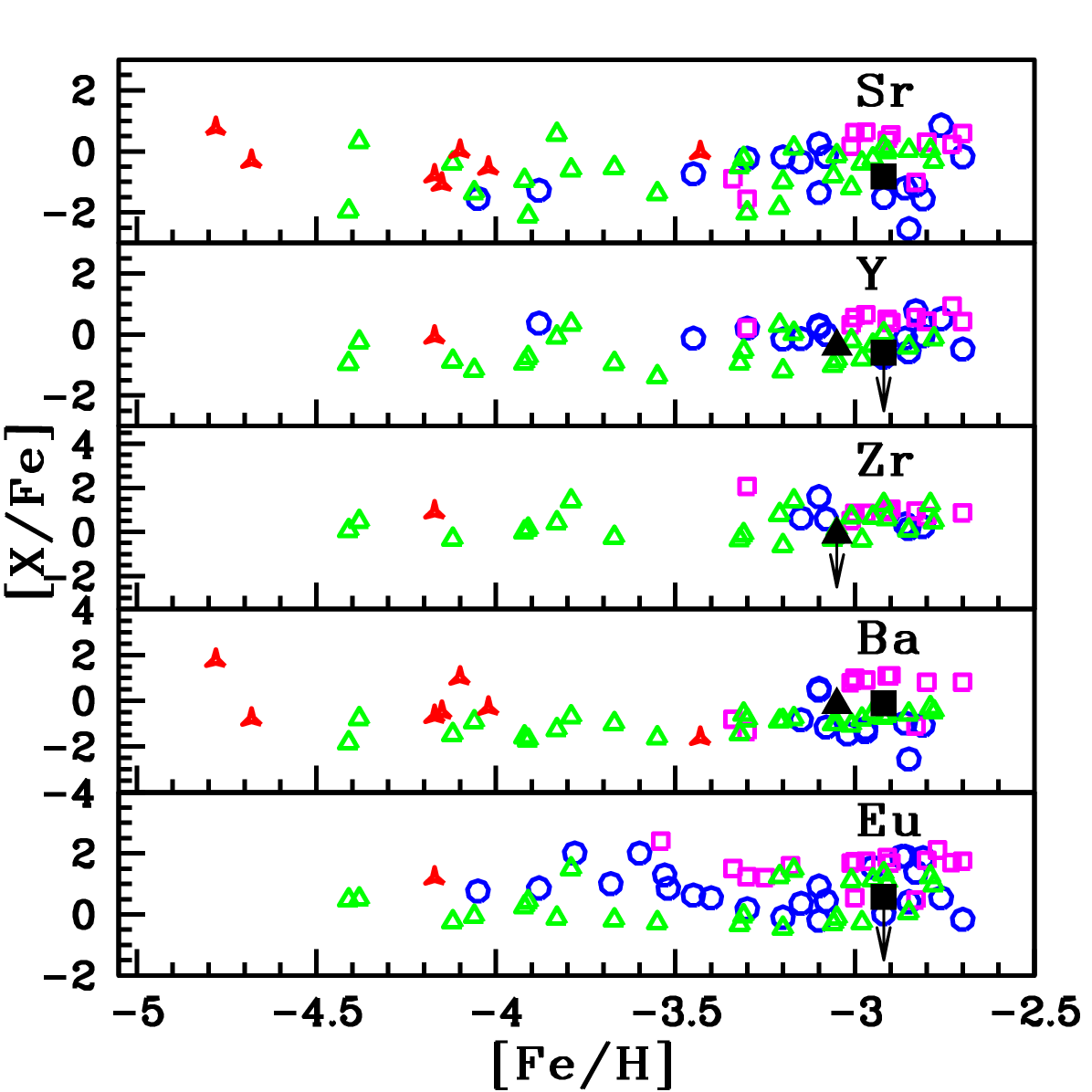}
\caption{Observed [X/Fe] ratios of the light elements (upper panel) and heavy elements (lower panel) 
in the program stars with respect to metallicity [Fe/H]. 
The blue open circles represent the CEMP-no stars in the dSph satellite galaxies and magenta open squares are 
the CEMP-no stars in the UFD satellite galaxies from the literature (\url{http://sagadatabase.jp/}).
Other symbols have same meaning as in Figure \ref{elements_abundance_comparison}.
Our program stars HE~0038$-$0345 (filled square), and HE~1243$-$2408 (filled triangle) are shown in black.
The symbols with downward arrow indicate the derived upper limit for the abundances.  
} \label{dwarf_halo_elements_abundance_comparison}
\end{figure}

We have compared the abundances of the program stars with those in the CEMP-no stars of the dSPhs and UFDs 
from the literature (SAGA database; \url{http://sagadatabase.jp/}). This comparison is shown in 
Figure \ref{dwarf_halo_elements_abundance_comparison}. From the figure, it is obvious that the 
abundance pattern of Group II CEMP-no stars of the Galaxy matches with those in the dSphs. 
As seen from the Figure \ref{Mg_C}, majority of Group II CEMP-no stars are multi-enriched, reinforcing their origin in dSphs.
Also, the elemental abundances of the program stars match closely with those in the CEMP-no stars of the 
dSphs. From this, along with the observations discussed earlier in this section, we infer that our program stars are 
likely accreted from the dSph satellite galaxies. 

In the next section, we have discussed the accretion origin of the program stars on the 
ground of kinematic analysis.

\section{Kinematic analysis} \label{section_kinematic_analysis}
The components of spatial velocity, total spatial velocity, and the Galactic population of the
program stars are estimated following the procedures described in \cite{Johnson_1987}, \cite{Bensby_2003, Bensby_2004}, 
\cite{Mishenina_2004}, and \cite{Reddy_2006}. The detailed procedure is discussed in our previous paper \cite{Shejeelammal_2021a}. 

The components of spatial velocity with respect to the Local Standard of Rest (LSR) is calculated as, \\
\begin{center}
  $(U, V, W)_{LSR} =(U,V,W)+(U, V, W)_{\odot}$ km/s.
  \end{center}
(U, V, W)$_{\odot}$ = (11.1, 12.2, 7.3) km/s is the solar motion with respect to LSR \citep{Schonrich_2010}. \\

\noindent Total space velocity with respect to LSR is estimated using; \\
\begin{center}
 $V\rm_{spa}^{2}=U\rm_{LSR}^{2}+V\rm_{LSR}^{2}+W\rm_{LSR}^{2}$. 
 \end{center}
 
\noindent The proper motions in RA ($\mu_{\alpha}$) and DEC ($\mu_{\delta}$), and the parallax ($\pi$) values 
 are taken from SIMBAD astronomical database and Gaia DR3 (\citealt{Gaia_2016b, Gaia_2022K, Babusiaux_2022}, \url{https://gea.esac.esa.int/archive/}).

\begin{figure}
\centering
\includegraphics[width=\columnwidth]{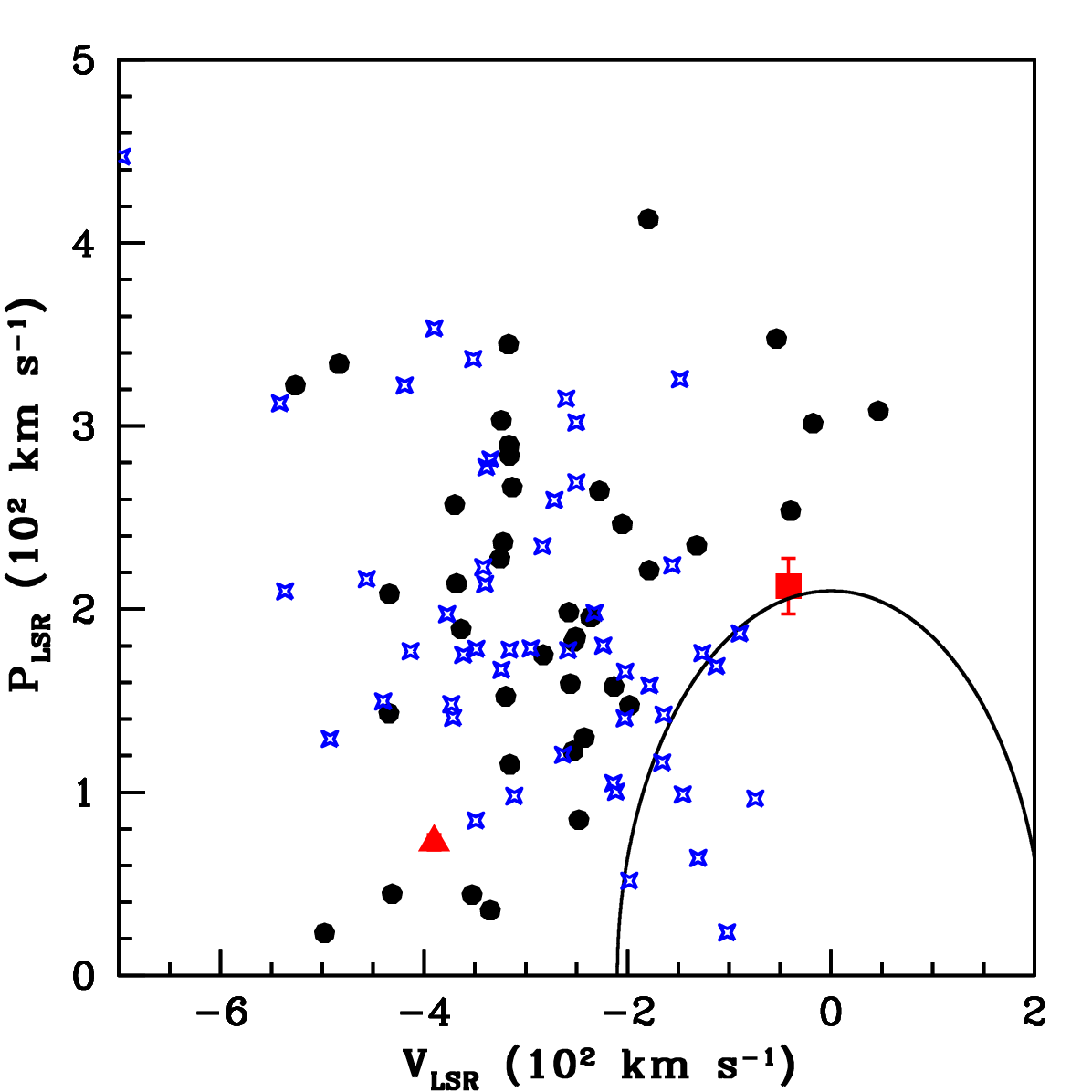}
\caption{The Toomre diagram for the program stars, HE~0038$-$0345 (filled square) 
and HE~1243$-$2408 (filled triangle). The black curve represents the loci of all points 
with V$_{spa}$ = 210 km s$^{-1}$, 
separating halo and disk objects \citep{Koppelman_2018}. The black filled circles are kinematically selected halo stars 
from the literature (\citealt{Hansen_2019} and references therein). Blue stars represent 
the dwarf satellite galaxies of the MW. } \label{Toomre_diagram}
\end{figure}

The estimated spatial velocities, along with the probabilities of the objects to belong to the Galactic thin/thick disk and halo, 
are presented in Table \ref{kinematic analysis}. While the object HE~0038$-$0345 is found to belong 
to the thick disk with a probability of 0.82,  the object HE~0038$-$0345 is found to be a member of the halo with a probability of 1. 
However, according to \cite{Koppelman_2018}, the stars with V$_{spa}$ $>$ 210 km s$^{-1}$ are bonafide halo objects. 
Both our program stars have total spatial velocities greater than this value (Table \ref{kinematic analysis}). 
If we follow this criterion, both the program stars belong to the Galactic halo.
A Toomre diagram (P$\rm_{LSR}$ vs V$\rm_{LSR}$) demonstrating the location of the program stars in the Galaxy is shown in Figure \ref{Toomre_diagram}. 
In this plot, the P$\rm_{LSR}$ = $\sqrt{U\rm_{LSR}^{2}+W\rm_{LSR}^{2}}$ is the perpendicular component of V$\rm_{LSR}$. 
As seen from the figure, the object HE~0038$-$0345 lies marginally in the disk (with the lower limit), while HE~1243$-$2408 is a bonafide halo object. 
We have also calculated the spatial velocities of the stars from \cite{Hansen_2019}, and 
the MW satellite galaxies. 
These are also shown in Figure \ref{Toomre_diagram}, for a comparison.  
Here, we have included only the kinematically selected halo stars (V$_{spa}$ $>$ 210 km s$^{-1}$) from \cite{Hansen_2019}. 
We note that, a few satellite galaxies have velocities typical of the Galactic disk.

{\footnotesize
\begin{table*}
\caption{Estimated spatial velocity and probabilities of membership 
to the Galactic population of the program stars} \label{kinematic analysis} 
\begin{tabular}{lcccccccc} 
\hline                       
Star name             & U$\rm_{LSR}$             & V$\rm_{LSR}$           & W$\rm_{LSR}$           &  P$\rm_{LSR}$             & V$_{spa}$    & P$_{thin}$ & P$_{thick}$ & P$_{halo}$ \\
                      & (km s$^{-1}$)          & (km s$^{-1}$)        & (km s$^{-1}$)     & (km s$^{-1}$)          & (km s$^{-1}$) &            &             &       \\
\hline
HE~0038$-$0345        & $-$199.22$\pm$15.07   & $-$41.72$\pm$2.70   & 74.15$\pm$2.07      & 212.57$\pm$15.19    & 216.63$\pm$13.54   & 0.00  & 0.82   & 0.18 \\
HE~1243$-$2408        & $-$6.01$\pm$1.89      & $-$390.06$\pm$4.70  & $-$72.57$\pm$3.85   & 72.81$\pm$4.28      & 396.80$\pm$5.37    & 0.00  & 0.00   & 1.00 \\
\hline
\end{tabular} 
\end{table*}
}

\subsection{Orbital properties}
We have computed the orbital characteristics of the program stars to understand more about their origin. 
The orbital properties such as Galactocentric distances, Galactocentric velocities, angular momenta, orbital energies, are 
estimated using the \texttt{galpy} (\url{http://github.com/jobovy/galpy}, \citealt{Bovy_2015}) python package for Galactic dynamics. 
We used the high-precision astrometric data - $\alpha$, $\delta$, $\mu_{\alpha}$, $\mu_{\delta}$, distance, 
and V$_{r}$, from Gaia DR3 \citep{Gaia_2022K} listed in Tables \ref{basic data of program stars} and \ref{orbital_parameters}. 
We set the Galactocentric distance of the sun at r$_{o}$ = 8 kpc, velocity at this distance at V$_{o}$ = 220 km s$^{-1}$ \citep{Bovy_2012}, 
and the vertical distance of the sun above the plane at z$_{o}$ = 20.8 pc \citep{Bennett_2019}. 
The adopted solar motion with respect to LSR is (U, V, W)$_{\odot}$ = (11.1, 12.2, 7.3) km/s \citep{Schonrich_2010}.

Using these parameters as initial conditions, we have integrated the orbits 
12 Gyrs backwards in a Milky Way-like potential. 
In the default \texttt{MWPotential2014} \citep{Bovy_2015} in \texttt{galpy}, the 
bulge is parametrized by a power-law density profile that exponentially cut-off at 1.9 kpc
with an exponent $-$1.8, the disk by Miyamoto-Nagai potential with vertical scale height of 
280 pc and radial scale-length of 3 kpc \citep{Miyamoto_1975}, and halo by 
Navarro-Frenk-White (NFW) potential with a scale-length of 16 kpc \citep{Navarro_1996}. 
This potential is characterized by a virial radius, r$_{vir}$ = 245 kpc and a virial mass, 
M$_{vir}$ = 0.8$\times$10$^{12}$ M$_{\odot}$. However, according to recent studies 
based on the Gaia proper motions, the virial mass of the Galaxy ranges 
from 1.1$\times$10$^{12}$ - 1.5$\times$10$^{12}$ M$_{\odot}$ (eg., \citealt{Watkins_2019, Wang_2020}). 
The lower mass of the halo in the default \texttt{MWPotential2014} lead to the 
inaccurate estimations of orbital parameters for Galactic halo objects, especially the 
orbital energies are highly over-estimated \citep{Kim_2019, Woody_2021}.
\cite{Kim_2019} noted that, many high-energy, retrograde stars in the outer halo are 
unbound in the default \texttt{MWPotential2014}. 
Hence, in our analysis, we have  used a scaled version of the default \texttt{MWPotential2014},
where the halo mass is increased by 50\%. In the scaled \texttt{MWPotential2014}, 
M$_{vir}$$\sim$1.2 $\times$ 10$^{12}$ M$_{\odot}$, in agreement with the recent observations. 

\begin{figure}
\centering
\includegraphics[width=\columnwidth]{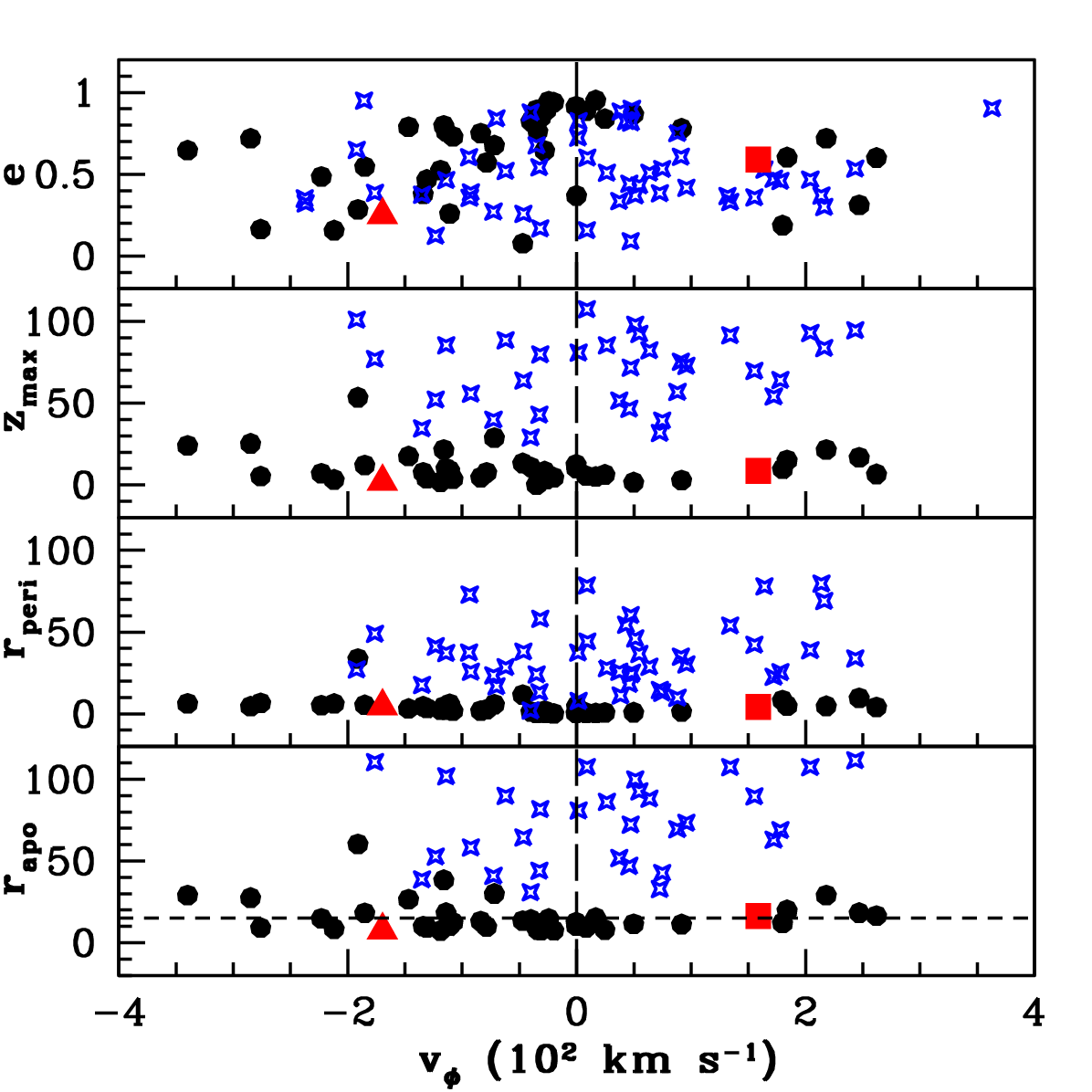}
\caption{The computed orbital parameters, v$\rm_{\phi}$, r$\rm_{apo}$, r$\rm_{peri}$, z$\rm_{max}$, and 
eccentricity, for the same objects as in Figure \ref{Toomre_diagram}. 
The Y-axes of the bottom three panels are in kpc.  
The horizontal short-dashed line in the bottom panel represents the 
apocentric distance (r$\rm_{apo}$$\sim$15 kpc, \citealt{Carollo_2007}) 
that separates outer halo from the inner halo. 
The vertical long-dashed line at v$\rm_{\phi}$=0 separates 
retrograde (v$\rm_{\phi}$$<$0) and prograde objects (v$\rm_{\phi}$$>$0). 
HE~0038$-$0345 (filled square) and HE~1243$-$2408 (filled triangle). } \label{orbital_parameters_diagram}
\end{figure}

\begin{figure}
\centering
\includegraphics[width=\columnwidth]{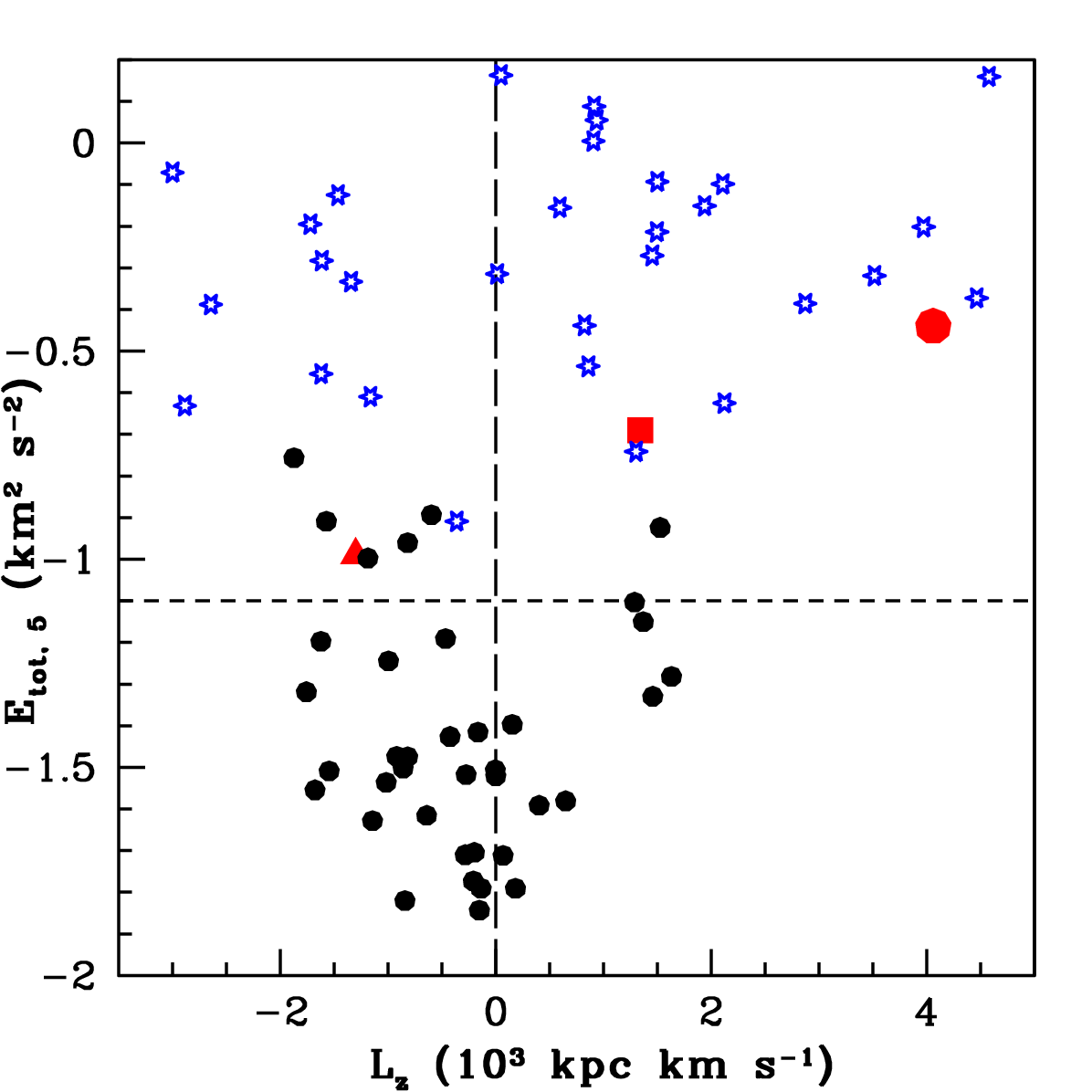}
\caption{The Lindblad diagram for the same objects as in Figure \ref{Toomre_diagram}. 
The Y-axis is the total orbital energy in 10$^{5}$ km$^{2}$ s$^{-2}$ (E$\rm_{tot, 5}$). 
The horizontal short-dashed line represents the high-energy limit 
E$\rm_{tot}$=$-$1.1$\times$10$^{5}$ km$^{2}$ s$^{-2}$ \citep{Myeong_2019}.  
The vertical long-dashed line at L$\rm_{z}$=0 separates 
retrograde (L$\rm_{z}$$<$0) and prograde objects (L$\rm_{z}$$>$0). 
HE~0038$-$0345 (filled square) and HE~1243$-$2408 (filled triangle). 
The filled red circle is LAMOSTJ045019.27 + 394758.7, a sculptor 
dwarf galaxy escapee from the literature \citep{Purandardas_2022}. } \label{Lindblad_diagram}
\end{figure}

The computed orbital parameters - apocentric distance (r$\rm_{apo}$), pericentric distance (r$\rm_{peri}$), 
maximum height from the Galactic plane (z$\rm_{max}$), eccentricity (e), total orbital energy (E$\rm_{tot}$), 
components of angular momentum (L$\rm_{x}$, L$\rm_{y}$, L$\rm_{z}$), Galactocentric tangential velocity (v$\rm_{\phi}$),  
and 3D Galactocentric distance, of the program stars are given in Table \ref{orbital_parameters}. 
For a comparison purpose, we have calculated these parameters for the halo stars that were kinematically chosen from 
\cite{Hansen_2019}, as well as for the 50 MW satellite galaxies, using the same procedure and potential as the program stars.
The resulting parameters are shown in Figures \ref{orbital_parameters_diagram} and \ref{Lindblad_diagram}, however, we 
have excluded the unbound (E$\rm_{tot}$$>$0) objects.  
In the Lindblad diagram (E$\rm_{tot}$ - L$\rm_{z}$; Figure \ref{Lindblad_diagram}), the star LAMOSTJ045019.27 + 394758.7, 
a sculptor dwarf galaxy escapee from the literature \citep{Purandardas_2022}, is also shown. 

We note  from Figure \ref{Lindblad_diagram} that our program stars and the escapee star 
LAMOSTJ045019.27 + 394758.7 have energies on par with a few satellite galaxies. 
The program stars and the halo stars have similar orbital parameters, as seen in Figure \ref{orbital_parameters_diagram}. 
We note that, in contrast to halo stars and program stars, satellite galaxies exhibit greater
r$\rm_{apo}$, r$\rm_{peri}$, and z$\rm_{max}$ values, yet all three samples' orbital eccentricities 
fall within the same range.

From Figure \ref{Lindblad_diagram}, we can see that both our program stars are bound to the Galaxy (E$\rm_{tot}$$<$0),
and belong to the high-energy ($>$$-$1.1$\times$10$^{5}$ km$^{2}$ s$^{-2}$, \citealt{Myeong_2019}) region. 
The object HE~0038$-$0345 has a prograde (L$\rm_{z}$, v$\rm_{\phi}$$>$ 0) orbit, whereas the object HE1243$-$2408 
has a retrograde (L$\rm_{z}$, v$\rm_{\phi}$$<$ 0) motion. While HE~0038$-$0345 belongs to the outer 
halo (outer halo starts from r$\rm_{apo}$$\sim$15 - 20 kpc, \citealt{Carollo_2007}) with r$\rm_{apo}$$\sim$16.27 kpc,  
HE~1243$-$2408 belongs to the inner halo (6$\leq$r$\rm_{apo}$$\leq$13 kpc, \citealt{Roderer_2018}) 
with r$\rm_{apo}$$\sim$7.70 kpc.  
To sum up, the star HE~0038$-$0345 is found to be a high-energy, prograde, outer halo object, 
and HE~1243$-$2408 is found to be a high-energy, retrograde, inner halo object.

{\footnotesize
\begin{table*}
\caption{Galactic orbital parameters of the program stars computed using \texttt{galpy} in a scaled \texttt{MWPotential2014}.} \label{orbital_parameters} 
\begin{tabular}{lcc} 
\hline                       
                                                                                                & HE~0038$-$0345      & HE~1243$-$2408            \\
\hline
\textit{Astrometric data from Gaia DR3:}                                                                  &                      &                    \\
Galactic longitude, l (degrees)                                                                 & 116.56                & 301.32    \\
Galactic latitude, b (degrees)                                                                  & $-$66.23              & +38.44   \\
Proper motions, ($\mu_{\alpha}$, $\mu_{\delta}$) (mas/yr)                                       & (19.26, 8.46)         & ($-$51.21, $-$58.59)  \\
parallax, $\pi$ (mas)                                                                           & 0.45                  & 1.05          \\
Radial velocity, V$_{r}$ (km s$^{-1}$)                                                          & $-$42.66              & 213.35     \\
\hline
3D Galactocentric distance, r$\rm_{gal}$ (kpc)                                                  & 8.69                  & 7.67     \\
Apocentric distance, r$\rm_{apo}$ (kpc)                                                         & 16.27                 & 7.70   \\
Pericentric distance, r$\rm_{peri}$ (kpc)                                                       & 4.25                  & 4.60   \\
Maximum height from the Galactic plane, z$\rm_{max}$ (kpc)                             & 8.65                  & 1.99   \\
Orbital eccentricity, e                                                                         & 0.59                  & 0.25      \\
Total orbital energy, E$\rm_{tot}$ (10$^{5}$ km$^{2}$ s$^{-2}$)                                 & $-$0.69               & $-$0.99  \\
Angular momentum, (L$\rm_{x}$, L$\rm_{y}$, L$\rm_{z}$) (10$^{3}$ kpc km s$^{-1}$)               & (0.42, $-$1.03, 1.34) & (0.15, 0.55, $-$ 1.30)    \\
Galactocentric tangential velocity, v$\rm_{\phi}$ (km s$^{-1}$)                                 & 158.70                & $-$169.74   \\
\hline
\end{tabular} 
\end{table*}
}

\section{Association of program stars to recognized Milky Way structures} \label{section_structure_association}

\noindent The likely-hood of accretion origin of the stellar halo with higher apocenters and/or retrograde orbits have been
discussed by several authors \citep{Roderer_2018, Myeong_2019, Dietz_2020, Aguado_2021}. 
The chemodynamical analysis of samples of globular clusters, stars, stellar streams, satellite galaxies etc., 
have shown that the Galactic halo shows different substructures in the energy - action 
space (E$\rm_{tot}$ - J$\rm_{\phi}$; J$\rm_{\phi}$$\equiv$L$\rm_{z}$), owing to various 
major and minor accretion events in the past \citep{Myeong_2019, Naidu_2020, Helmi_2020, Aguado_2021, Dietz_2021, Woody_2021, Malhan_2022, Zepeda_2023}. 
In addition to the substructures associated with the major massive accretion 
events such as Gaia-Sausage-Enceladus (GSE), Kraken, and sequoia, several other 
structures such as LMS-1/Wukong, Thamnos, Metal-weak Thick Disk (MWTD), I’itoi, Splashed Disk (SD), 
Sagittarius (Sgr), In Situ Halo, Arjuna, Aleph, Helmi Streams etc. associated with 
minor mergers (massive/low-mass) are found 
to contribute significantly to the hierarchical Galactic halo 
assembly \citep{Helmi_2018, Helmi_2020, Naidu_2020, Naidu_2022, Malhan_2022, Zepeda_2023}.  
Each of these substructures show different chemical and dynamical properties. 
\cite{Zepeda_2023} grouped a sample of 644 CEMP stars to different 
Chemodynamically Tagged Groups (CDTGs) from their chemical 
abundances and orbital properties, and associated these CTDGs to various aforementioned Milky Way substructures.  

We have performed a chemodynamical analysis of our program stars, and compared their properties with those of  
Milky Way substructures to see if they could be associated with any of these accretion events.
The properties of the MW structures relevant to this study are listed in Table \ref{MW_structure_properties}. 
We found that none of our object meet the criteria of In situ halo (Table \ref{MW_structure_properties}), 
and hence they belong to the ex situ halo component. 
According to \cite{Dietz_2020}, while the high-energy, retrograde, outer halo, metal-poor 
objects are donated by the Sequoia dwarf progenitor galaxy, 
the EMP ([FeH]$<$$-$2), high-energy, prograde, outer halo objects are 
accreted onto the halo from a series of minor mergers. 
Recent studies have shown that the high-energy, retrograde halo is associated with 
Arjuna, Sequoia, and I'itoi events, depending on the 
metallicity \citep{Naidu_2020, Malhan_2022} \footnote{\citealt{Naidu_2020} 
and \citealt{Malhan_2022} used a coordinate system where
prograde orbits are those with v$\rm_{\phi}$, L$\rm_{z}$ $<$ 0.}. 
The Metallicity Distribution Function (MDF) of this component of halo has shown 
that the stars with [Fe/H]$>$$-$1.5 belong to Arjuna, those with $-$2$<$[Fe/H]$<$$-$1.5 to Sequoia, and  
with [Fe/H]$<$$-$2 to I'itoi structures. I'itoi, on the other hand, might be thought of as a metal-poor tail
of Sequoia and/or Arjuna \citep{Naidu_2020}.
We found from our inspection that our high-energy, retrograde object HE~1243$-$2408 satisfies the
orbital as well as the chemical properties ([Fe/H], [$\alpha$/Fe] \footnote{The mean of [Mg/Fe] and [Ca/Fe] 
values is taken as the [$\alpha$/Fe] of program stars, that are 0.57 and 0.58 respectively in 
HE~0038$-$0345 and HE~1243$-$2408.}) of the I'itoi substructure.  

Our high-energy, prograde object HE~0038$-$0345 does not fully satisfy the requirement of any specific MW structure; 
it falls under the unclassified debris class of \cite{Naidu_2020}. However, they found that certain classes of this unclassified debris
clustered around the selection boxes of various structures. These stars may be plausible members of that 
particular structure in whose close proximity they tend to cluster. 
If we consider such a scenario, the object HE~0038$-$0345 may plausibly belong to either the Sagittarius or 
the GSE. We note that even though the MDF of the GSE and the Sgr peaks at higher metallicities ([Fe/H]$\sim$$-$1.3 and $-$1.0 respectively; \citealt{Myeong_2018, Dietz_2020, Naidu_2020}), 
they could contribute a few stars at metallicities as low as $-$3 (\citealt{Zepeda_2023},  Figures 8 and 11 of \citealt{Naidu_2020}). 
The star HE~0038$-$0345 satisfies all the properties of the Sgr except for L$\rm_{y}$ $<$ $-$0.3L$\rm_{z}$ $-$ 2.5. 
Alternatively, it satisfies all the properties of the GSE structure except for the eccentricity (e$>$0.7); but with an estimated eccentricity of $\sim$0.6, this may be a possible member of the GSE as well.

Finally, we wind up this section with a discussion on the mass of the possible progenitor galaxies of the program stars. 
From the above discussion, we infer the likely association of our program stars with structures such as I'itoi, 
metal-poor extension of Arjuna/Sequoia (Arjuna may be a highly retrograde extension of GSE itself, as noted by \citealt{Naidu_2020}), 
Sgr, or GSE. The progenitor satellite galaxies of all these events have 
stellar masses M$_{\star}$ $\geq$ 10$^{8}$ - 10$^{9}$ M$_{\odot}$ \citep{Myeong_2019, Kruijssen_2020, Naidu_2020, Naidu_2022}.
This mass of the progenitor galaxies is in agreement with the typical masses of dSphs, and hence 
this may justify the origin of program stars in massive ($\geq$ 10$^{9}$ M$_{\odot}$) dSph galaxies 
as discussed in Section \ref{Section_possible_origin}.

{\footnotesize
\begin{table*}
\caption{Chemodynamical properties of MW substructures relevant to this study$^{\dagger}$.} \label{MW_structure_properties} 
\begin{tabular}{lcccc} 
\hline                       
                                            & In situ halo        & I'itoi         & Gaia-Sausage-Enceladus (GSE)   & Sagittarius (Sgr)  \\
\hline
 E$\rm_{tot}$ (10$^{5}$ km$^{2}$ s$^{-2}$)  & $<$$-$0.8           & $>$$-$1.25     & $<$$-$0.5                      & $\sim$$-$0.9 - $-$0.4    \\
 L$\rm_{z}$ (10$^{3}$ kpc km s$^{-1}$)      & $>$1.0              & $<$$-$0.7      & $\sim$$-$150 - +150            & $>$0          \\
 v$\rm_{\phi}$ (km s$^{-1}$)                & $>$200              & $<$$-$100      & $\sim$$-$150 - +200            & $>$ 50 \\
 Eccentricity, e                            & $<$0.3              & 0.1 - 0.7      & $>$0.7                         & $\sim$0.4 - 0.7      \\
 r$\rm_{gal}$ (kpc)                         & $\leq$25            & $\leq$40       & $\leq$50$^{\ddagger}$          & $>$8          \\
Metallicty, [Fe/H]$^{\ast}$                 & $>$$-$1             & $<$$-$2        & $\sim$$-$1.3                   & $\sim$$-$1.0    \\
$$\rm[$\alpha$/Fe]$$                        & $>$0.25 - 0.5       & 0.2 - 0.6      & $\sim$0.3 - 0.6$^{\star}$      & $\sim$0.4 - 0.6$^{\star}$  \\
L$\rm_{y}$ (10$^{3}$ kpc km s$^{-1}$)       & --                  & --             & --                          & $<$ $-$0.3L$\rm_{z}$ $-$ 2.5 \\
\hline
\end{tabular} \\
$^{\dagger}$ Adopted from \cite{Naidu_2020} \\
$^{\ddagger}$ The sample is limited up to 50 kpc; $^{\ast}$ Peak of MDF; $^{\star}$ At [Fe/H]$\sim$$-$3.0 \\
\end{table*} 
}

\section{conclusions} \label{section_conclusion}
The high-resolution spectroscopic analysis of two CEMP star candidates, 
HE~0038$-$0345 and HE~1243$-$2408, selected from the HES is carried out. 
We present the first-time detailed chemodynamical analysis of these stars. 
Both the objects are found to be extremely metal-poor stars with [Fe/H]$\sim$$-$3.0. 
The estimated [C/Fe] values, 1.05 and 1.03 (after applying the correction), and 
[Ba/Fe] values, $-$0.11 and $-$0.18, for HE~0038$-$0345 and HE~1243$-$2408 respectively, 
show that both the objects are CEMP-no stars. From the estimated [C/N] values, the stars are 
found to have undergone mixing that altered the surface carbon abundances. 

The location of the program stars on the A(C) - [Fe/H] diagram shows that both the objects 
belong to the Group II CEMP-no stars of \cite{Yoon_2016}. The observed abundances of Na, Mg, and Ba 
also confirms the assignment of the program stars to the Group II. The [Mg/C] and [Ca/Fe] ratios of 
the objects reveal that  they are multi-enriched. Based on the multi-enrichment and other 
abundance profile analysis, we found that the program stars may have formed in the dSph satellite galaxies. 
 This result is in strong agreement with the Galactic assembly and  accretion of halo CEMP-no stars from mini-halos, 
as suggested by \cite{Yoon_2018}, \cite{Yoon_2019}, and \cite{Yoon_2020}.  
Furthermore, from the finding that the program stars were likely born in 
the dSph-like systems, and based on the [C/Fe], [Fe/H], and [Sr/Ba] values, we found that 
the likely progenitors of the program stars are Pop II core-collapse SNe. 

We have performed a detailed kinematic analysis and derived various orbital 
and dynamical properties of the program stars to derive non-abundance-based conclusions 
about their origin. The kinematic analysis shows that the objects HE~0038$-$0345 and 
HE~1243$-$2408 are members of the Galactic halo with V$_{spa}$$>$210 km s$^{-1}$. 
From the dynamical and orbital parameters, the object HE~0038$-$0345 is found to be 
a high-energy, prograde, outer halo object, and HE~1243$-$2408 is found to be a 
high-energy, retrograde, inner halo object. Our detailed chemodynamical analysis shows 
that HE~1243$-$2408 belongs to the I'itoi structure (which can be likely a metal-poor 
tail of the Sequoia and/or Arjuna), where as HE~0038$-$0345 is a plausible member 
of the Sgr or the GSE accretion events. The mass of the progenitor galaxies of the program 
stars inferred from their dynamics (M$_{\star}$ $\geq$ 10$^{8}$ - 10$^{9}$ M$_{\odot}$), 
agrees with our speculations based on the chemical abundances that the stars may 
have originated in massive ($\geq$ 10$^{9}$ M$_{\odot}$) dSph like systems.

 \section{ACKNOWLEDGMENT}
Based on observations obtained with the HERMES spectrograph, supported by the Fund for
Scientific Research of Flanders (FWO), the Research Council of K.U.Leuven,
the Fonds National de la Recherche Scientifique (F.R.S.- FNRS), Belgium, the
Royal Observatory of Belgium, the Observatoire de Gen\'eve, Switzerland and the
Th\"uringer Landessternwarte Tautenburg, Germany.
We thank Alain Jorissen for sharing the HERMES spectra used in this study.
This work made use of the SIMBAD astronomical database, operated
at CDS, Strasbourg, France, and the NASA ADS, USA.
This work has made use of data from the European Space Agency (ESA) mission
{\it Gaia} (\url{https://www.cosmos.esa.int/gaia}), processed by the {\it Gaia}
Data Processing and Analysis Consortium (DPAC,
\url{https://www.cosmos.esa.int/web/gaia/dpac/consortium}). Funding for the DPAC
has been provided by national institutions, in particular the institutions
participating in the {\it Gaia} Multilateral Agreement.
We are grateful for the support received from the Belgo-Indian Network for Astronomy \&
Astrophysics  project BINA - 2  (DST/INT/Belg/P-02 (India) and BL/11/IN07 (Belgium)).\\

\noindent
{\bf Data Availability}\\
The data underlying this article will be shared on reasonable request to the authors.

\bibliographystyle{mnras}
\bibliography{Bibliography}

\begin{thebibliography}{}
\makeatletter
\relax
\def\mn@urlcharsother{\let\do\@makeother \do\$\do\&\do\#\do\^\do\_\do\%\do\~}
\def\mn@doi{\begingroup\mn@urlcharsother \@ifnextchar [ {\mn@doi@}
  {\mn@doi@[]}}
\def\mn@doi@[#1]#2{\def\@tempa{#1}\ifx\@tempa\@empty \href
  {http://dx.doi.org/#2} {doi:#2}\else \href {http://dx.doi.org/#2} {#1}\fi
  \endgroup}
\def\mn@eprint#1#2{\mn@eprint@#1:#2::\@nil}
\def\mn@eprint@arXiv#1{\href {http://arxiv.org/abs/#1} {{\tt arXiv:#1}}}
\def\mn@eprint@dblp#1{\href {http://dblp.uni-trier.de/rec/bibtex/#1.xml}
  {dblp:#1}}
\def\mn@eprint@#1:#2:#3:#4\@nil{\def\@tempa {#1}\def\@tempb {#2}\def\@tempc
  {#3}\ifx \@tempc \@empty \let \@tempc \@tempb \let \@tempb \@tempa \fi \ifx
  \@tempb \@empty \def\@tempb {arXiv}\fi \@ifundefined
  {mn@eprint@\@tempb}{\@tempb:\@tempc}{\expandafter \expandafter \csname
  mn@eprint@\@tempb\endcsname \expandafter{\@tempc}}}

\bibitem[\protect\citeauthoryear{{Aguado} et~al.,}{{Aguado}
  et~al.}{2021}]{Aguado_2021}
{Aguado} D.~S.,  et~al., 2021, \mn@doi [\apjl] {10.3847/2041-8213/abdbb8},
  \href {https://ui.adsabs.harvard.edu/abs/2021ApJ...908L...8A} {908, L8}

\bibitem[\protect\citeauthoryear{{Amorisco}}{{Amorisco}}{2017}]{Amorisco_2017}
{Amorisco} N.~C.,  2017, \mn@doi [\mnras] {10.1093/mnras/stw2229}, \href
  {https://ui.adsabs.harvard.edu/abs/2017MNRAS.464.2882A} {464, 2882}

\bibitem[\protect\citeauthoryear{{Andrievsky}, {Spite}, {Korotin}, {Spite},
  {Bonifacio}, {Cayrel}, {Hill}  \& {Fran{\c{c}}ois}}{{Andrievsky}
  et~al.}{2007}]{Andrievsky_2007}
{Andrievsky} S.~M.,  {Spite} M.,  {Korotin} S.~A.,  {Spite} F.,  {Bonifacio}
  P.,  {Cayrel} R.,  {Hill} V.,   {Fran{\c{c}}ois} P.,  2007, \mn@doi [\aap]
  {10.1051/0004-6361:20066232}, \href
  {https://ui.adsabs.harvard.edu/abs/2007A&A...464.1081A} {464, 1081}

\bibitem[\protect\citeauthoryear{{Aoki}, {Ryan}, {Norris}, {Beers}, {Ando}  \&
  {Tsangarides}}{{Aoki} et~al.}{2002}]{Aoki_2002}
{Aoki} W.,  {Ryan} S.~G.,  {Norris} J.~E.,  {Beers} T.~C.,  {Ando} H.,
  {Tsangarides} S.,  2002, \mn@doi [\apj] {10.1086/343885}, \href
  {https://ui.adsabs.harvard.edu/abs/2002ApJ...580.1149A} {580, 1149}

\bibitem[\protect\citeauthoryear{{Aoki} et~al.,}{{Aoki}
  et~al.}{2005}]{Aoki_2005}
{Aoki} W.,  et~al., 2005, \mn@doi [\apj] {10.1086/432862}, \href
  {https://ui.adsabs.harvard.edu/abs/2005ApJ...632..611A} {632, 611}

\bibitem[\protect\citeauthoryear{{Aoki}, {Beers}, {Christlieb}, {Norris},
  {Ryan}  \& {Tsangarides}}{{Aoki} et~al.}{2007}]{Aoki_2007}
{Aoki} W.,  {Beers} T.~C.,  {Christlieb} N.,  {Norris} J.~E.,  {Ryan} S.~G.,
  {Tsangarides} S.,  2007, \mn@doi [\apj] {10.1086/509817}, \href
  {https://ui.adsabs.harvard.edu/abs/2007ApJ...655..492A} {655, 492}

\bibitem[\protect\citeauthoryear{{Aoki}, {Matsuno}, {Honda}, {Ishigaki}, {Li},
  {Suda}  \& {Kumar}}{{Aoki} et~al.}{2018}]{Aoki_2018}
{Aoki} W.,  {Matsuno} T.,  {Honda} S.,  {Ishigaki} M.~N.,  {Li} H.,  {Suda} T.,
    {Kumar} Y.~B.,  2018, \mn@doi [\pasj] {10.1093/pasj/psy092}, \href
  {https://ui.adsabs.harvard.edu/abs/2018PASJ...70...94A} {70, 94}

\bibitem[\protect\citeauthoryear{{Arentsen}, {Starkenburg}, {Shetrone}, {Venn},
  {Depagne}  \& {McConnachie}}{{Arentsen} et~al.}{2019}]{Arentsen_2019b}
{Arentsen} A.,  {Starkenburg} E.,  {Shetrone} M.~D.,  {Venn} K.~A.,  {Depagne}
  {\'E}.,   {McConnachie} A.~W.,  2019, \mn@doi [\aap]
  {10.1051/0004-6361/201834146}, \href
  {https://ui.adsabs.harvard.edu/abs/2019A&A...621A.108A} {621, A108}

\bibitem[\protect\citeauthoryear{{Asplund}, {Grevesse}, {Sauval}  \&
  {Scott}}{{Asplund} et~al.}{2009}]{Asplund_2009}
{Asplund} M.,  {Grevesse} N.,  {Sauval} A.~J.,   {Scott} P.,  2009, \mn@doi
  [\araa] {10.1146/annurev.astro.46.060407.145222}, \href
  {https://ui.adsabs.harvard.edu/abs/2009ARA&A..47..481A} {47, 481}

\bibitem[\protect\citeauthoryear{{Babusiaux} et~al.,}{{Babusiaux}
  et~al.}{2022}]{Babusiaux_2022}
{Babusiaux} C.,  et~al., 2022, \mn@doi [arXiv e-prints]
  {10.48550/arXiv.2206.05989}, \href
  {https://ui.adsabs.harvard.edu/abs/2022arXiv220605989B} {p. arXiv:2206.05989}

\bibitem[\protect\citeauthoryear{{Banerjee}, {Qian}  \& {Heger}}{{Banerjee}
  et~al.}{2018}]{Banerjee_2018}
{Banerjee} P.,  {Qian} Y.-Z.,   {Heger} A.,  2018, \mn@doi [\apj]
  {10.3847/1538-4357/aadb8c}, \href
  {https://ui.adsabs.harvard.edu/abs/2018ApJ...865..120B} {865, 120}

\bibitem[\protect\citeauthoryear{{Beers}}{{Beers}}{1999}]{Beers_1999}
{Beers} T.~C.,  1999, in {Gibson} B.~K.,  {Axelrod} R.~S.,   {Putman} M.~E.,
  eds,  Astronomical Society of the Pacific Conference Series Vol. 165, The
  Third Stromlo Symposium: The Galactic Halo. p.~202

\bibitem[\protect\citeauthoryear{{Beers} \& {Christlieb}}{{Beers} \&
  {Christlieb}}{2005}]{Beers_2005}
{Beers} T.~C.,  {Christlieb} N.,  2005, \mn@doi [\araa]
  {10.1146/annurev.astro.42.053102.134057}, \href
  {https://ui.adsabs.harvard.edu/abs/2005ARA&A..43..531B} {43, 531}

\bibitem[\protect\citeauthoryear{{Beers} et~al.,}{{Beers}
  et~al.}{2007}]{Beers_2007}
{Beers} T.~C.,  et~al., 2007, \mn@doi [\apjs] {10.1086/509324}, \href
  {https://ui.adsabs.harvard.edu/abs/2007ApJS..168..128B} {168, 128}

\bibitem[\protect\citeauthoryear{{Beers} et~al.,}{{Beers}
  et~al.}{2017}]{Beers_2017}
{Beers} T.~C.,  et~al., 2017, \mn@doi [\apj] {10.3847/1538-4357/835/1/81},
  \href {https://ui.adsabs.harvard.edu/abs/2017ApJ...835...81B} {835, 81}

\bibitem[\protect\citeauthoryear{{Belokurov}, {Erkal}, {Evans}, {Koposov}  \&
  {Deason}}{{Belokurov} et~al.}{2018}]{Belokurov_2018}
{Belokurov} V.,  {Erkal} D.,  {Evans} N.~W.,  {Koposov} S.~E.,   {Deason}
  A.~J.,  2018, \mn@doi [\mnras] {10.1093/mnras/sty982}, \href
  {https://ui.adsabs.harvard.edu/abs/2018MNRAS.478..611B} {478, 611}

\bibitem[\protect\citeauthoryear{{Bennett} \& {Bovy}}{{Bennett} \&
  {Bovy}}{2019}]{Bennett_2019}
{Bennett} M.,  {Bovy} J.,  2019, \mn@doi [\mnras] {10.1093/mnras/sty2813},
  \href {https://ui.adsabs.harvard.edu/abs/2019MNRAS.482.1417B} {482, 1417}

\bibitem[\protect\citeauthoryear{{Bensby}, {Feltzing}  \&
  {Lundstr{\"o}m}}{{Bensby} et~al.}{2003}]{Bensby_2003}
{Bensby} T.,  {Feltzing} S.,   {Lundstr{\"o}m} I.,  2003, \mn@doi [\aap]
  {10.1051/0004-6361:20031213}, \href
  {https://ui.adsabs.harvard.edu/abs/2003A&A...410..527B} {410, 527}

\bibitem[\protect\citeauthoryear{{Bensby}, {Feltzing}  \&
  {Lundstr{\"o}m}}{{Bensby} et~al.}{2004}]{Bensby_2004}
{Bensby} T.,  {Feltzing} S.,   {Lundstr{\"o}m} I.,  2004, \mn@doi [\aap]
  {10.1051/0004-6361:20031655}, \href
  {https://ui.adsabs.harvard.edu/abs/2004A&A...415..155B} {415, 155}

\bibitem[\protect\citeauthoryear{{Bessell}, {Castelli}  \& {Plez}}{{Bessell}
  et~al.}{1998}]{Bessel_1998}
{Bessell} M.~S.,  {Castelli} F.,   {Plez} B.,  1998, \aap, \href
  {https://ui.adsabs.harvard.edu/abs/1998A&A...333..231B} {333, 231}

\bibitem[\protect\citeauthoryear{{Bonifacio} et~al.,}{{Bonifacio}
  et~al.}{2015}]{Bonifacio_2015}
{Bonifacio} P.,  et~al., 2015, \mn@doi [\aap] {10.1051/0004-6361/201425266},
  \href {https://ui.adsabs.harvard.edu/abs/2015A&A...579A..28B} {579, A28}

\bibitem[\protect\citeauthoryear{{Bovy}}{{Bovy}}{2015}]{Bovy_2015}
{Bovy} J.,  2015, \mn@doi [\apjs] {10.1088/0067-0049/216/2/29}, \href
  {https://ui.adsabs.harvard.edu/abs/2015ApJS..216...29B} {216, 29}

\bibitem[\protect\citeauthoryear{{Bovy} et~al.,}{{Bovy}
  et~al.}{2012}]{Bovy_2012}
{Bovy} J.,  et~al., 2012, \mn@doi [\apj] {10.1088/0004-637X/759/2/131}, \href
  {https://ui.adsabs.harvard.edu/abs/2012ApJ...759..131B} {759, 131}

\bibitem[\protect\citeauthoryear{{Buder} et~al.,}{{Buder}
  et~al.}{2018}]{Buder_2018}
{Buder} S.,  et~al., 2018, \mn@doi [\mnras] {10.1093/mnras/sty1281}, \href
  {https://ui.adsabs.harvard.edu/abs/2018MNRAS.478.4513B} {478, 4513}

\bibitem[\protect\citeauthoryear{{Carollo} et~al.,}{{Carollo}
  et~al.}{2007}]{Carollo_2007}
{Carollo} D.,  et~al., 2007, \mn@doi [\nat] {10.1038/nature06460}, \href
  {https://ui.adsabs.harvard.edu/abs/2007Natur.450.1020C} {450, 1020}

\bibitem[\protect\citeauthoryear{{Carollo} et~al.,}{{Carollo}
  et~al.}{2012}]{Carollo_2012}
{Carollo} D.,  et~al., 2012, \mn@doi [\apj] {10.1088/0004-637X/744/2/195},
  \href {https://ui.adsabs.harvard.edu/abs/2012ApJ...744..195C} {744, 195}

\bibitem[\protect\citeauthoryear{{Castelli} \& {Kurucz}}{{Castelli} \&
  {Kurucz}}{2003}]{Castelli_2003}
{Castelli} F.,  {Kurucz} R.~L.,  2003, in {Piskunov} N.,  {Weiss} W.~W.,
  {Gray} D.~F.,  eds,  Proceedings of the IAU Symp. No 210 Vol. 210, Modelling
  of Stellar Atmospheres. p.~A20 (\mn@eprint {arXiv} {astro-ph/0405087})

\bibitem[\protect\citeauthoryear{{Chiaki}, {Tominaga}  \& {Nozawa}}{{Chiaki}
  et~al.}{2017}]{Chiaki_2017}
{Chiaki} G.,  {Tominaga} N.,   {Nozawa} T.,  2017, \mn@doi [\mnras]
  {10.1093/mnrasl/slx163}, \href
  {https://ui.adsabs.harvard.edu/abs/2017MNRAS.472L.115C} {472, L115}

\bibitem[\protect\citeauthoryear{{Chiaki}, {Susa}  \& {Hirano}}{{Chiaki}
  et~al.}{2018}]{Chiaki_2018}
{Chiaki} G.,  {Susa} H.,   {Hirano} S.,  2018, \mn@doi [\mnras]
  {10.1093/mnras/sty040}, \href
  {https://ui.adsabs.harvard.edu/abs/2018MNRAS.475.4378C} {475, 4378}

\bibitem[\protect\citeauthoryear{{Choplin}, {Hirschi}, {Meynet}  \&
  {Ekstr{\"o}m}}{{Choplin} et~al.}{2017}]{Choplin_2017}
{Choplin} A.,  {Hirschi} R.,  {Meynet} G.,   {Ekstr{\"o}m} S.,  2017, \mn@doi
  [\aap] {10.1051/0004-6361/201731948}, \href
  {https://ui.adsabs.harvard.edu/abs/2017A&A...607L...3C} {607, L3}

\bibitem[\protect\citeauthoryear{{Clarkson}, {Herwig}  \&
  {Pignatari}}{{Clarkson} et~al.}{2018}]{Clarkson_2018}
{Clarkson} O.,  {Herwig} F.,   {Pignatari} M.,  2018, \mn@doi [\mnras]
  {10.1093/mnrasl/slx190}, \href
  {https://ui.adsabs.harvard.edu/abs/2018MNRAS.474L..37C} {474, L37}

\bibitem[\protect\citeauthoryear{{Collet}, {Asplund}  \& {Trampedach}}{{Collet}
  et~al.}{2007}]{Collet_2007}
{Collet} R.,  {Asplund} M.,   {Trampedach} R.,  2007, \mn@doi [\aap]
  {10.1051/0004-6361:20066321}, \href
  {https://ui.adsabs.harvard.edu/abs/2007A&A...469..687C} {469, 687}

\bibitem[\protect\citeauthoryear{{Cooke} \& {Madau}}{{Cooke} \&
  {Madau}}{2014}]{Cooke_2014}
{Cooke} R.~J.,  {Madau} P.,  2014, \mn@doi [\apj]
  {10.1088/0004-637X/791/2/116}, \href
  {https://ui.adsabs.harvard.edu/abs/2014ApJ...791..116C} {791, 116}

\bibitem[\protect\citeauthoryear{{Deason}, {Belokurov}, {Evans}  \&
  {Johnston}}{{Deason} et~al.}{2013}]{Deason_2013}
{Deason} A.~J.,  {Belokurov} V.,  {Evans} N.~W.,   {Johnston} K.~V.,  2013,
  \mn@doi [\apj] {10.1088/0004-637X/763/2/113}, \href
  {https://ui.adsabs.harvard.edu/abs/2013ApJ...763..113D} {763, 113}

\bibitem[\protect\citeauthoryear{{Dietz}, {Yoon}, {Beers}  \& {Placco}}{{Dietz}
  et~al.}{2020}]{Dietz_2020}
{Dietz} S.~E.,  {Yoon} J.,  {Beers} T.~C.,   {Placco} V.~M.,  2020, \mn@doi
  [\apj] {10.3847/1538-4357/ab7fa4}, \href
  {https://ui.adsabs.harvard.edu/abs/2020ApJ...894...34D} {894, 34}

\bibitem[\protect\citeauthoryear{{Dietz}, {Yoon}, {Beers}, {Placco}  \&
  {Lee}}{{Dietz} et~al.}{2021}]{Dietz_2021}
{Dietz} S.~E.,  {Yoon} J.,  {Beers} T.~C.,  {Placco} V.~M.,   {Lee} Y.~S.,
  2021, \mn@doi [\apj] {10.3847/1538-4357/abefd6}, \href
  {https://ui.adsabs.harvard.edu/abs/2021ApJ...914..100D} {914, 100}

\bibitem[\protect\citeauthoryear{{Dobrovolskas}, {Ku{\v{c}}inskas}, {Steffen},
  {Ludwig}, {Prakapavi{\v{c}}ius}, {Klevas}, {Caffau}  \&
  {Bonifacio}}{{Dobrovolskas} et~al.}{2013}]{Dobrovolskas_2013}
{Dobrovolskas} V.,  {Ku{\v{c}}inskas} A.,  {Steffen} M.,  {Ludwig} H.~G.,
  {Prakapavi{\v{c}}ius} D.,  {Klevas} J.,  {Caffau} E.,   {Bonifacio} P.,
  2013, \mn@doi [\aap] {10.1051/0004-6361/201321036}, \href
  {https://ui.adsabs.harvard.edu/abs/2013A&A...559A.102D} {559, A102}

\bibitem[\protect\citeauthoryear{{Ezzeddine} et~al.,}{{Ezzeddine}
  et~al.}{2019}]{Ezzedine_2019}
{Ezzeddine} R.,  et~al., 2019, \mn@doi [\apj] {10.3847/1538-4357/ab14e7}, \href
  {https://ui.adsabs.harvard.edu/abs/2019ApJ...876...97E} {876, 97}

\bibitem[\protect\citeauthoryear{{Forbes}}{{Forbes}}{2020}]{Forbes_2020}
{Forbes} D.~A.,  2020, \mn@doi [\mnras] {10.1093/mnras/staa245}, \href
  {https://ui.adsabs.harvard.edu/abs/2020MNRAS.493..847F} {493, 847}

\bibitem[\protect\citeauthoryear{{Frebel} \& {Ji}}{{Frebel} \&
  {Ji}}{2023}]{Frebel_2023}
{Frebel} A.,  {Ji} A.~P.,  2023, \mn@doi [arXiv e-prints]
  {10.48550/arXiv.2302.09188}, \href
  {https://ui.adsabs.harvard.edu/abs/2023arXiv230209188F} {p. arXiv:2302.09188}

\bibitem[\protect\citeauthoryear{{Frebel} \& {Norris}}{{Frebel} \&
  {Norris}}{2015}]{Frebel_2015}
{Frebel} A.,  {Norris} J.~E.,  2015, \mn@doi [\araa]
  {10.1146/annurev-astro-082214-122423}, \href
  {https://ui.adsabs.harvard.edu/abs/2015ARA&A..53..631F} {53, 631}

\bibitem[\protect\citeauthoryear{{Frebel} et~al.,}{{Frebel}
  et~al.}{2006}]{Frebel_2006}
{Frebel} A.,  et~al., 2006, \mn@doi [\apj] {10.1086/508506}, \href
  {https://ui.adsabs.harvard.edu/abs/2006ApJ...652.1585F} {652, 1585}

\bibitem[\protect\citeauthoryear{{Frebel}, {Collet}, {Eriksson}, {Christlieb}
  \& {Aoki}}{{Frebel} et~al.}{2008}]{Frebel_2008}
{Frebel} A.,  {Collet} R.,  {Eriksson} K.,  {Christlieb} N.,   {Aoki} W.,
  2008, \mn@doi [\apj] {10.1086/590327}, \href
  {https://ui.adsabs.harvard.edu/abs/2008ApJ...684..588F} {684, 588}

\bibitem[\protect\citeauthoryear{{Frischknecht}, {Hirschi}  \&
  {Thielemann}}{{Frischknecht} et~al.}{2012}]{Frischknecht_2012}
{Frischknecht} U.,  {Hirschi} R.,   {Thielemann} F.~K.,  2012, \mn@doi [\aap]
  {10.1051/0004-6361/201117794}, \href
  {https://ui.adsabs.harvard.edu/abs/2012A&A...538L...2F} {538, L2}

\bibitem[\protect\citeauthoryear{{Gaia Collaboration} et~al.,}{{Gaia
  Collaboration} et~al.}{2016}]{Gaia_2016b}
{Gaia Collaboration} et~al., 2016, \mn@doi [\aap]
  {10.1051/0004-6361/201629272}, \href
  {https://ui.adsabs.harvard.edu/abs/2016A&A...595A...1G} {595, A1}

\bibitem[\protect\citeauthoryear{{Gaia Collaboration} et~al.,}{{Gaia
  Collaboration} et~al.}{2022}]{Gaia_2022K}
{Gaia Collaboration} et~al., 2022, \mn@doi [arXiv e-prints]
  {10.48550/arXiv.2208.00211}, \href
  {https://ui.adsabs.harvard.edu/abs/2022arXiv220800211G} {p. arXiv:2208.00211}

\bibitem[\protect\citeauthoryear{{Girardi}, {Bressan}, {Bertelli}  \&
  {Chiosi}}{{Girardi} et~al.}{2000}]{Girardi_2000}
{Girardi} L.,  {Bressan} A.,  {Bertelli} G.,   {Chiosi} C.,  2000, \mn@doi
  [\aaps] {10.1051/aas:2000126}, \href
  {https://ui.adsabs.harvard.edu/abs/2000A&AS..141..371G} {141, 371}

\bibitem[\protect\citeauthoryear{{Goswami}, {Rathour}  \& {Goswami}}{{Goswami}
  et~al.}{2021}]{Goswami_2021}
{Goswami} P.~P.,  {Rathour} R.~S.,   {Goswami} A.,  2021, \mn@doi [\aap]
  {10.1051/0004-6361/202038258}, \href
  {https://ui.adsabs.harvard.edu/abs/2021A&A...649A..49G} {649, A49}

\bibitem[\protect\citeauthoryear{{Gratton}, {Sneden}, {Carretta}  \&
  {Bragaglia}}{{Gratton} et~al.}{2000}]{Gratton_2000}
{Gratton} R.~G.,  {Sneden} C.,  {Carretta} E.,   {Bragaglia} A.,  2000, \aap,
  \href {https://ui.adsabs.harvard.edu/abs/2000A&A...354..169G} {354, 169}

\bibitem[\protect\citeauthoryear{{Hansen} et~al.,}{{Hansen}
  et~al.}{2015}]{Hansen_2015}
{Hansen} T.,  et~al., 2015, \mn@doi [\apj] {10.1088/0004-637X/807/2/173}, \href
  {https://ui.adsabs.harvard.edu/abs/2015ApJ...807..173H} {807, 173}

\bibitem[\protect\citeauthoryear{{Hansen}, {Andersen}, {Nordstr{\"o}m},
  {Beers}, {Placco}, {Yoon}  \& {Buchhave}}{{Hansen}
  et~al.}{2016a}]{Hansen_2016a}
{Hansen} T.~T.,  {Andersen} J.,  {Nordstr{\"o}m} B.,  {Beers} T.~C.,  {Placco}
  V.~M.,  {Yoon} J.,   {Buchhave} L.~A.,  2016a, \mn@doi [\aap]
  {10.1051/0004-6361/201527235}, \href
  {https://ui.adsabs.harvard.edu/abs/2016A&A...586A.160H} {586, A160}

\bibitem[\protect\citeauthoryear{{Hansen}, {Andersen}, {Nordstr{\"o}m},
  {Beers}, {Placco}, {Yoon}  \& {Buchhave}}{{Hansen}
  et~al.}{2016b}]{Hansen_2016b}
{Hansen} T.~T.,  {Andersen} J.,  {Nordstr{\"o}m} B.,  {Beers} T.~C.,  {Placco}
  V.~M.,  {Yoon} J.,   {Buchhave} L.~A.,  2016b, \mn@doi [\aap]
  {10.1051/0004-6361/201527409}, \href
  {https://ui.adsabs.harvard.edu/abs/2016A&A...588A...3H} {588, A3}

\bibitem[\protect\citeauthoryear{{Hansen} et~al.,}{{Hansen}
  et~al.}{2016c}]{Hansen_2016c}
{Hansen} C.~J.,  et~al., 2016c, \mn@doi [\aap] {10.1051/0004-6361/201526895},
  \href {https://ui.adsabs.harvard.edu/abs/2016A&A...588A..37H} {588, A37}

\bibitem[\protect\citeauthoryear{{Hansen}, {Hansen}, {Koch}, {Beers},
  {Nordstr{\"o}m}, {Placco}  \& {Andersen}}{{Hansen}
  et~al.}{2019}]{Hansen_2019}
{Hansen} C.~J.,  {Hansen} T.~T.,  {Koch} A.,  {Beers} T.~C.,  {Nordstr{\"o}m}
  B.,  {Placco} V.~M.,   {Andersen} J.,  2019, \mn@doi [\aap]
  {10.1051/0004-6361/201834601}, \href
  {https://ui.adsabs.harvard.edu/abs/2019A&A...623A.128H} {623, A128}

\bibitem[\protect\citeauthoryear{{Hartwig}, {Bromm}, {Klessen}  \&
  {Glover}}{{Hartwig} et~al.}{2015}]{Hartwig_2015}
{Hartwig} T.,  {Bromm} V.,  {Klessen} R.~S.,   {Glover} S. C.~O.,  2015,
  \mn@doi [\mnras] {10.1093/mnras/stu2740}, \href
  {https://ui.adsabs.harvard.edu/abs/2015MNRAS.447.3892H} {447, 3892}

\bibitem[\protect\citeauthoryear{{Hartwig} et~al.,}{{Hartwig}
  et~al.}{2018}]{Hartwig_2018}
{Hartwig} T.,  et~al., 2018, \mn@doi [\mnras] {10.1093/mnras/sty1176}, \href
  {https://ui.adsabs.harvard.edu/abs/2018MNRAS.478.1795H} {478, 1795}

\bibitem[\protect\citeauthoryear{{Hayek}, {Asplund}, {Collet}  \&
  {Nordlund}}{{Hayek} et~al.}{2011}]{Hayek_2011}
{Hayek} W.,  {Asplund} M.,  {Collet} R.,   {Nordlund} {\r{A}}.,  2011, \mn@doi
  [\aap] {10.1051/0004-6361/201015782}, \href
  {https://ui.adsabs.harvard.edu/abs/2011A&A...529A.158H} {529, A158}

\bibitem[\protect\citeauthoryear{{Heger} \& {Woosley}}{{Heger} \&
  {Woosley}}{2010}]{Heger_2010}
{Heger} A.,  {Woosley} S.~E.,  2010, \mn@doi [\apj]
  {10.1088/0004-637X/724/1/341}, \href
  {https://ui.adsabs.harvard.edu/abs/2010ApJ...724..341H} {724, 341}

\bibitem[\protect\citeauthoryear{{Helmi}}{{Helmi}}{2020}]{Helmi_2020}
{Helmi} A.,  2020, \mn@doi [\araa] {10.1146/annurev-astro-032620-021917}, \href
  {https://ui.adsabs.harvard.edu/abs/2020ARA&A..58..205H} {58, 205}

\bibitem[\protect\citeauthoryear{{Helmi}, {Babusiaux}, {Koppelman}, {Massari},
  {Veljanoski}  \& {Brown}}{{Helmi} et~al.}{2018}]{Helmi_2018}
{Helmi} A.,  {Babusiaux} C.,  {Koppelman} H.~H.,  {Massari} D.,  {Veljanoski}
  J.,   {Brown} A. G.~A.,  2018, \mn@doi [\nat] {10.1038/s41586-018-0625-x},
  \href {https://ui.adsabs.harvard.edu/abs/2018Natur.563...85H} {563, 85}

\bibitem[\protect\citeauthoryear{{Holmbeck} et~al.,}{{Holmbeck}
  et~al.}{2020}]{Holmbeck_2020}
{Holmbeck} E.~M.,  et~al., 2020, \mn@doi [\apjs] {10.3847/1538-4365/ab9c19},
  \href {https://ui.adsabs.harvard.edu/abs/2020ApJS..249...30H} {249, 30}

\bibitem[\protect\citeauthoryear{{Honda}, {Aoki}, {Kajino}, {Ando}, {Beers},
  {Izumiura}, {Sadakane}  \& {Takada-Hidai}}{{Honda} et~al.}{2004}]{Honda_2004}
{Honda} S.,  {Aoki} W.,  {Kajino} T.,  {Ando} H.,  {Beers} T.~C.,  {Izumiura}
  H.,  {Sadakane} K.,   {Takada-Hidai} M.,  2004, \mn@doi [\apj]
  {10.1086/383406}, \href
  {https://ui.adsabs.harvard.edu/abs/2004ApJ...607..474H} {607, 474}

\bibitem[\protect\citeauthoryear{{Horta} et~al.,}{{Horta}
  et~al.}{2021}]{Horta_2021}
{Horta} D.,  et~al., 2021, \mn@doi [\mnras] {10.1093/mnras/staa2987}, \href
  {https://ui.adsabs.harvard.edu/abs/2021MNRAS.500.1385H} {500, 1385}

\bibitem[\protect\citeauthoryear{{Jablonka} et~al.,}{{Jablonka}
  et~al.}{2015}]{Jablonka_2015}
{Jablonka} P.,  et~al., 2015, \mn@doi [\aap] {10.1051/0004-6361/201525661},
  \href {https://ui.adsabs.harvard.edu/abs/2015A&A...583A..67J} {583, A67}

\bibitem[\protect\citeauthoryear{{Jeon}, {Besla}  \& {Bromm}}{{Jeon}
  et~al.}{2017}]{Jeon_2017}
{Jeon} M.,  {Besla} G.,   {Bromm} V.,  2017, \mn@doi [\apj]
  {10.3847/1538-4357/aa8c80}, \href
  {https://ui.adsabs.harvard.edu/abs/2017ApJ...848...85J} {848, 85}

\bibitem[\protect\citeauthoryear{{Jeon}, {Bromm}, {Besla}, {Yoon}  \&
  {Choi}}{{Jeon} et~al.}{2021}]{Jeon_2021}
{Jeon} M.,  {Bromm} V.,  {Besla} G.,  {Yoon} J.,   {Choi} Y.,  2021, \mn@doi
  [\mnras] {10.1093/mnras/staa4017}, \href
  {https://ui.adsabs.harvard.edu/abs/2021MNRAS.502....1J} {502, 1}

\bibitem[\protect\citeauthoryear{{Johnson} \& {Soderblom}}{{Johnson} \&
  {Soderblom}}{1987}]{Johnson_1987}
{Johnson} D. R.~H.,  {Soderblom} D.~R.,  1987, \mn@doi [\aj] {10.1086/114370},
  \href {https://ui.adsabs.harvard.edu/abs/1987AJ.....93..864J} {93, 864}

\bibitem[\protect\citeauthoryear{{Johnson}, {Herwig}, {Beers}  \&
  {Christlieb}}{{Johnson} et~al.}{2007}]{Johnson_2007}
{Johnson} J.~A.,  {Herwig} F.,  {Beers} T.~C.,   {Christlieb} N.,  2007,
  \mn@doi [\apj] {10.1086/510114}, \href
  {https://ui.adsabs.harvard.edu/abs/2007ApJ...658.1203J} {658, 1203}

\bibitem[\protect\citeauthoryear{{Kim}, {Lee}  \& {Beers}}{{Kim}
  et~al.}{2019}]{Kim_2019}
{Kim} Y.~K.,  {Lee} Y.~S.,   {Beers} T.~C.,  2019, \mn@doi [\apj]
  {10.3847/1538-4357/ab3660}, \href
  {https://ui.adsabs.harvard.edu/abs/2019ApJ...882..176K} {882, 176}

\bibitem[\protect\citeauthoryear{{Koch-Hansen}, {Hansen}, {Lombardo},
  {Bonifacio}, {Hanke}  \& {Caffau}}{{Koch-Hansen}
  et~al.}{2021}]{Koch-Hansen_2021}
{Koch-Hansen} A.~J.,  {Hansen} C.~J.,  {Lombardo} L.,  {Bonifacio} P.,  {Hanke}
  M.,   {Caffau} E.,  2021, \mn@doi [\aap] {10.1051/0004-6361/202039711}, \href
  {https://ui.adsabs.harvard.edu/abs/2021A&A...645A..64K} {645, A64}

\bibitem[\protect\citeauthoryear{{Komiya}, {Suda}, {Minaguchi}, {Shigeyama},
  {Aoki}  \& {Fujimoto}}{{Komiya} et~al.}{2007}]{Komiya_2007}
{Komiya} Y.,  {Suda} T.,  {Minaguchi} H.,  {Shigeyama} T.,  {Aoki} W.,
  {Fujimoto} M.~Y.,  2007, \mn@doi [\apj] {10.1086/510826}, \href
  {https://ui.adsabs.harvard.edu/abs/2007ApJ...658..367K} {658, 367}

\bibitem[\protect\citeauthoryear{{Koppelman}, {Helmi}  \&
  {Veljanoski}}{{Koppelman} et~al.}{2018}]{Koppelman_2018}
{Koppelman} H.,  {Helmi} A.,   {Veljanoski} J.,  2018, \mn@doi [\apjl]
  {10.3847/2041-8213/aac882}, \href
  {https://ui.adsabs.harvard.edu/abs/2018ApJ...860L..11K} {860, L11}

\bibitem[\protect\citeauthoryear{{Koppelman}, {Helmi}, {Massari},
  {Price-Whelan}  \& {Starkenburg}}{{Koppelman} et~al.}{2019}]{Koppelman_2019}
{Koppelman} H.~H.,  {Helmi} A.,  {Massari} D.,  {Price-Whelan} A.~M.,
  {Starkenburg} T.~K.,  2019, \mn@doi [\aap] {10.1051/0004-6361/201936738},
  \href {https://ui.adsabs.harvard.edu/abs/2019A&A...631L...9K} {631, L9}

\bibitem[\protect\citeauthoryear{{Korn}, {Grundahl}, {Richard}, {Mashonkina},
  {Barklem}, {Collet}, {Gustafsson}  \& {Piskunov}}{{Korn}
  et~al.}{2007}]{Korn_2007}
{Korn} A.~J.,  {Grundahl} F.,  {Richard} O.,  {Mashonkina} L.,  {Barklem}
  P.~S.,  {Collet} R.,  {Gustafsson} B.,   {Piskunov} N.,  2007, \mn@doi [\apj]
  {10.1086/523098}, \href
  {https://ui.adsabs.harvard.edu/abs/2007ApJ...671..402K} {671, 402}

\bibitem[\protect\citeauthoryear{{Kruijssen}, {Pfeffer}, {Reina-Campos},
  {Crain}  \& {Bastian}}{{Kruijssen} et~al.}{2019}]{Kruijssen_2019}
{Kruijssen} J.~M.~D.,  {Pfeffer} J.~L.,  {Reina-Campos} M.,  {Crain} R.~A.,
  {Bastian} N.,  2019, \mn@doi [\mnras] {10.1093/mnras/sty1609}, \href
  {https://ui.adsabs.harvard.edu/abs/2019MNRAS.486.3180K} {486, 3180}

\bibitem[\protect\citeauthoryear{{Kruijssen} et~al.,}{{Kruijssen}
  et~al.}{2020}]{Kruijssen_2020}
{Kruijssen} J.~M.~D.,  et~al., 2020, \mn@doi [\mnras] {10.1093/mnras/staa2452},
  \href {https://ui.adsabs.harvard.edu/abs/2020MNRAS.498.2472K} {498, 2472}

\bibitem[\protect\citeauthoryear{{Lee} et~al.,}{{Lee} et~al.}{2013}]{Lee_2013}
{Lee} Y.~S.,  et~al., 2013, \mn@doi [\aj] {10.1088/0004-6256/146/5/132}, \href
  {https://ui.adsabs.harvard.edu/abs/2013AJ....146..132L} {146, 132}

\bibitem[\protect\citeauthoryear{{Lombardo} et~al.,}{{Lombardo}
  et~al.}{2022}]{Lombardo_2022}
{Lombardo} L.,  et~al., 2022, \mn@doi [\aap] {10.1051/0004-6361/202243932},
  \href {https://ui.adsabs.harvard.edu/abs/2022A&A...665A..10L} {665, A10}

\bibitem[\protect\citeauthoryear{{Maeder} \& {Meynet}}{{Maeder} \&
  {Meynet}}{2015}]{Maeder_2015b}
{Maeder} A.,  {Meynet} G.,  2015, \mn@doi [\aap] {10.1051/0004-6361/201526234},
  \href {https://ui.adsabs.harvard.edu/abs/2015A&A...580A..32M} {580, A32}

\bibitem[\protect\citeauthoryear{{Malhan} et~al.,}{{Malhan}
  et~al.}{2022}]{Malhan_2022}
{Malhan} K.,  et~al., 2022, \mn@doi [\apj] {10.3847/1538-4357/ac4d2a}, \href
  {https://ui.adsabs.harvard.edu/abs/2022ApJ...926..107M} {926, 107}

\bibitem[\protect\citeauthoryear{{Mardini} et~al.,}{{Mardini}
  et~al.}{2022}]{Mardini_2022}
{Mardini} M.~K.,  et~al., 2022, \mn@doi [\mnras] {10.1093/mnras/stac2783},
  \href {https://ui.adsabs.harvard.edu/abs/2022MNRAS.517.3993M} {517, 3993}

\bibitem[\protect\citeauthoryear{{Mashonkina}, {Jablonka}, {Pakhomov},
  {Sitnova}  \& {North}}{{Mashonkina} et~al.}{2017}]{Mashonkina_2017}
{Mashonkina} L.,  {Jablonka} P.,  {Pakhomov} Y.,  {Sitnova} T.,   {North} P.,
  2017, \mn@doi [\aap] {10.1051/0004-6361/201730779}, \href
  {https://ui.adsabs.harvard.edu/abs/2017A&A...604A.129M} {604, A129}

\bibitem[\protect\citeauthoryear{{Matsuno}, {Aoki}  \& {Suda}}{{Matsuno}
  et~al.}{2019}]{Matsuno_2019}
{Matsuno} T.,  {Aoki} W.,   {Suda} T.,  2019, \mn@doi [\apjl]
  {10.3847/2041-8213/ab0ec0}, \href
  {https://ui.adsabs.harvard.edu/abs/2019ApJ...874L..35M} {874, L35}

\bibitem[\protect\citeauthoryear{{Meynet}, {Ekstr{\"o}m}  \& {Maeder}}{{Meynet}
  et~al.}{2006}]{Meynet_2006}
{Meynet} G.,  {Ekstr{\"o}m} S.,   {Maeder} A.,  2006, \mn@doi [\aap]
  {10.1051/0004-6361:20053070}, \href
  {https://ui.adsabs.harvard.edu/abs/2006A&A...447..623M} {447, 623}

\bibitem[\protect\citeauthoryear{{Mishenina}, {Soubiran}, {Kovtyukh}  \&
  {Korotin}}{{Mishenina} et~al.}{2004}]{Mishenina_2004}
{Mishenina} T.~V.,  {Soubiran} C.,  {Kovtyukh} V.~V.,   {Korotin} S.~A.,  2004,
  \mn@doi [\aap] {10.1051/0004-6361:20034454}, \href
  {https://ui.adsabs.harvard.edu/abs/2004A&A...418..551M} {418, 551}

\bibitem[\protect\citeauthoryear{{Miyamoto} \& {Nagai}}{{Miyamoto} \&
  {Nagai}}{1975}]{Miyamoto_1975}
{Miyamoto} M.,  {Nagai} R.,  1975, \pasj, \href
  {https://ui.adsabs.harvard.edu/abs/1975PASJ...27..533M} {27, 533}

\bibitem[\protect\citeauthoryear{{Mucciarelli} \& {Bonifacio}}{{Mucciarelli} \&
  {Bonifacio}}{2020}]{Mucciarelli_2020}
{Mucciarelli} A.,  {Bonifacio} P.,  2020, \mn@doi [\aap]
  {10.1051/0004-6361/202037703}, \href
  {https://ui.adsabs.harvard.edu/abs/2020A&A...640A..87M} {640, A87}

\bibitem[\protect\citeauthoryear{{Myeong}, {Evans}, {Belokurov}, {Sanders}  \&
  {Koposov}}{{Myeong} et~al.}{2018}]{Myeong_2018}
{Myeong} G.~C.,  {Evans} N.~W.,  {Belokurov} V.,  {Sanders} J.~L.,   {Koposov}
  S.~E.,  2018, \mn@doi [\apjl] {10.3847/2041-8213/aab613}, \href
  {https://ui.adsabs.harvard.edu/abs/2018ApJ...856L..26M} {856, L26}

\bibitem[\protect\citeauthoryear{{Myeong}, {Vasiliev}, {Iorio}, {Evans}  \&
  {Belokurov}}{{Myeong} et~al.}{2019}]{Myeong_2019}
{Myeong} G.~C.,  {Vasiliev} E.,  {Iorio} G.,  {Evans} N.~W.,   {Belokurov} V.,
  2019, \mn@doi [\mnras] {10.1093/mnras/stz1770}, \href
  {https://ui.adsabs.harvard.edu/abs/2019MNRAS.488.1235M} {488, 1235}

\bibitem[\protect\citeauthoryear{{Naidu}, {Conroy}, {Bonaca}, {Johnson},
  {Ting}, {Caldwell}, {Zaritsky}  \& {Cargile}}{{Naidu}
  et~al.}{2020}]{Naidu_2020}
{Naidu} R.~P.,  {Conroy} C.,  {Bonaca} A.,  {Johnson} B.~D.,  {Ting} Y.-S.,
  {Caldwell} N.,  {Zaritsky} D.,   {Cargile} P.~A.,  2020, \mn@doi [\apj]
  {10.3847/1538-4357/abaef4}, \href
  {https://ui.adsabs.harvard.edu/abs/2020ApJ...901...48N} {901, 48}

\bibitem[\protect\citeauthoryear{{Naidu} et~al.,}{{Naidu}
  et~al.}{2022}]{Naidu_2022}
{Naidu} R.~P.,  et~al., 2022, \mn@doi [\apjl] {10.3847/2041-8213/ac5589}, \href
  {https://ui.adsabs.harvard.edu/abs/2022ApJ...926L..36N} {926, L36}

\bibitem[\protect\citeauthoryear{{Navarro}, {Frenk}  \& {White}}{{Navarro}
  et~al.}{1996}]{Navarro_1996}
{Navarro} J.~F.,  {Frenk} C.~S.,   {White} S. D.~M.,  1996, \mn@doi [\apj]
  {10.1086/177173}, \href
  {https://ui.adsabs.harvard.edu/abs/1996ApJ...462..563N} {462, 563}

\bibitem[\protect\citeauthoryear{{Nomoto}, {Kobayashi}  \& {Tominaga}}{{Nomoto}
  et~al.}{2013}]{Nomoto_2013}
{Nomoto} K.,  {Kobayashi} C.,   {Tominaga} N.,  2013, \mn@doi [\araa]
  {10.1146/annurev-astro-082812-140956}, \href
  {https://ui.adsabs.harvard.edu/abs/2013ARA&A..51..457N} {51, 457}

\bibitem[\protect\citeauthoryear{{Norris} \& {Yong}}{{Norris} \&
  {Yong}}{2019}]{Noris_2019}
{Norris} J.~E.,  {Yong} D.,  2019, \mn@doi [\apj] {10.3847/1538-4357/ab1f84},
  \href {https://ui.adsabs.harvard.edu/abs/2019ApJ...879...37N} {879, 37}

\bibitem[\protect\citeauthoryear{{Norris} et~al.,}{{Norris}
  et~al.}{2013}]{Norris_2013b}
{Norris} J.~E.,  et~al., 2013, \mn@doi [\apj] {10.1088/0004-637X/762/1/25},
  \href {https://ui.adsabs.harvard.edu/abs/2013ApJ...762...25N} {762, 25}

\bibitem[\protect\citeauthoryear{{Pfeffer}, {Lardo}, {Bastian}, {Saracino}  \&
  {Kamann}}{{Pfeffer} et~al.}{2021}]{Pfeffer_2021}
{Pfeffer} J.,  {Lardo} C.,  {Bastian} N.,  {Saracino} S.,   {Kamann} S.,  2021,
  \mn@doi [\mnras] {10.1093/mnras/staa3407}, \href
  {https://ui.adsabs.harvard.edu/abs/2021MNRAS.500.2514P} {500, 2514}

\bibitem[\protect\citeauthoryear{{Pillepich} et~al.,}{{Pillepich}
  et~al.}{2014}]{Phillepich_2014}
{Pillepich} A.,  et~al., 2014, \mn@doi [\mnras] {10.1093/mnras/stu1408}, \href
  {https://ui.adsabs.harvard.edu/abs/2014MNRAS.444..237P} {444, 237}

\bibitem[\protect\citeauthoryear{{Placco}, {Frebel}, {Beers}  \&
  {Stancliffe}}{{Placco} et~al.}{2014}]{Placco_2014}
{Placco} V.~M.,  {Frebel} A.,  {Beers} T.~C.,   {Stancliffe} R.~J.,  2014,
  \mn@doi [\apj] {10.1088/0004-637X/797/1/21}, \href
  {https://ui.adsabs.harvard.edu/abs/2014ApJ...797...21P} {797, 21}

\bibitem[\protect\citeauthoryear{{Placco} et~al.,}{{Placco}
  et~al.}{2016}]{Placco_2016}
{Placco} V.~M.,  et~al., 2016, \mn@doi [\apj] {10.3847/0004-637X/833/1/21},
  \href {https://ui.adsabs.harvard.edu/abs/2016ApJ...833...21P} {833, 21}

\bibitem[\protect\citeauthoryear{{Placco}, {Sneden}, {Roederer}, {Lawler}, {Den
  Hartog}, {Hejazi}, {Maas}  \& {Bernath}}{{Placco} et~al.}{2021}]{Placco_2021}
{Placco} V.~M.,  {Sneden} C.,  {Roederer} I.~U.,  {Lawler} J.~E.,  {Den Hartog}
  E.~A.,  {Hejazi} N.,  {Maas} Z.,   {Bernath} P.,  2021, \mn@doi [Research
  Notes of the American Astronomical Society] {10.3847/2515-5172/abf651}, \href
  {https://ui.adsabs.harvard.edu/abs/2021RNAAS...5...92P} {5, 92}

\bibitem[\protect\citeauthoryear{{Pols}, {Izzard}, {Stancliffe}  \&
  {Glebbeek}}{{Pols} et~al.}{2012}]{Pols_2012}
{Pols} O.~R.,  {Izzard} R.~G.,  {Stancliffe} R.~J.,   {Glebbeek} E.,  2012,
  \mn@doi [\aap] {10.1051/0004-6361/201219597}, \href
  {https://ui.adsabs.harvard.edu/abs/2012A&A...547A..76P} {547, A76}

\bibitem[\protect\citeauthoryear{{Purandardas} \& {Goswami}}{{Purandardas} \&
  {Goswami}}{2021a}]{Purandardas_2021a}
{Purandardas} M.,  {Goswami} A.,  2021a, \mn@doi [\apj]
  {10.3847/1538-4357/abec45}, \href
  {https://ui.adsabs.harvard.edu/abs/2021ApJ...912...74P} {912, 74}

\bibitem[\protect\citeauthoryear{{Purandardas} \& {Goswami}}{{Purandardas} \&
  {Goswami}}{2021b}]{Purandardas_2021b}
{Purandardas} M.,  {Goswami} A.,  2021b, \mn@doi [\apj]
  {10.3847/1538-4357/ac1d4d}, \href
  {https://ui.adsabs.harvard.edu/abs/2021ApJ...922...28P} {922, 28}

\bibitem[\protect\citeauthoryear{{Purandardas}, {Goswami}, {Goswami},
  {Shejeelammal}  \& {Masseron}}{{Purandardas} et~al.}{2019}]{Purandardas_2019}
{Purandardas} M.,  {Goswami} A.,  {Goswami} P.~P.,  {Shejeelammal} J.,
  {Masseron} T.,  2019, \mn@doi [\mnras] {10.1093/mnras/stz759}, \href
  {https://ui.adsabs.harvard.edu/abs/2019MNRAS.486.3266P} {486, 3266}

\bibitem[\protect\citeauthoryear{{Purandardas}, {Goswami}, {Shejeelammal},
  {Sonamben}, {Pawar}, {Mkrtichian}, {Doddamani}  \& {Joshi}}{{Purandardas}
  et~al.}{2022}]{Purandardas_2022}
{Purandardas} M.,  {Goswami} A.,  {Shejeelammal} J.,  {Sonamben} M.,  {Pawar}
  G.,  {Mkrtichian} D.,  {Doddamani} V.~H.,   {Joshi} S.,  2022, \mn@doi
  [\mnras] {10.1093/mnras/stac1169}, \href
  {https://ui.adsabs.harvard.edu/abs/2022MNRAS.513.4696P} {513, 4696}

\bibitem[\protect\citeauthoryear{{Raskin} et~al.,}{{Raskin}
  et~al.}{2011}]{Raskin_2011}
{Raskin} G.,  et~al., 2011, \mn@doi [\aap] {10.1051/0004-6361/201015435}, \href
  {https://ui.adsabs.harvard.edu/abs/2011A&A...526A..69R} {526, A69}

\bibitem[\protect\citeauthoryear{{Reddy}, {Lambert}  \& {Allende
  Prieto}}{{Reddy} et~al.}{2006}]{Reddy_2006}
{Reddy} B.~E.,  {Lambert} D.~L.,   {Allende Prieto} C.,  2006, \mn@doi [\mnras]
  {10.1111/j.1365-2966.2006.10148.x}, \href
  {https://ui.adsabs.harvard.edu/abs/2006MNRAS.367.1329R} {367, 1329}

\bibitem[\protect\citeauthoryear{{Roederer}, {Preston}, {Thompson}, {Shectman},
  {Sneden}, {Burley}  \& {Kelson}}{{Roederer} et~al.}{2014}]{Roederer_2014}
{Roederer} I.~U.,  {Preston} G.~W.,  {Thompson} I.~B.,  {Shectman} S.~A.,
  {Sneden} C.,  {Burley} G.~S.,   {Kelson} D.~D.,  2014, \mn@doi [\aj]
  {10.1088/0004-6256/147/6/136}, \href
  {https://ui.adsabs.harvard.edu/abs/2014AJ....147..136R} {147, 136}

\bibitem[\protect\citeauthoryear{{Roederer}, {Sneden}, {Lawler}, {Sobeck},
  {Cowan}  \& {Boesgaard}}{{Roederer} et~al.}{2018}]{Roderer_2018}
{Roederer} I.~U.,  {Sneden} C.,  {Lawler} J.~E.,  {Sobeck} J.~S.,  {Cowan}
  J.~J.,   {Boesgaard} A.~M.,  2018, \mn@doi [\apj] {10.3847/1538-4357/aac6df},
  \href {https://ui.adsabs.harvard.edu/abs/2018ApJ...860..125R} {860, 125}

\bibitem[\protect\citeauthoryear{{Salvadori} \& {Ferrara}}{{Salvadori} \&
  {Ferrara}}{2009}]{Salvadori_2009}
{Salvadori} S.,  {Ferrara} A.,  2009, \mn@doi [\mnras]
  {10.1111/j.1745-3933.2009.00627.x}, \href
  {https://ui.adsabs.harvard.edu/abs/2009MNRAS.395L...6S} {395, L6}

\bibitem[\protect\citeauthoryear{{Salvadori}, {Schneider}  \&
  {Ferrara}}{{Salvadori} et~al.}{2007}]{Salvadori_2007}
{Salvadori} S.,  {Schneider} R.,   {Ferrara} A.,  2007, \mn@doi [\mnras]
  {10.1111/j.1365-2966.2007.12133.x}, \href
  {https://ui.adsabs.harvard.edu/abs/2007MNRAS.381..647S} {381, 647}

\bibitem[\protect\citeauthoryear{{Salvadori}, {Sk{\'u}lad{\'o}ttir}  \&
  {Tolstoy}}{{Salvadori} et~al.}{2015}]{Salvadori_2015}
{Salvadori} S.,  {Sk{\'u}lad{\'o}ttir} {\'A}.,   {Tolstoy} E.,  2015, \mn@doi
  [\mnras] {10.1093/mnras/stv1969}, \href
  {https://ui.adsabs.harvard.edu/abs/2015MNRAS.454.1320S} {454, 1320}

\bibitem[\protect\citeauthoryear{{Sch{\"o}nrich}, {Binney}  \&
  {Dehnen}}{{Sch{\"o}nrich} et~al.}{2010}]{Schonrich_2010}
{Sch{\"o}nrich} R.,  {Binney} J.,   {Dehnen} W.,  2010, \mn@doi [\mnras]
  {10.1111/j.1365-2966.2010.16253.x}, \href
  {https://ui.adsabs.harvard.edu/abs/2010MNRAS.403.1829S} {403, 1829}

\bibitem[\protect\citeauthoryear{{Shejeelammal} \& {Goswami}}{{Shejeelammal} \&
  {Goswami}}{2021}]{Shejeelammal_2021b}
{Shejeelammal} J.,  {Goswami} A.,  2021, \mn@doi [\apj]
  {10.3847/1538-4357/ac1ac9}, \href
  {https://ui.adsabs.harvard.edu/abs/2021ApJ...921...77S} {921, 77}

\bibitem[\protect\citeauthoryear{{Shejeelammal} \& {Goswami}}{{Shejeelammal} \&
  {Goswami}}{2022}]{Shejeelammal_2022}
{Shejeelammal} J.,  {Goswami} A.,  2022, \mn@doi [\apj]
  {10.3847/1538-4357/ac7aac}, \href
  {https://ui.adsabs.harvard.edu/abs/2022ApJ...934..110S} {934, 110}

\bibitem[\protect\citeauthoryear{{Shejeelammal}, {Goswami}, {Goswami},
  {Rathour}  \& {Masseron}}{{Shejeelammal} et~al.}{2020}]{Shejeelammal_2020}
{Shejeelammal} J.,  {Goswami} A.,  {Goswami} P.~P.,  {Rathour} R.~S.,
  {Masseron} T.,  2020, \mn@doi [\mnras] {10.1093/mnras/stz3518}, \href
  {https://ui.adsabs.harvard.edu/abs/2020MNRAS.492.3708S} {492, 3708}

\bibitem[\protect\citeauthoryear{{Shejeelammal}, {Goswami}  \&
  {Shi}}{{Shejeelammal} et~al.}{2021}]{Shejeelammal_2021a}
{Shejeelammal} J.,  {Goswami} A.,   {Shi} J.,  2021, \mn@doi [\mnras]
  {10.1093/mnras/staa3892}, \href
  {https://ui.adsabs.harvard.edu/abs/2021MNRAS.502.1008S} {502, 1008}

\bibitem[\protect\citeauthoryear{{Simpson} \& {Martell}}{{Simpson} \&
  {Martell}}{2019}]{Simpson_2019}
{Simpson} J.~D.,  {Martell} S.~L.,  2019, \mn@doi [\mnras]
  {10.1093/mnras/stz2611}, \href
  {https://ui.adsabs.harvard.edu/abs/2019MNRAS.490..741S} {490, 741}

\bibitem[\protect\citeauthoryear{{Sneden}}{{Sneden}}{1973}]{Sneden_1973}
{Sneden} C.~A.,  1973, PhD thesis, THE UNIVERSITY OF TEXAS AT AUSTIN.

\bibitem[\protect\citeauthoryear{{Spite} et~al.,}{{Spite}
  et~al.}{2005}]{Spite_2005}
{Spite} M.,  et~al., 2005, \mn@doi [\aap] {10.1051/0004-6361:20041274}, \href
  {https://ui.adsabs.harvard.edu/abs/2005A&A...430..655S} {430, 655}

\bibitem[\protect\citeauthoryear{{Spite} et~al.,}{{Spite}
  et~al.}{2006}]{Spite_2006}
{Spite} M.,  et~al., 2006, \mn@doi [\aap] {10.1051/0004-6361:20065209}, \href
  {https://ui.adsabs.harvard.edu/abs/2006A&A...455..291S} {455, 291}

\bibitem[\protect\citeauthoryear{{Spite}, {Caffau}, {Bonifacio}, {Spite},
  {Ludwig}, {Plez}  \& {Christlieb}}{{Spite} et~al.}{2013}]{Spite_2013}
{Spite} M.,  {Caffau} E.,  {Bonifacio} P.,  {Spite} F.,  {Ludwig} H.~G.,
  {Plez} B.,   {Christlieb} N.,  2013, \mn@doi [\aap]
  {10.1051/0004-6361/201220989}, \href
  {https://ui.adsabs.harvard.edu/abs/2013A&A...552A.107S} {552, A107}

\bibitem[\protect\citeauthoryear{{Spite}, {Spite}, {Fran{\c{c}}ois},
  {Bonifacio}, {Caffau}  \& {Salvadori}}{{Spite} et~al.}{2018}]{Spite_2018}
{Spite} M.,  {Spite} F.,  {Fran{\c{c}}ois} P.,  {Bonifacio} P.,  {Caffau} E.,
  {Salvadori} S.,  2018, \mn@doi [\aap] {10.1051/0004-6361/201833548}, \href
  {https://ui.adsabs.harvard.edu/abs/2018A&A...617A..56S} {617, A56}

\bibitem[\protect\citeauthoryear{{Starkenburg}, {Shetrone}, {McConnachie}  \&
  {Venn}}{{Starkenburg} et~al.}{2014}]{Starkenburg_2014}
{Starkenburg} E.,  {Shetrone} M.~D.,  {McConnachie} A.~W.,   {Venn} K.~A.,
  2014, \mn@doi [\mnras] {10.1093/mnras/stu623}, \href
  {https://ui.adsabs.harvard.edu/abs/2014MNRAS.441.1217S} {441, 1217}

\bibitem[\protect\citeauthoryear{{Starkenburg}, {Oman}, {Navarro}, {Crain},
  {Fattahi}, {Frenk}, {Sawala}  \& {Schaye}}{{Starkenburg}
  et~al.}{2017}]{Starkenburg_2017}
{Starkenburg} E.,  {Oman} K.~A.,  {Navarro} J.~F.,  {Crain} R.~A.,  {Fattahi}
  A.,  {Frenk} C.~S.,  {Sawala} T.,   {Schaye} J.,  2017, \mn@doi [\mnras]
  {10.1093/mnras/stw2873}, \href
  {https://ui.adsabs.harvard.edu/abs/2017MNRAS.465.2212S} {465, 2212}

\bibitem[\protect\citeauthoryear{{Tafelmeyer} et~al.,}{{Tafelmeyer}
  et~al.}{2010}]{Tafelmeyer_2010}
{Tafelmeyer} M.,  et~al., 2010, \mn@doi [\aap] {10.1051/0004-6361/201014733},
  \href {https://ui.adsabs.harvard.edu/abs/2010A&A...524A..58T} {524, A58}

\bibitem[\protect\citeauthoryear{{Tominaga}, {Iwamoto}  \& {Nomoto}}{{Tominaga}
  et~al.}{2014}]{Tominaga_2014}
{Tominaga} N.,  {Iwamoto} N.,   {Nomoto} K.,  2014, \mn@doi [\apj]
  {10.1088/0004-637X/785/2/98}, \href
  {https://ui.adsabs.harvard.edu/abs/2014ApJ...785...98T} {785, 98}

\bibitem[\protect\citeauthoryear{{Umeda} \& {Nomoto}}{{Umeda} \&
  {Nomoto}}{2003}]{Umeda_2003}
{Umeda} H.,  {Nomoto} K.,  2003, \mn@doi [\nat] {10.1038/nature01571}, \href
  {https://ui.adsabs.harvard.edu/abs/2003Natur.422..871U} {422, 871}

\bibitem[\protect\citeauthoryear{{Umeda} \& {Nomoto}}{{Umeda} \&
  {Nomoto}}{2005}]{Umeda_2005}
{Umeda} H.,  {Nomoto} K.,  2005, \mn@doi [\apj] {10.1086/426097}, \href
  {https://ui.adsabs.harvard.edu/abs/2005ApJ...619..427U} {619, 427}

\bibitem[\protect\citeauthoryear{{Venn}, {Irwin}, {Shetrone}, {Tout}, {Hill}
  \& {Tolstoy}}{{Venn} et~al.}{2004}]{Venn_2004}
{Venn} K.~A.,  {Irwin} M.,  {Shetrone} M.~D.,  {Tout} C.~A.,  {Hill} V.,
  {Tolstoy} E.,  2004, \mn@doi [\aj] {10.1086/422734}, \href
  {https://ui.adsabs.harvard.edu/abs/2004AJ....128.1177V} {128, 1177}

\bibitem[\protect\citeauthoryear{{Vincenzo}, {Matteucci}, {Vattakunnel}  \&
  {Lanfranchi}}{{Vincenzo} et~al.}{2014}]{Vincenzo_2014}
{Vincenzo} F.,  {Matteucci} F.,  {Vattakunnel} S.,   {Lanfranchi} G.~A.,  2014,
  \mn@doi [\mnras] {10.1093/mnras/stu710}, \href
  {https://ui.adsabs.harvard.edu/abs/2014MNRAS.441.2815V} {441, 2815}

\bibitem[\protect\citeauthoryear{{Wang}, {Han}, {Cautun}, {Li}  \&
  {Ishigaki}}{{Wang} et~al.}{2020}]{Wang_2020}
{Wang} W.,  {Han} J.,  {Cautun} M.,  {Li} Z.,   {Ishigaki} M.~N.,  2020,
  \mn@doi [Science China Physics, Mechanics, and Astronomy]
  {10.1007/s11433-019-1541-6}, \href
  {https://ui.adsabs.harvard.edu/abs/2020SCPMA..6309801W} {63, 109801}

\bibitem[\protect\citeauthoryear{{Watkins}, {van der Marel}, {Sohn}  \&
  {Evans}}{{Watkins} et~al.}{2019}]{Watkins_2019}
{Watkins} L.~L.,  {van der Marel} R.~P.,  {Sohn} S.~T.,   {Evans} N.~W.,  2019,
  \mn@doi [\apj] {10.3847/1538-4357/ab089f}, \href
  {https://ui.adsabs.harvard.edu/abs/2019ApJ...873..118W} {873, 118}

\bibitem[\protect\citeauthoryear{{Woody} \& {Schlaufman}}{{Woody} \&
  {Schlaufman}}{2021}]{Woody_2021}
{Woody} T.,  {Schlaufman} K.~C.,  2021, \mn@doi [\aj]
  {10.3847/1538-3881/abff5f}, \href
  {https://ui.adsabs.harvard.edu/abs/2021AJ....162...42W} {162, 42}

\bibitem[\protect\citeauthoryear{{Yong} et~al.,}{{Yong}
  et~al.}{2013}]{Yong_2013}
{Yong} D.,  et~al., 2013, \mn@doi [\apj] {10.1088/0004-637X/762/1/26}, \href
  {https://ui.adsabs.harvard.edu/abs/2013ApJ...762...26Y} {762, 26}

\bibitem[\protect\citeauthoryear{{Yoon} et~al.,}{{Yoon}
  et~al.}{2016}]{Yoon_2016}
{Yoon} J.,  et~al., 2016, \mn@doi [\apj] {10.3847/0004-637X/833/1/20}, \href
  {https://ui.adsabs.harvard.edu/abs/2016ApJ...833...20Y} {833, 20}

\bibitem[\protect\citeauthoryear{{Yoon} et~al.,}{{Yoon}
  et~al.}{2018}]{Yoon_2018}
{Yoon} J.,  et~al., 2018, \mn@doi [\apj] {10.3847/1538-4357/aaccea}, \href
  {https://ui.adsabs.harvard.edu/abs/2018ApJ...861..146Y} {861, 146}

\bibitem[\protect\citeauthoryear{{Yoon}, {Beers}, {Tian}  \& {Whitten}}{{Yoon}
  et~al.}{2019}]{Yoon_2019}
{Yoon} J.,  {Beers} T.~C.,  {Tian} D.,   {Whitten} D.~D.,  2019, \mn@doi [\apj]
  {10.3847/1538-4357/ab1ead}, \href
  {https://ui.adsabs.harvard.edu/abs/2019ApJ...878...97Y} {878, 97}

\bibitem[\protect\citeauthoryear{{Yoon}, {Whitten}, {Beers}, {Lee}, {Masseron}
  \& {Placco}}{{Yoon} et~al.}{2020}]{Yoon_2020}
{Yoon} J.,  {Whitten} D.~D.,  {Beers} T.~C.,  {Lee} Y.~S.,  {Masseron} T.,
  {Placco} V.~M.,  2020, \mn@doi [\apj] {10.3847/1538-4357/ab7daf}, \href
  {https://ui.adsabs.harvard.edu/abs/2020ApJ...894....7Y} {894, 7}

\bibitem[\protect\citeauthoryear{{Zepeda} et~al.,}{{Zepeda}
  et~al.}{2023}]{Zepeda_2023}
{Zepeda} J.,  et~al., 2023, \mn@doi [\apj] {10.3847/1538-4357/acbbcc}, \href
  {https://ui.adsabs.harvard.edu/abs/2023ApJ...947...23Z} {947, 23}

\bibitem[\protect\citeauthoryear{{de Bennassuti}, {Salvadori}, {Schneider},
  {Valiante}  \& {Omukai}}{{de Bennassuti} et~al.}{2017}]{Bennassuti_2017}
{de Bennassuti} M.,  {Salvadori} S.,  {Schneider} R.,  {Valiante} R.,
  {Omukai} K.,  2017, \mn@doi [\mnras] {10.1093/mnras/stw2687}, \href
  {https://ui.adsabs.harvard.edu/abs/2017MNRAS.465..926D} {465, 926}

\makeatother
\end{thebibliography}

\appendix

\section{Line list} \label{section_linelist}
The lines used to derive the elemental abundances are listed in Tables \ref{linelist1} - \ref{linelist2}. 
{\footnotesize
\begin{table*}
\caption{Equivalent widths (in m\r{A}) of Fe lines used for deriving 
atmospheric parameters.} \label{linelist1}
\resizebox{0.5\textwidth}{!}{\begin{tabular}{lccccc}
\hline                       
Wavelength(\r{A}) & El & $E_{low}$(eV) & log gf & HE~0038$-$0345 & HE~1243$-$2408    \\ 
\hline 
4132.9		& Fe I  &	2.84	&	$-$0.920	&	28.5(4.59)  &	-      	\\
4153.90     &       &   3.40    &   $-$0.280    &   -           &   18.4(4.26)          \\
4337.05		&		&	1.56	&	$-$1.7		&	-       	&	63.4(4.63)		   			\\
4489.74 	&		&	0.12	&	$-$3.90		&	40.9(4.54)	&	-				\\
4547.847	&		&	3.546	&	$-$0.820	&	-	    	&	8.2(4.51)					\\
5005.711	&		&	3.883	&	$-$0.120	&	17.5(4.59) 	&	-   	\\
5006.119	&		&	2.833	&	$-$0.61		&	-          	&	38.1(4.33)	         	\\
5151.911    &       &   1.010   &   $-$3.320    &   -           &   22.2(4.45)                      \\ 
5198.711	&		&	2.222	&	$-$2.135	&	-	    	&	14.8(4.55)		    			\\
5215.179	&		&	3.266	&	$-$0.933	&	-		    &	14.2(4.48)		   		\\
5226.862	&		&	3.038	&	$-$0.667	&   -       	&	28.5(4.28)		\\
5263.31		&		&	3.260	&	$-$0.870	&	-		    &	13.4(4.42)	  	\\
5339.93		&		&	3.27	&	$-$0.680	&	27.3(4.64)	&	-	    	  	\\
5364.870    &       &   4.44    &   0.220       &   -           &   10.4(4.56)      \\
5393.17		&		&	3.24	&	$-$0.720	&	24.8(4.64)  &	-		    		\\
5569.62		&		&	3.42	&	$-$0.490	&	-		    &	22.5(4.52)		   		\\
5586.76     &       &   3.370   &   $-$0.110    &   36.8(4.45)  &   -              \\ 
6137.694	&		&	2.588	&	$-$1.403	&	-	    	&	23.1(4.39)		    			\\
6219.28     &       &   2.200   &   $-$2.450    &   -           &   8.8(4.46)       \\
6252.555    &       &   2.402   &   $-$1.690    &   -           &   23.0(4.45)                     \\ 
6393.60     &       &   2.430   &   $-$1.620    &   30.7(4.65)  &   -           \\
6430.85		&		&	2.18	&	$-$2.010	&	29.1(4.63)  &	22.4(4.40)			\\
4233.16     & Fe II &   2.580   &   $-$2.020    &   -           &   32.8(4.42)      \\
4416.820    &       &   2.780   &   $-$2.570    &   11.2(4.61)  &   -               \\
4491.405	&   	&	2.855	&	$-$2.700	&	6.6(4.56)	&	-		  		\\
4508.288	&		&	2.855	&	$-$2.210	&	-       	&	16.1(4.59)	   			\\
4520.224	&		&	2.81	&	$-$2.600	&	8.4(4.56)	&	7.4(4.35)	    		\\
\hline
\end{tabular}}

The numbers in the parenthesis in columns 5 \& 6 give the derived absolute abundances from the respective line. 
  
\end{table*}
}

{\footnotesize
\begin{table*}
\caption{Equivalent widths (in m\r{A}) of lines used for deriving elemental abundances.} \label{linelist2}
\resizebox{0.5\textwidth}{!}{\begin{tabular}{lccccc}
\hline                       
Wavelength(\r{A}) & El & $E_{low}$(eV) & log gf & HE~0038$-$0345 & HE~1243$-$2408   \\ 
\hline 
5682.633	& Na I	&	2.102	&	$-$0.700	&	8.3(4.20)	&	-	 \\
4702.991	& Mg I	&	4.346	&	$-$0.666	&	44.3(5.26)	&	53.8(5.48)			\\
5528.405	&		&	4.346	&	$-$0.620	&	57.0(5.44)  &	50.8(5.30)  	\\
5948.541	& Si I	&	5.083	&	$-$1.23		&	6.9(5.41)   &	-			\\
4098.530	& Ca I	&	2.530	&	$-$0.520	&   12.9(4.07) 	&   -		\\
4283.011	&    	&	1.886	&	$-$0.224	&	-       	&	28.8(3.54)			\\
4455.890	&		&	1.899	&	$-$0.540	&	-		    &	23.9(3.73)			\\
5594.462	&		&	2.523	&	$-$0.050	&	23.3(3.70)	&	-	    		\\
6102.723	&		&	1.879	&	$-$0.890	&	-		    &	16.9(3.66)		\\
6162.173	&		&	1.899	&	0.1		    &	57.6(4.08)  &	-		        \\
6439.075	&   	&	2.525	&	0.47		&	43.4(3.90)  &	31.1(3.57)		\\
6493.781	&		&	2.521	&	0.14		&	17.8(3.70)	&   19.2(3.49)			\\
4533.239	& Ti I	&	0.848	&	0.476		&	-	        &	32.3(2.20)	    	\\
4999.50     &       &   0.820   &   0.170       &   29.3(2.40)  &   -             \\
5007.21		&		&	0.82	&	0.17		&	33.1(2.57)  &	33.3(2.50)	 	\\
5210.386	&		&	0.047	&	$-$0.884	&	-	    	&	22.1(2.32)		\\
4470.86		& Ti II	&	1.16	&	$-$2.28		&	-	    	&	20.1(2.35)			\\
4493.51		&		&	1.08	&	$-$2.73		&	15.3(2.70) 	&	-		   	\\
5185.9		&		&	1.89	&	$-$1.35		&	21.7(2.43)	&	-		\\
5226.543	&		&	1.566	&	$-$1.300	&	47.9(2.64)	&	33.8(2.12)		\\
5336.77     &       &   1.58    &   $-$1.700    &   -           &   29.9(2.43)   \\
5381.015	&		&	1.566	&	$-$2.08		&	20.7(2.56)  &	26.7(2.73)		 	\\
4289.73		& Cr I	&	0.00	&	$-$0.37		&	78.2(2.97)  &	67.1(2.48)			\\
5206.04     &       &   0.94    &   0.020       &   -           &   51.4(2.46)      \\
5345.801	&		&	1.003	&	$-$0.980	&	-		    &	12.8(2.39)		\\
5409.772	&		&	1.03	&	$-$0.720	&	30.3(2.69)  &   26.5(2.56)		\\
4855.41     &  Ni I &   3.54    &   0.000       &   9.7(3.46)  &   -                 \\
4904.41     &       &   3.54    &   $-$0.170    &   -           &   3.1(3.08)                \\
5084.08     &       &   3.68    &   0.030       &   -           &   -              \\
5115.39     &       &   3.83    &   $-$0.110    &   -           &   2.0(3.15)   \\
5146.48		&		&	3.706	&	0.12		&	7.5(3.38)	&	-		\\
4722.15		& Zn I	&	4.029	&	$-$0.370	&	-		    &	8.1(1.88)			\\
4810.53		&		&	4.08	&	$-$0.170	&   17.9(2.20)	&	-			\\
\hline
\end{tabular}}

The numbers in the parenthesis in columns 5 \& 6 give the derived absolute abundances from the respective line. \\
\end{table*}
}
\end{document}